\documentclass{article} % For LaTeX2e
\usepackage{iclr2020_conference,times}

\usepackage{hyperref}
\usepackage{url}
\usepackage{amsmath,amsfonts,bm}
\usepackage{graphicx}
\usepackage{comment}
\usepackage[bottom]{footmisc}
\usepackage{gensymb}
\usepackage{lscape}
\usepackage{booktabs}
\usepackage[para]{threeparttable}
\usepackage[super]{nth}

\title{Non-linear interlinkages and key objectives amongst the Paris Agreement and the Sustainable Development Goals}

\author{Felix Laumann \\
Department of Mathematics \\
Imperial College London \\
\texttt{fjl1218@ic.ac.uk}
\And
Julius von K{\"u}gelgen \\
Max Planck Institute for Intelligent Systems, T\"ubingen \\
Department of Engineering, University of Cambridge
\And 
Mauricio Barahona \\
Department of Mathematics \\
Imperial College London \\
}

\newcommand{\z}{\mathbf{z}}
\newcommand{\x}{\mathbf{x}}
\newcommand{\y}{\mathbf{y}}

\iclrfinalcopy % Uncomment for camera-ready version, but NOT for submission.
\begin{document}

\maketitle

\begin{abstract}
The United Nations' ambitions to combat climate change and prosper human development are manifested in the Paris Agreement and the Sustainable Development Goals (SDGs), respectively. These are inherently inter-linked as progress towards some of these objectives may accelerate or hinder progress towards others. We investigate how these two agendas influence each other by defining networks of 18 nodes, consisting of the 17 SDGs and climate change, for various groupings of countries. We compute a non-linear measure of conditional dependence, the partial distance correlation, given any subset of the remaining 16 variables. These correlations are treated as weights on edges, and weighted eigenvector centralities are calculated to determine the most important nodes. 

We find that SDG 6, \textit{clean water and sanitation}, and SDG 4, \textit{quality education}, are most central across nearly all groupings of countries. In developing regions, SDG 17, \textit{partnerships for the goals}, is strongly connected to the progress of other objectives in the two agendas whilst, somewhat surprisingly, SDG 8, \textit{decent work and economic growth}, is not as important in terms of eigenvector centrality.
\end{abstract}

\vspace{-0.1cm}
\section{Inter-linked human and natural worlds}
\vspace{-0.1cm}
The state-of-the-art in sustainability is described by two United Nations (UN) landmark agendas, the Paris Agreement \citep{UNParisAgreement} and the Sustainable Development Goals (SDGs) \citep{UNSDG}.
Whilst the former focuses on preventing a global climate crisis with far reaching consequences by limiting global warming to 1.5 to 2\degree C above pre-industrial levels, the purpose of the latter is to end poverty, protect the planet and ensure that all people enjoy peace and prosperity by 2030. 
Any action for the progress on either agenda often has an influence on the other \citep{UNCopenhagen}, reflecting the complexity of the human and natural worlds. 

This inter-linked nature gives rise to opportunities for the creation of synergistic interventions: civil, corporate and institutional actions can efficiently create impact across both agendas, thereby improving the world profoundly. 
On the other hand, this inter-linked construct can also be subject to trade-offs between objectives, i.e., progress towards one agenda constrains progress towards the other.
In this work, we aim to discover how climate change, as measured by local temperature rises, and the 17 SDGs are inter-linked by learning the structure of undirected graphs over these variables from their (conditional) dependencies. 
% Ultimately, we strive to inform principled policy making to maximise impact across both agendas via a better understanding of these interactions, and our work constitutes a pre-requisite for capturing such causal influences.

Adding climate change as an \nth{18} variable is motivated by the observation that temperature rises (or any other direct metrics of climate change) are not actually tracked within SDG 13 (\textit{climate action}). Indicators of SDG 13  only track inputs (such as investment), means (such as plans and strategies), and impacts (number of people affected by disasters), but they do not account for outputs, such as changes in temperature or green house gas emissions.\footnote{Only recently (and after performing the present analysis) have "total greenhouse gas emissions" been added as an output-quantifying indicator (13.2.2).}

We use distance correlation \citep{szekely2007measuring} as a measure of non-linear dependence between variables of possibly varying dimensions. To account for possible interactions, each pair of variables is conditioned on any subset of the remaining variables, and the minimum resulting distance correlation is taken as the weight on an edge between these two variables. Subsequently, the weighted eigenvector centrality of every node is calculated to measure its importance within the network.

In summary, the contributions of this paper include: first, the application of a \textit{non-linear} measure of (conditional) dependence to SDG data, thereby relaxing the linearity assumption on the nature of interlinkages between the SDGs, compared to the work of \citet{lusseau2019income}; and secondly, the use of eigenvector centrality as a \textit{relative} measure which also takes the importance of a node's neighbours into account, as opposed to simple degree centrality as used by \citet{mcgowan2019imperfect}.
% Doing so, we measure of importance not only on the strength of dependence to direct neighbours, how degree centrality does, but additionally consider the importances of these neighbours.
% thus enhancing interpretability and applicability to environmental and socio-economic policy recommendations.
\vspace{-0.1cm}
\section{Methodology}
\vspace{-0.1cm}
We use data provided by the World Bank (\citeyear{WBdata}) and the UN (\citeyear{UNdata}) in form of time-series for various indicators, which measure progress towards their associated SDGs, in conjunction with temperature recordings \citep{WBclimatedata}.\footnote{For detailed descriptions of indicators, see \url{https://sustainabledevelopment.un.org/sdgs}} In total, these three sources provide 379 time-series, which are available on a country-level with annual measurements from 2000 to 2016\footnote{We impute missing values (especially for the time 2000-2005) using a weighted average across countries (where data is available) with weights inversely proportional to the Euclidean distance between indicators.}. Apart from measurements for the 17 SDGs, we introduce climate change as an additional variable which we define by annual average temperature per country. We consider these 18 variables as the set of nodes $\boldsymbol{V}$ of an undirected graph $\mathcal{G}$. We learn the graph structure by computing partial distance covariances \citep{szekely2014partial} between any pair $(X, Y)$ of nodes, given any subset $\mathbf{Z} \subseteq \boldsymbol{V} \setminus (X, Y)$ of the remaining 16 nodes. This yields a sparsely-connected undirected graph with weighted edges $\boldsymbol{E}$ capturing non-linear dependencies between variables. Using these weights, we compute weighted eigenvector centralities \citep[p.159; Appendix \ref{eigvect}]{newman2018networks} to find the most important nodes. Code to reproduce our findings and visualisations of networks may be found online at \url{https://github.com/felix-laumann/SDG-dataset}.
 \vspace{-0.2cm}
\subsection{Distance covariance}
\vspace{-0.1cm}
Let $X \in \mathbb{R}^{d_X}$ and $Y \in \mathbb{R}^{d_Y}$ be two random vectors with finite first moments, i.e., $\mathbb{E}[X],\mathbb{E}[Y] < \infty$.
The \textit{distance covariance} between $X$ and $Y$, denoted by $\mathcal{V}^2(X, Y)$, is a measure of dependence between $X$ and $Y$ with the following important properties: 
(i) $\mathcal{V}^2(X, Y) \geq 0$, with equality if and only if $X$ and $Y$ are independent, i.e., it is a non-parametric measure that---unlike, e.g., standard correlation---is able to pick up complex non-linear dependencies;
(ii) $\mathcal{V}^2(X, Y) = \mathcal{V}^2(Y, X)$, i.e., it is symmetric;
and (iii) unlike many other dependence measures $\mathcal{V}^2(X, Y)$ is well-defined even for $d_X \neq d_Y$.
This last point makes it particularly useful for our setting where, due to the different numbers of indicators per SDG, dimensionality varies considerably between variables. 

Formally, the distance covariance between $X$ and $Y$ is defined as
\begin{equation}
    \mathcal{V}^2 (X, Y) = \| f_{X, Y}(t, s) - f_{X}(t) f_{Y}(s) \|^2  := \int | f_{X, Y}(t, s) - f_{X}(t) f_{Y}(s) |^2 w(t,s)dt \ ds
\end{equation}
where $w(t,s) := (|t|_{d_X}^{1+d_X} |s|_{d_Y}^{1+d_Y})^{-1}$, and where the \textit{characteristic function}  $f$ of a random variable $Z$ is denoted as $f_{Z} (t) = \mathbb{E}[e^{itZ}]$ with $i^2 = -1$.

 The corresponding \textit{distance correlation} $ \mathcal{R}^2$ is the normalised distance covariance, computed by 
\begin{equation}
    \mathcal{R}^2(X, Y) = \begin{cases}
    \frac{\mathcal{V}^2 (X, Y)}{\sqrt{\mathcal{V}^2 (X, X)\mathcal{V}^2 (Y, Y)}}, &\text{\quad if \quad $\mathcal{V}^2 (X, X)\mathcal{V}^2 (Y, Y) > 0$} \\
    0, &\text{\quad if \quad $\mathcal{V}^2 (X, X)\mathcal{V}^2 (Y, Y) = 0$}.
    \end{cases}
\end{equation}
Properties of $\mathcal{R}^2$ include: (i) $0 \leq  \mathcal{R}^2 (X, Y) \leq 1$; and (ii) $\mathcal{R}^2 (X, Y) = 1$ if and only if there exists a vector $\boldsymbol{a}$, a non-zero real number $b$, and an orthogonal matrix $\mathcal{C}$ such that $Y = \boldsymbol{a} + b \mathcal{C} X$.

Since $\mathcal{V}^2(X, Y)$ and therefore $\mathcal{R}^2(X, Y)$ are defined in terms of the underlying joint distribution of $(X,Y)$  which is usually not known, we require a way to estimate them from data. Definitions of biased and unbiased estimators, referred to as $A$ and $\tilde{A}$, can be found in Appendix \ref{biasedest} and \ref{unbiasedest}.
\vspace{-0.2cm}
\subsection{Partial distance covariance}
\vspace{-0.1cm}
As we deal with graphs of 18 nodes, any pairwise covariance may occur through the remaining 16 nodes. Thus, we condition any pair of nodes $(X, Y)$ on any subset $\mathbf{Z} \subseteq \boldsymbol{V} \setminus (X, Y)$ of the remaining 16 nodes. The pairwise distances $c_{ij} = \|Z_i - Z_j \|$ and the distance matrix $C_{ij}$ for $\mathbf{Z}$ are computed equivalently to $A_{ij}$ and $B_{ij}$ for $X$ and $Y$ as explained in Appendix \ref{biasedest}. 
For any number $n \geq 4$ of samples $\{(x_i, y_i, \mathbf{z}_i)\}_{i=1}^n$ from $(X,Y,\mathbf{Z})$, we define a Hilbert space $\mathcal{H}_n$ over distance matrices computed on these $n$ points, with inner products $\langle \cdot, \cdot \rangle$ as defined in Appendix \ref{unbiasedest} \citep{szekely2014partial}.
With this, we can compute partial distance covariances for random vectors of varying dimensions as follows. 

Let $\tilde{A}(\x)$, $\tilde{B}(\y)$ and $\tilde{C}(\z)$ be elements of the Hilbert space $\mathcal{H}_n$ corresponding to the distance matrices computed using the samples $\x = (x_1, ...,x_n)$, $\y=(y_1,...,y_n)$, and $\mathbf{z}=(\z_1, ..., \z_n)$, respectively. The projection $P_\z(\x)$ of $\tilde{A}(\x)$ onto $\tilde{C}(\z)$ and the complementary orthogonal projection $P_{\z^{\bot}} (\x)$ are defined by
\begin{equation}
    P_\z (\x) := \frac{\langle \tilde{A}(\x), \tilde{C}(\z) \rangle}{\langle \tilde{C}(\z), \tilde{C}(\z) \rangle} \tilde{C}(\z), \ \ \ \text{and} \ \ \ 
    P_{\z^{\bot}} (\x) := \tilde{A}(\x) - P_\z (\x) = \tilde{A}(\x) - \frac{\langle \tilde{A}(\x), \tilde{C}(\z) \rangle}{\langle \tilde{C}(\z), \tilde{C}(\z) \rangle} \tilde{C}(\z),
\end{equation}
respectively. The \textit{sample partial distance covariance} is then defined as
\begin{equation}
    \textstyle
    \mathcal{V}^2_n(X,Y \ | \ \mathbf{Z}) = \langle P_{\z^{\bot}} (\x), P_{\z^{\bot}} (\y) \rangle =  \frac{1}{n(n-3)} \sum_{i \neq j}^n \big( P_{\z^{\bot}} (\x) \big)_{ij} \ \big( P_{\z^{\bot}} (\y) \big)_{ij}.
\end{equation}
Finally, we can normalise these covariances to arrive at the \textit{sample partial distance correlations}
\begin{equation}
    \mathcal{R}^2_n(X,Y \ | \ \mathbf{Z}) = \begin{cases}
    \frac{\langle P_{\z^{\bot}} (\x), P_{\z^{\bot}} (\y) \rangle}{\| P_{\z^{\bot}} (\x) \| \ \| P_{\z^{\bot}} (\y) \|}, &\text{if} \ \| P_{\z^{\bot}} (\x) \| \ \| P_{\z^{\bot}} (\y) \| \neq 0 \\
    0, &\text{if} \ \| P_{\z^{\bot}} (\x) \| \ \| P_{\z^{\bot}} (\y) \| = 0
    \end{cases},
\end{equation}
which serve as weights on edges between any two nodes.
\vspace{-0.1cm}
\section{Results}
\vspace{-0.1cm}
We apply this methodology to the data set of the aforementioned 379 indicators for various groupings of countries, for which countries are assumed to be independent samples. This assumption allows us to see the indicators' non-stationary time-series as $d$-dimensional probability distributions, where $d = \# \text{indicators} \times \# \text{years}$. Whilst we only describe the networks of a few groupings in this section, we would like to refer to Appendix \ref{appendixB} for results on all groupings. 

\begin{table}[t]
\caption{Comparison of eigenvector centralities between the Global South and the Global North (\textit{left}), and between Western Asia and Northern Europe (\textit{right}). Results for all groupings can be found in \ref{app_central}.}
\centering
\vspace{0.2cm}
\renewcommand{\arraystretch}{1.2}
\scalebox{0.9}{
\begin{tabular}{ l l l l l l l l }
\toprule
 \multicolumn{2}{c}{Global South} &
 \multicolumn{2}{c}{Global North} \\ 
 \midrule
 SDG 6 & 0.48 & SDG 6 & 0.43 \\  
 SDG 4 & 0.42 & SDG 4 & 0.40 \\
 SDG 7 & 0.38 & SDG 9 & 0.33 \\ 
 SDG 17 & 0.37 & SDG 3 & 0.32 \\
 SDG 3 & 0.26 & SDG 17 & 0.29 \\
 SDG 15 & 0.25 & SDG 7 & 0.27 \\
 \bottomrule
 \end{tabular}
 \quad
 \begin{tabular}{ l l l l }
 \toprule
  \multicolumn{2}{c}{Western Asia} & 
  \multicolumn{2}{c}{Northern Europe} \\ 
 \midrule
 SDG 6 & 0.48 & SDG 4 & 0.38 \\  
 SDG 4 & 0.36 & SDG 3 & 0.35 \\
 SDG 17 & 0.34 & SDG 6 & 0.30 \\ 
 SDG 3 & 0.33 & SDG 16 & 0.30 \\
 SDG 16 & 0.32 & SDG 7 & 0.29 \\
 SDG 7 & 0.26 & SDG 9 & 0.28 \\
 \bottomrule
 \end{tabular}}
\label{tab:Eigencentr}
\end{table} 

\begin{figure}[t]
\centering
\begin{minipage}{.47\textwidth}
  \centering
  \includegraphics[width=\linewidth]{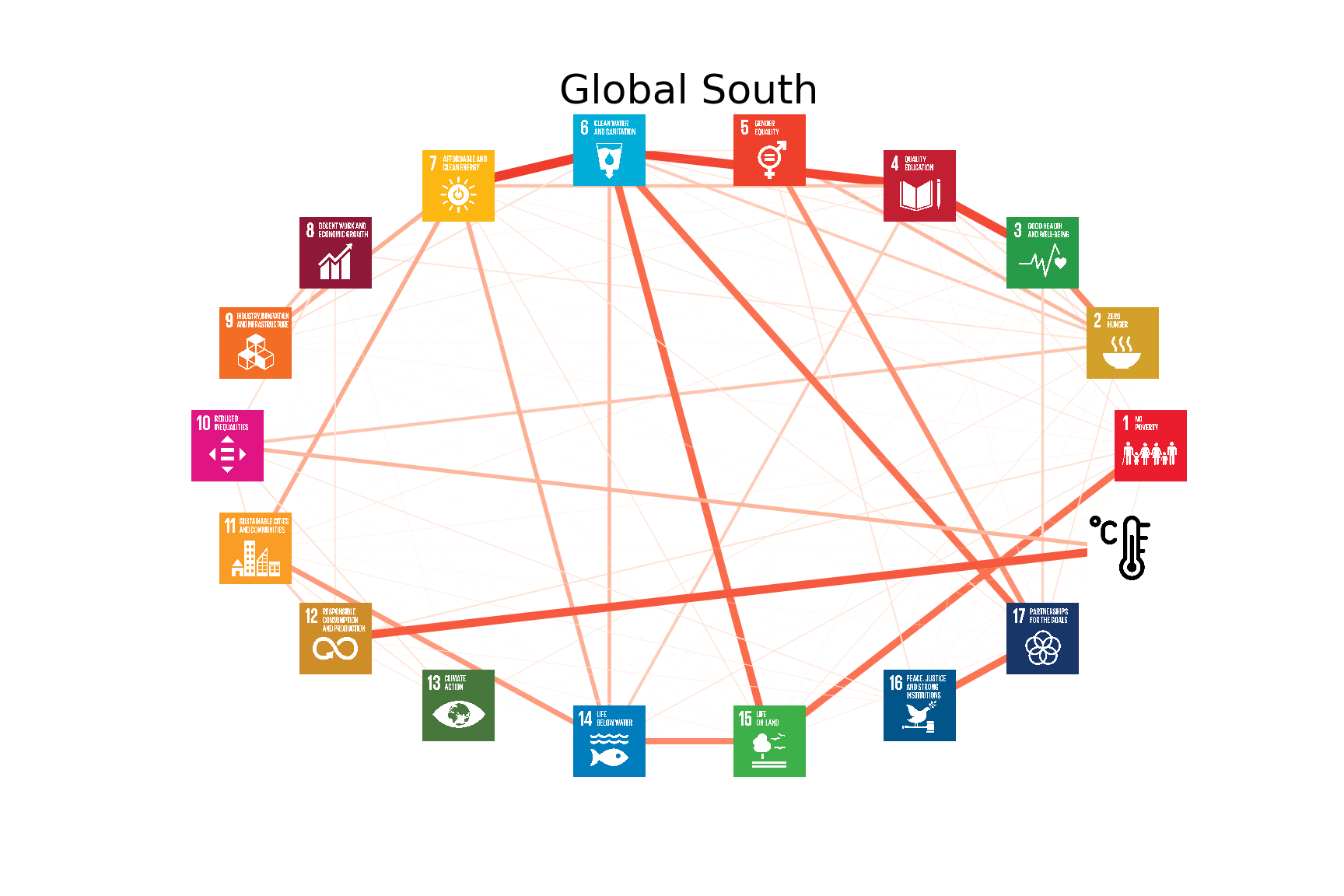}
\end{minipage}
\begin{minipage}{.47\textwidth}
  \centering
  \includegraphics[width=\linewidth]{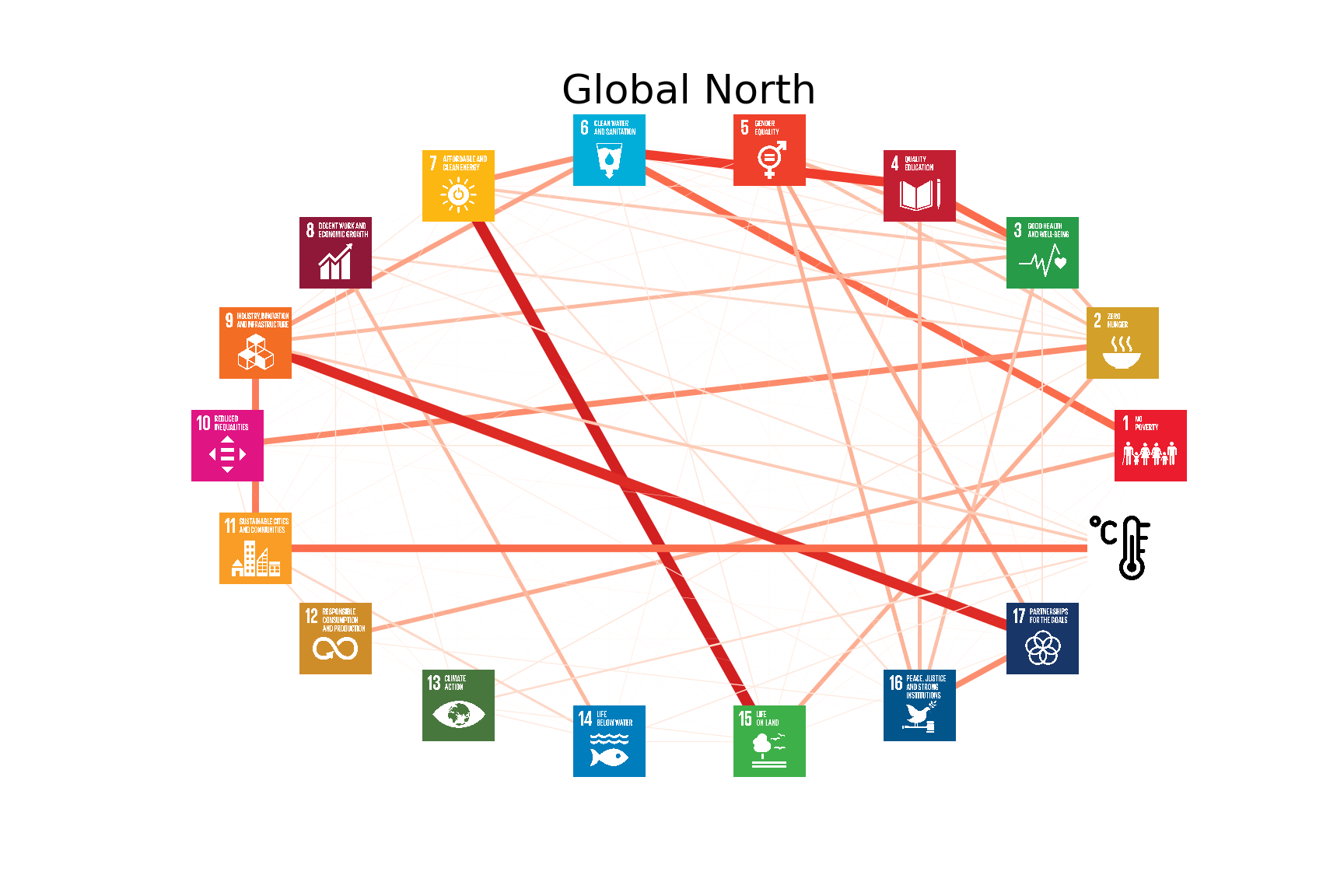}
\end{minipage}
\begin{minipage}{.04\textwidth}
  \centering
  \includegraphics[width=\linewidth]{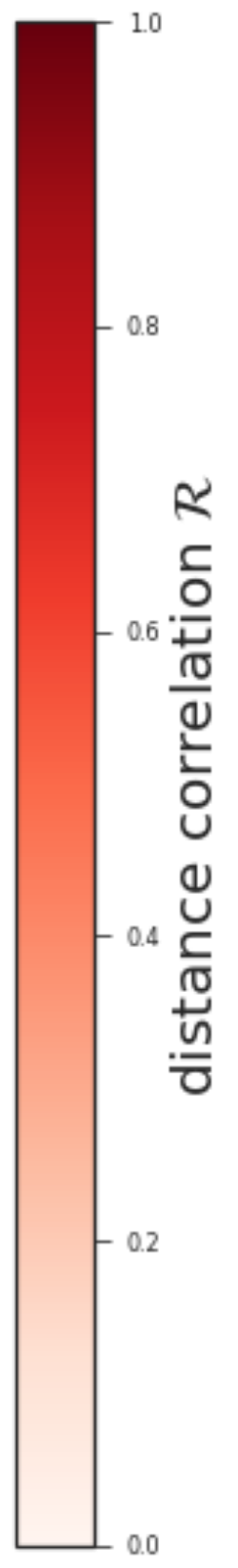}
\end{minipage}
\caption{Networks with weighted edges of (\textit{left}) the Global South and (\textit{right}) the Global North. The minimum partial distance correlations between the two adjacent nodes $X$ and $Y$, given any subset $\mathbf{Z} \subseteq \boldsymbol{V} \setminus (X, Y)$ are weights on edges.}
\label{fig:GlobalSouthNorth}
\end{figure}

Firstly, we compare the Global South and the Global North (see Figure \ref{fig:GlobalSouthNorth}). The accompanied eigenvector centralities are shown in Table \ref{tab:Eigencentr}.
In both groupings, SDG 6, \textit{clean water and sanitation}, followed closely by SDG 4, \textit{quality education}, are the most central objectives of the 18 variables.
In the Global South, temperature rises are more strongly dependent on variables than in the Global North, which broadly aligns with \citet{king2018inequality} who find that geographical areas in the Global South are more vulnerable to climate change than regions in the Global North. Further, SDG 1, \textit{no poverty}, is strongly linked to SDG 14, \textit{life below water}, in the Global South. This may be explained by the dependence of small island developing states (SIDS)---all of which  lie in the Global South---on marine life to provide for their citizens' living. 

Contrarily, the Global North strongly depends on SDG 9, \textit{industry, innovation and infrastructure}, to maintain its citizens' high levels of living standards and to further progress towards other SDGs, as well as climate change mitigation and adaptation. Moreover, SDG 7, \textit{clean and affordable energy}, is closely related to SDG 15, \textit{life on land}, which could result from the increasing area of biodiverse land populated by wind turbines, solar panels, or water dams (e.g., \citeauthor{hernandez2015solar}, \citeyear{hernandez2015solar}).

\begin{figure}[b]
\centering
\begin{minipage}{.47\textwidth}
  \centering
  \includegraphics[width=\linewidth]{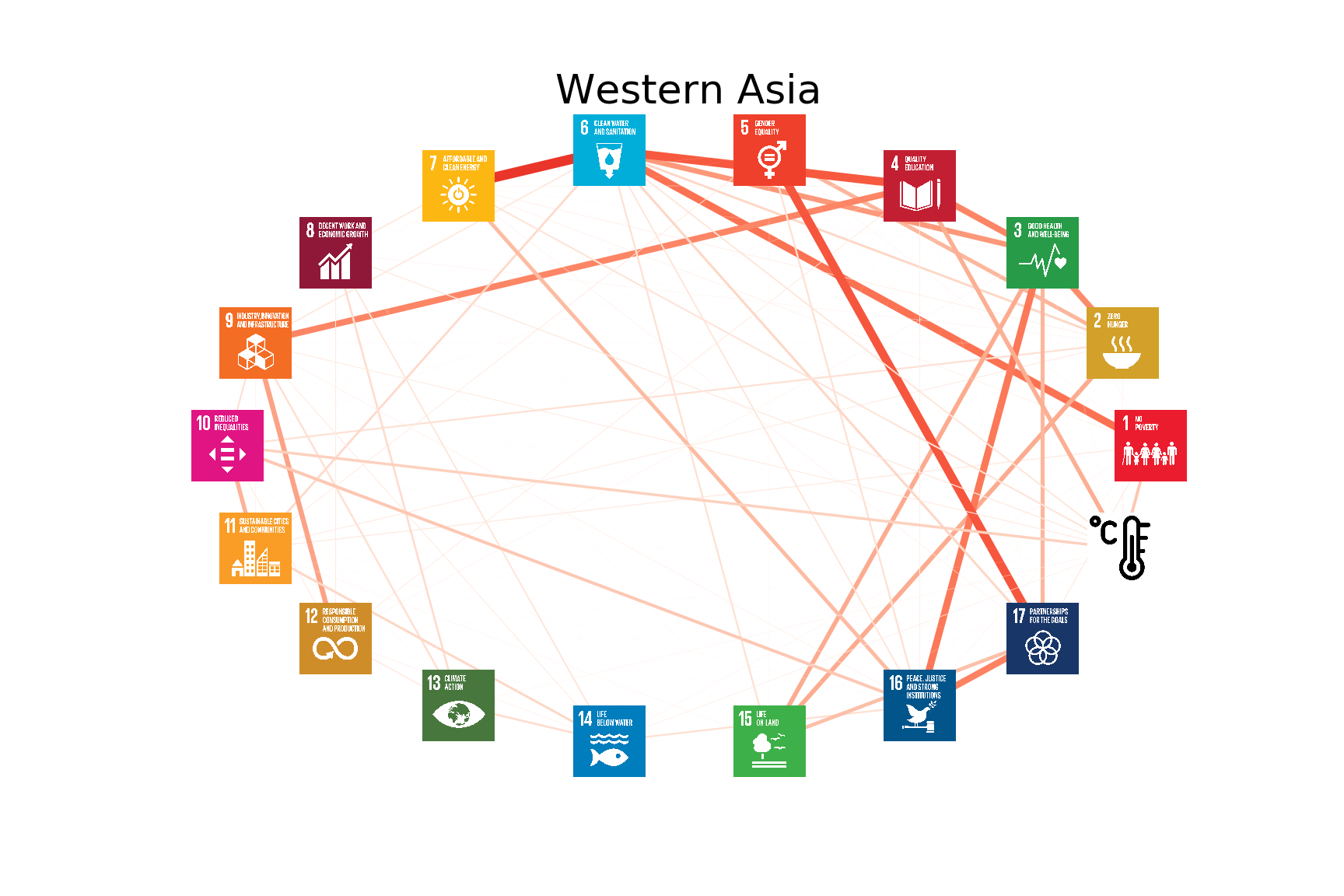}
\end{minipage}
\begin{minipage}{.47\textwidth}
  \centering
  \includegraphics[width=\linewidth]{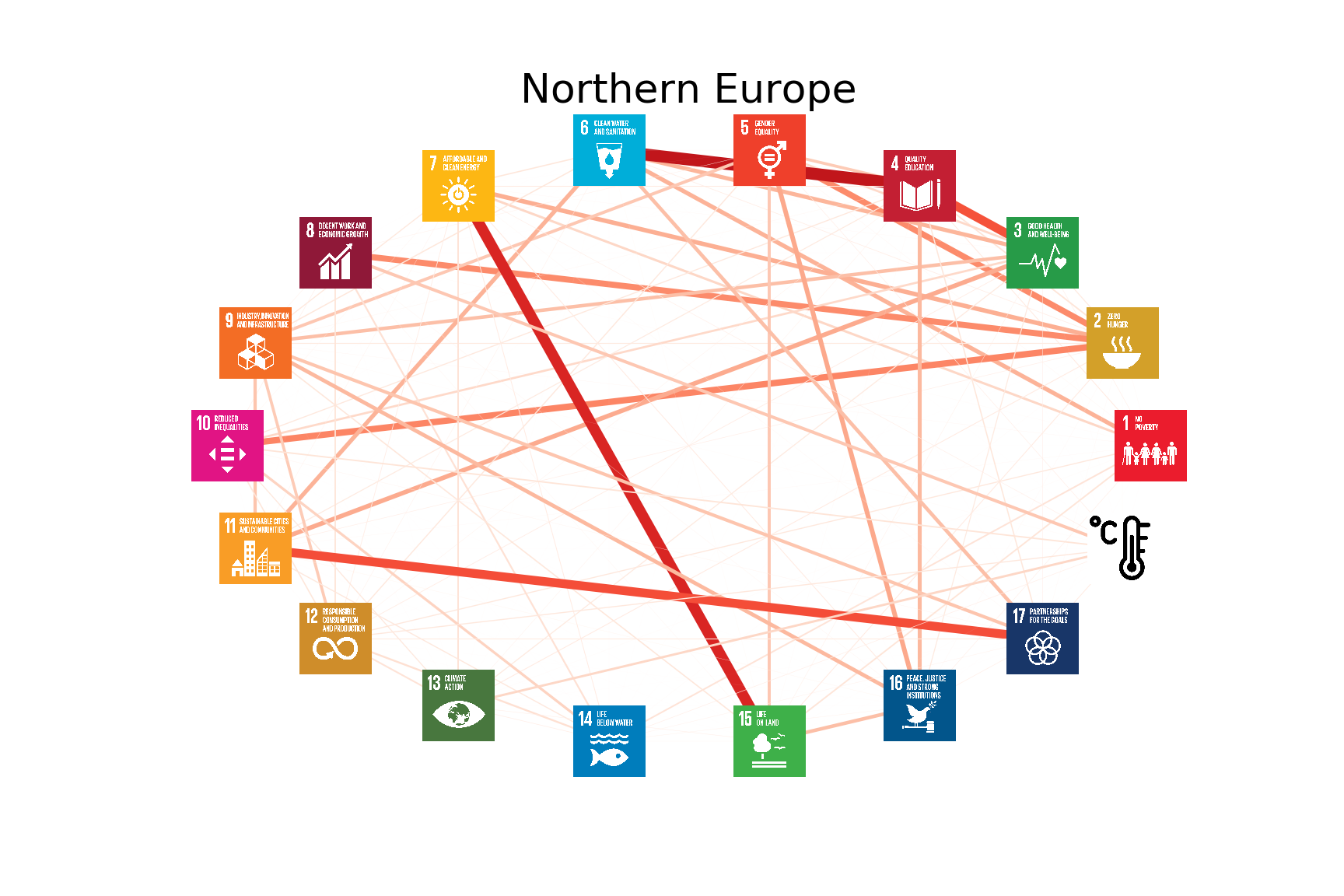}
\end{minipage}
\begin{minipage}{.04\textwidth}
  \centering
  \includegraphics[width=\linewidth]{networks/index.png}
\end{minipage}
\caption{Networks with weighted edges of (\textit{left}) Western Asia and (\textit{right}) Northern Europe. The minimum partial distance correlations between the two adjacent nodes $X$ and $Y$, given any subset $\mathbf{Z} \subseteq \boldsymbol{V} \setminus (X, Y)$ are weights on edges.}
\label{fig:WestAsiaNorthEurope}
\end{figure}

Next, we compare two geographical regions, Western Asia and Northern Europe, shown in Figure \ref{fig:WestAsiaNorthEurope} with accompanied eigenvector centralities in Table \ref{tab:Eigencentr}.
In Western Asia, SDG 6 together with SDG 4 are again the two most central nodes, but SDG 16, \textit{peace, justice and strong institutions}, is also important, likely to be associated with the unstable political circumstances in this area during the period of recorded measurements. Additionally, SDG 5, \textit{gender equality}, is strongly linked to SDG 17, \textit{partnerships for the goals}, which coincides with the remarkably low percentage of women in managerial positions in Western Asia.\footnote{In Saudi Arabia, for example,   only 5 to 9\% of managerial positions were held by women from 2000 to 2015, whereas this number fluctuated between 32 and 36\% in the United Kingdom in the same period (\citet{UNdata}, indicator 5.5.2)} 

In contrast, Northern Europe does not see a remarkable difference between the centralities of SDGs 6 and 4 to all others, but finds SDGs 4, 6, 3, and 17 with almost equivalently high centralities. As in the Global North, \textit{industry, innovation and infrastructure} are of particular importance to progress towards the SDGs, and we fine that \textit{clean and affordable energy} is closely linked to \textit{life on land}.

We note, however, that most edges found in our network analysis are not statistically significant at $p=0.05$, using the test of \cite{szekely2014partial}. This is likely linked to the high dimensionality of the data and the short recording period.
The present work is thus only a first step, and further analysis is needed to better understand non-linear interlinkages between the SDGs and climate change. 

% conduct statistical significance tests according to \cite{szekely2014partial} and find that nearly none of the edges in any network rejects the null hypothesis of partial independence, likely to be caused by the high dimensionality and short time period data are recorded for.
\vspace{-0.1cm}
\section{Conclusions}
\vspace{-0.1cm}
We report findings of our work in progress towards discovering dependencies amongst the Sustainable Development Goals (SDGs) and climate change.
As a first step, we compute partial distance correlations between the 17 SDGs and climate change, as measured by indicators associated to the SDGs and annual average temperature, respectively.
Using these measurements of non-linear dependence as edge weights in a network over these variables, we determine eigenvector centralities to unveil which variables are of particular importance, given the available data. Our results indicate that SDG 6, \textit{clean water and sanitation}, together with SDG 4, \textit{quality education}, are the most central nodes in nearly all continents and other groupings of countries. 
%However, SDG 16, peace, justice and strong institutions, becomes also important in regions which recently suffered from unstable governments or wars, such as Western Asia. 
In contrast to many contemporary policies, our preliminary results suggest that \textit{economic growth}, as measured by SDG 8, appears not to play as central of a role for sustainable development or mitigating climate change as other SDGs.
% Whilst the causal direction of edges remains to be discovered in future work, our findings may already be of interest to policy-makers to set priorities.

\bibliographystyle{iclr2020_conference}
\bibliography{iclr2020_conference}

\newpage

\appendix

\section{Appendix A} \label{appendixA}

\subsection{Distance covariance estimators}
\subsubsection{Biased estimators} \label{biasedest}
Suppose that we have access to a sample of pairs  $(x_1,y_1), ..., (x_n,y_n)\overset{\text{i.i.d}}{\sim}P_{X,Y}$. First, define the \textit{pairwise distances}: $a_{ij} := \|x_i - x_j \|$ and $b_{ij} = \|y_i - y_j \| \quad \forall i,j = 1, ..., n$. 
Next, define the corresponding \textit{distance matrices}, denoted by $(A_{ij})^n_{i,j=1}$ and $(B_{ij})^n_{i,j=1}$, as follows:
\begin{equation}
    A_{ij} = \begin{cases}
    a_{ij} - \frac{1}{n} \sum_{l=1}^n a_{il} - \frac{1}{n} \sum_{k=1}^n a_{kj} + \frac{1}{n^2} \sum_{k,l=1}^n a_{kl}, &\quad \text{if} \quad i \neq j \\
    0, & \quad \text{if} \quad i = j
    \end{cases}
\end{equation}
and
\begin{equation}
    B_{ij} = \begin{cases}
    b_{ij} - \frac{1}{n} \sum_{l=1}^n b_{il} - \frac{1}{n} \sum_{k=1}^n b_{kj} + \frac{1}{n^2} \sum_{k,l=1}^n b_{kl}, &\quad \text{if} \quad i \neq j \\
    0, &\quad \text{if} \quad i = j.
\end{cases}
\end{equation}
Having computed these, the sample distance covariance $\mathcal{V}^2_n(X,Y)$ can be estimated by
\begin{equation}
    \mathcal{V}^2_n(X,Y) = \frac{1}{n^2} \sum_{i,j=1}^n A_{ij} \ B_{ij},
\end{equation}
which converges almost surely to the population distance covariance $\mathcal{V}^2(X,Y)$ as $n \rightarrow \infty$ \citep{szekely2014partial}.

\subsubsection{Unbiased estimators} \label{unbiasedest}
Unbiased estimators of the distance covariance are denoted as $\Omega_n(x,y)$. Firstly, we must redefine our distance matrices $(A_{ij})^n_{i,j=1}$ and $(B_{ij})^n_{i,j=1}$, which we call $(\tilde{A}_{ij})^n_{i,j=1}$ and $(\tilde{B}_{ij})^n_{i,j=1}$ as
\begin{equation}
    \tilde{A}_{ij} = \begin{cases}
    a_{ij} - \frac{1}{n-2} \sum_{l=1}^n a_{il} - \frac{1}{n-2} \sum_{k=1}^n a_{kj} + \frac{1}{(n-1)(n-2)} \sum_{k,l=1}^n a_{kl}, & \text{if} \quad i \neq j; \\
    0, & \text{if} \quad i = j
    \end{cases}
\end{equation}
and
\begin{equation}
    \tilde{B}_{ij} = \begin{cases}
    b_{ij} - \frac{1}{n-2} \sum_{l=1}^n b_{il} - \frac{1}{n-2} \sum_{k=1}^n b_{kj} + \frac{1}{(n-1)(n-2)} \sum_{k,l=1}^n b_{kl}, & \text{if} \quad i \neq j; \\
    0, & \text{if} \quad i = j.
\end{cases}
\end{equation}
Finally, we can compute the unbiased estimator $\Omega_n(X,Y)$ for $\mathcal{V}^2(X,Y)$ as the dot product $\langle \tilde{A}, \tilde{B} \rangle$:
\begin{equation}
    \Omega_n(X,Y) = \langle \tilde{A}, \tilde{B} \rangle = \frac{1}{n(n-3)} \sum_{i,j=1}^n \tilde{A}_{ij} \ \tilde{B}_{ij}
\end{equation}

\subsection{Eigenvector centrality} \label{eigvect}
For any graph $\mathcal{G} := (\boldsymbol{V, E})$, let $K$ be the adjacency matrix of graph $\mathcal{G}$ with $k_{v,t}$ equal to the weight on the edge between node $v$ and $t$. The eigenvector centrality $x$ of node $v$ is a measure relative to all other nodes in $\mathcal{G}$, defined as
\begin{equation}
    x_v = \frac{1}{\lambda} \sum_{t \in \mathcal{G}} k_{v,t} x_t ,
\end{equation}
where $\lambda$ is the greatest eigenvalue in the eigenvector equation $K \mathbf{x} = \lambda \mathbf{x}$, subject to $\x\neq 0$. Consequently, this centrality measure is an extension of the widely used degree centrality by considering the centrality of its neighbours besides its own.

\newpage

\section{Appendix B} \label{appendixB}
\vspace{-1cm}
\subsection{Networks of groupings}
% CONTINENTS

%\begin{figure}[h]
%\centering
\begin{minipage}{.47\textwidth}
  \centering
  \includegraphics[width=\linewidth]{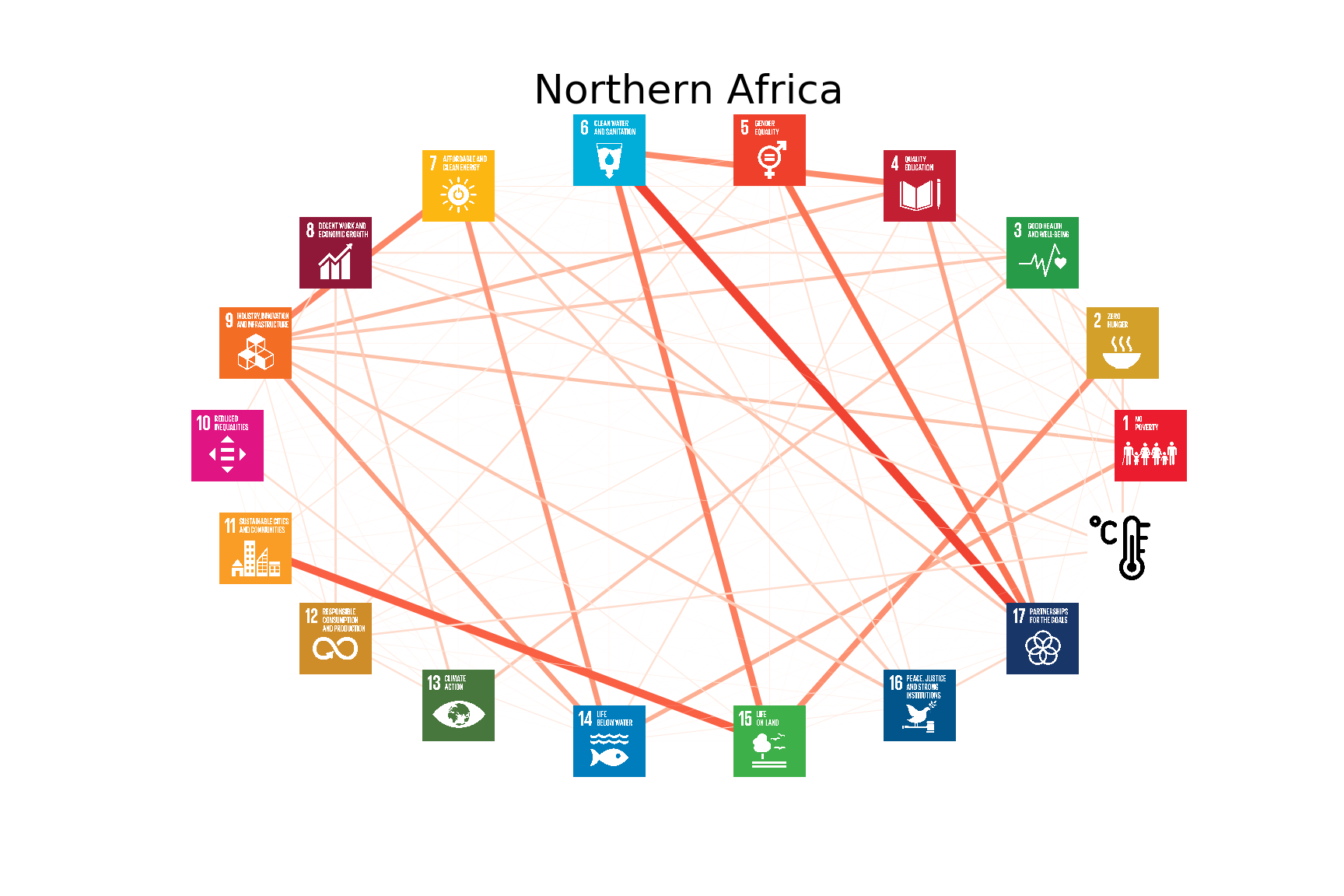}
\end{minipage}
\begin{minipage}{.47\textwidth}
  \centering
  \includegraphics[width=\linewidth]{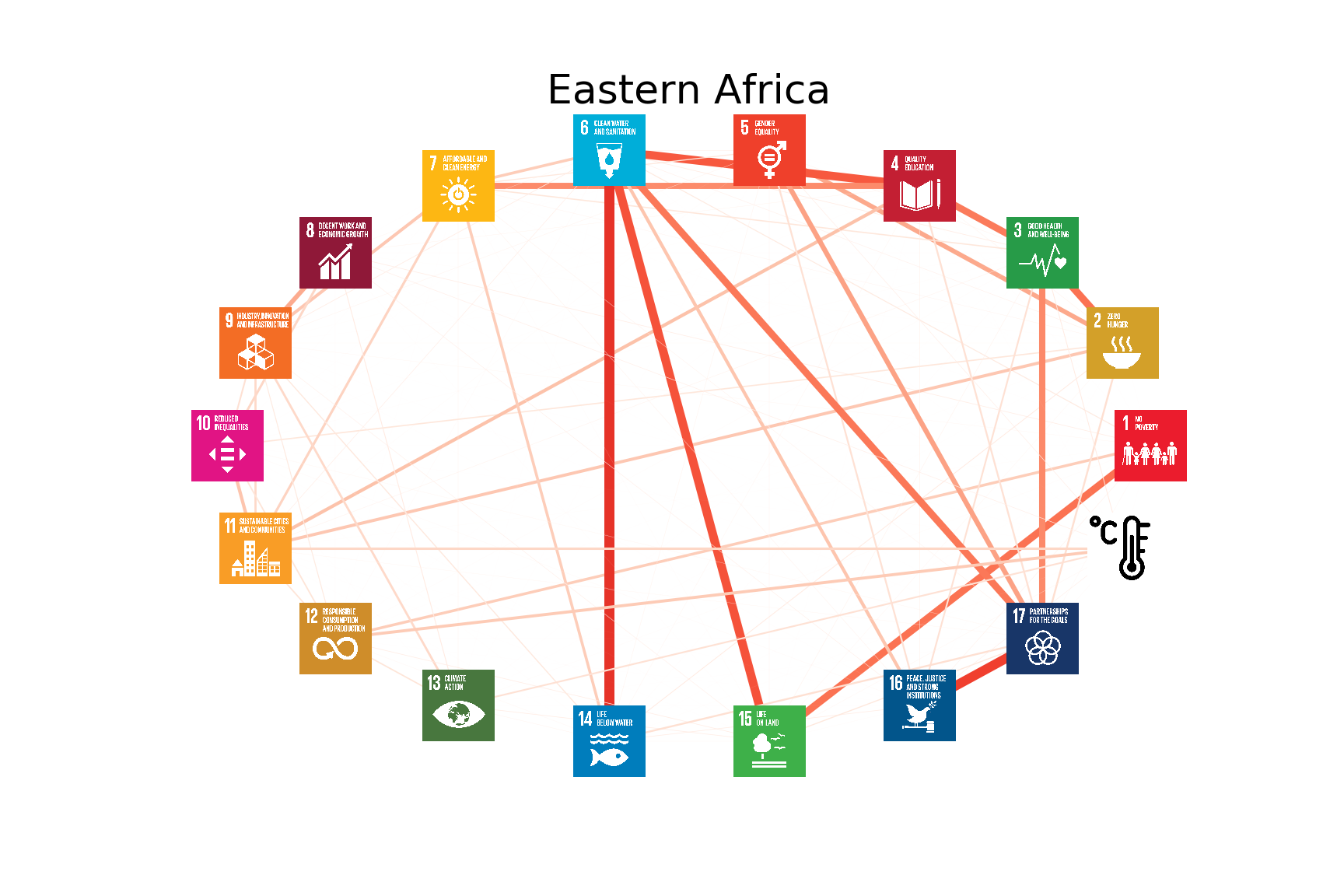}
\end{minipage}
\begin{minipage}{.04\textwidth}
  \centering
  \includegraphics[width=\linewidth]{networks/index.png}
\end{minipage}
%\end{figure}

%\begin{figure}[h]
%\centering
\begin{minipage}{.47\textwidth}
  \centering
  \includegraphics[width=\linewidth]{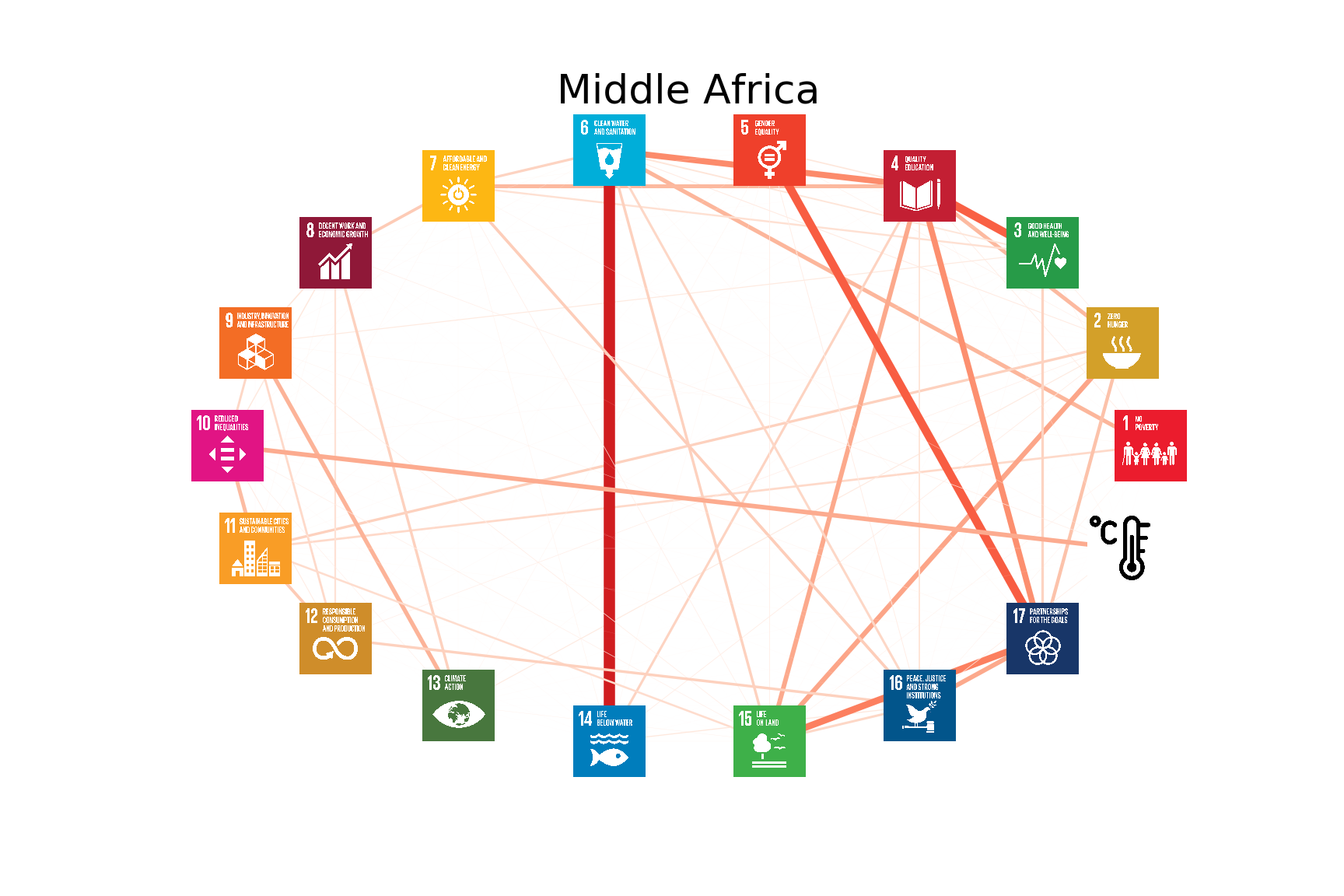}
\end{minipage}
\begin{minipage}{.47\textwidth}
  \centering
  \includegraphics[width=\linewidth]{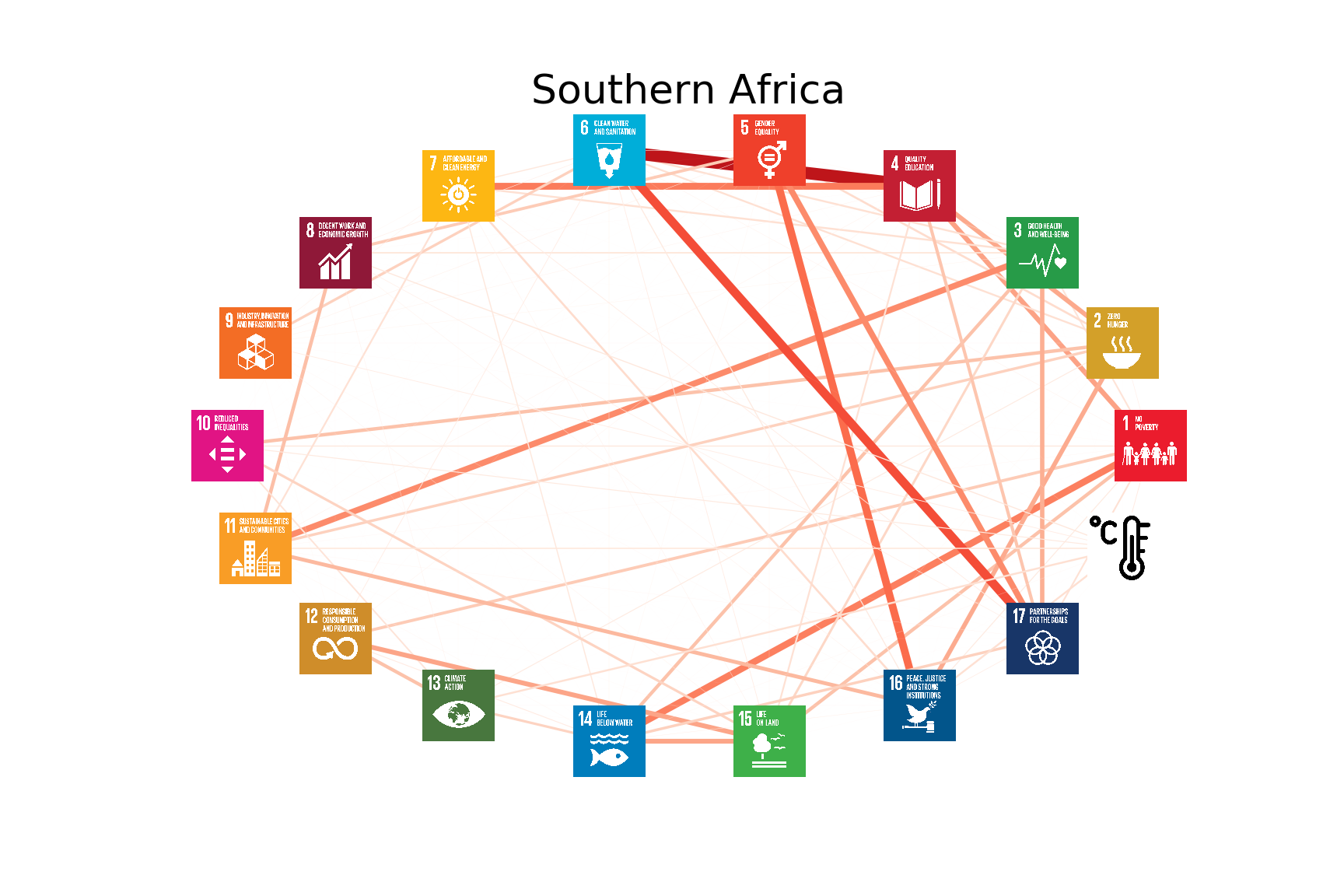}
\end{minipage}
\begin{minipage}{.04\textwidth}
  \centering
  \includegraphics[width=\linewidth]{networks/index.png}
\end{minipage}
%\end{figure}

%\begin{figure}[h]
%\centering
\begin{minipage}{.47\textwidth}
  \centering
  \includegraphics[width=\linewidth]{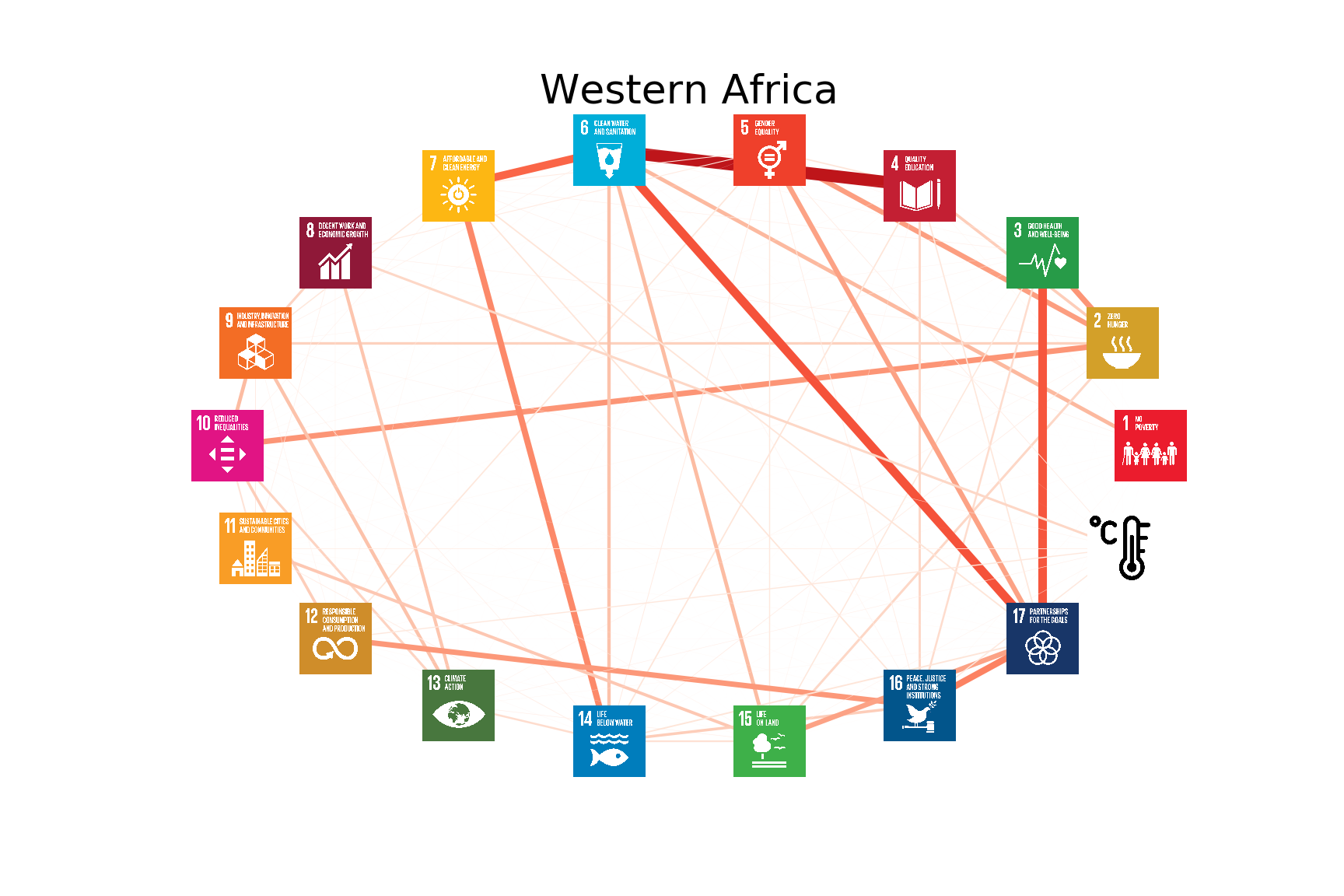}
\end{minipage}
\begin{minipage}{.47\textwidth}
  \centering
  \includegraphics[width=\linewidth]{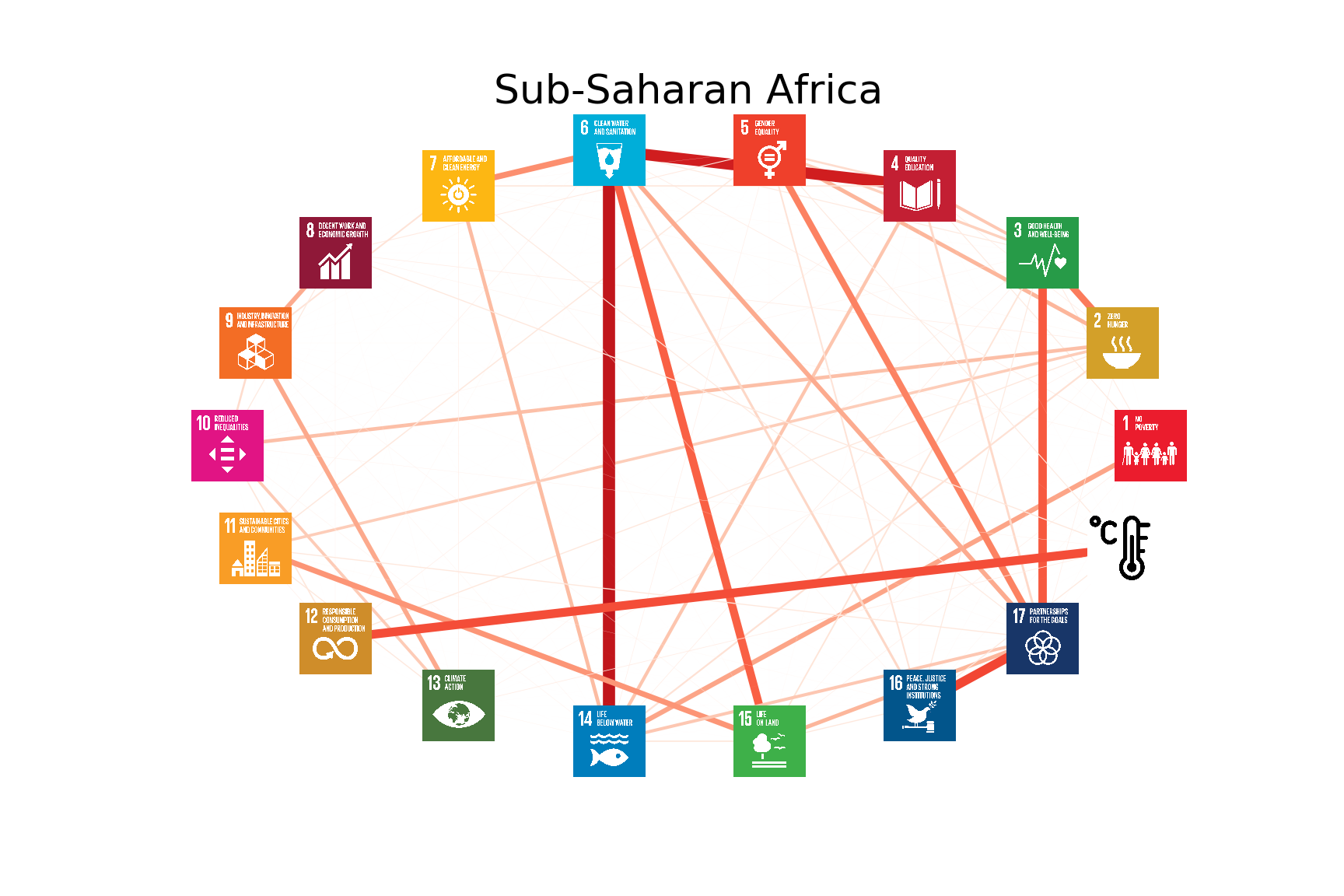}
\end{minipage}
\begin{minipage}{.04\textwidth}
  \centering
  \includegraphics[width=\linewidth]{networks/index.png}
\end{minipage}
%\end{figure}

%\begin{figure}[h]
\centering
\begin{minipage}{.47\textwidth}
  \centering
  \includegraphics[width=\linewidth]{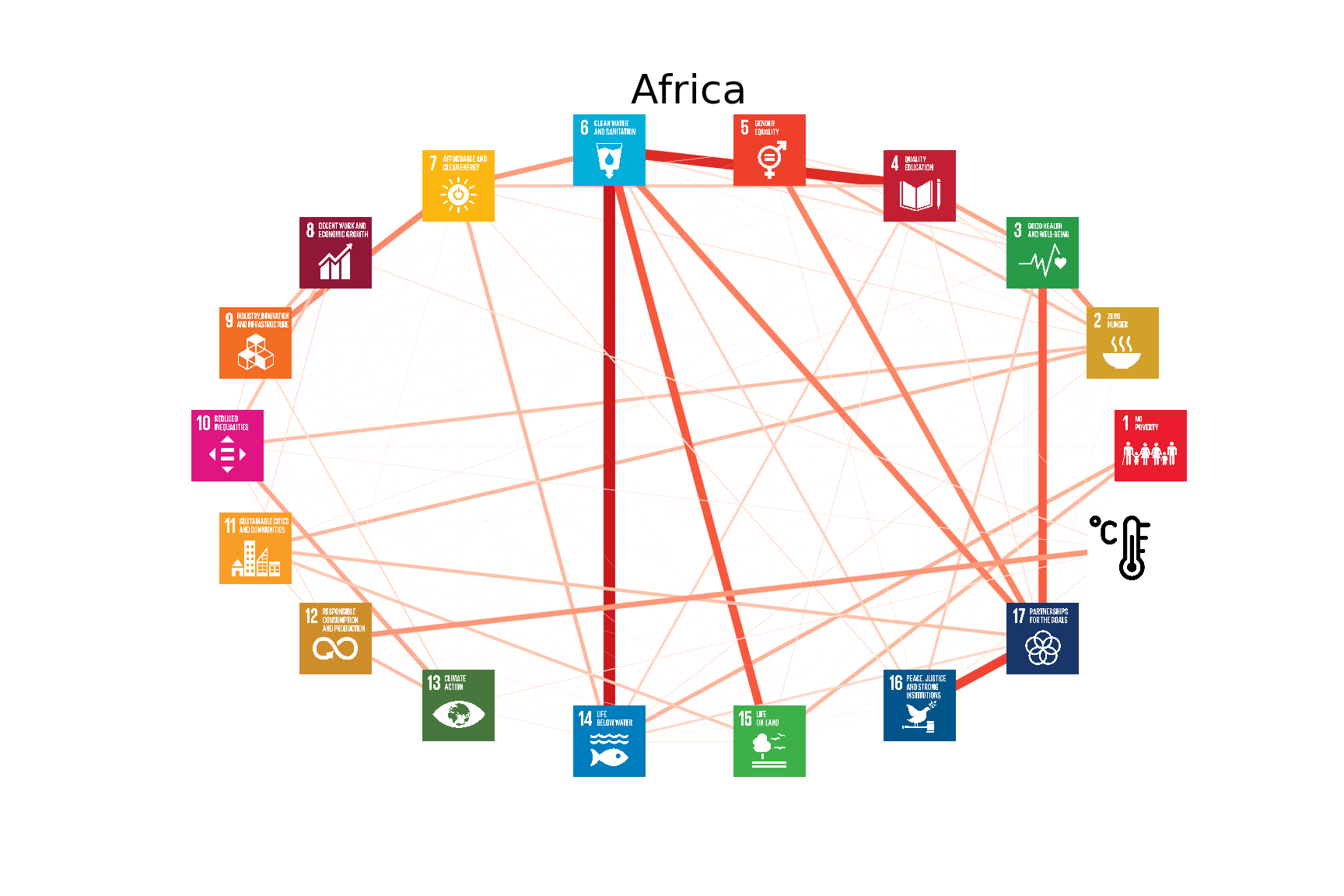}
\end{minipage}
\begin{minipage}{.04\textwidth}
  \centering
  \includegraphics[width=\linewidth]{networks/index.png}
\end{minipage}
%\end{figure}

%\begin{figure}[h]
%\centering
\begin{minipage}{.47\textwidth}
  \centering
  \includegraphics[width=\linewidth]{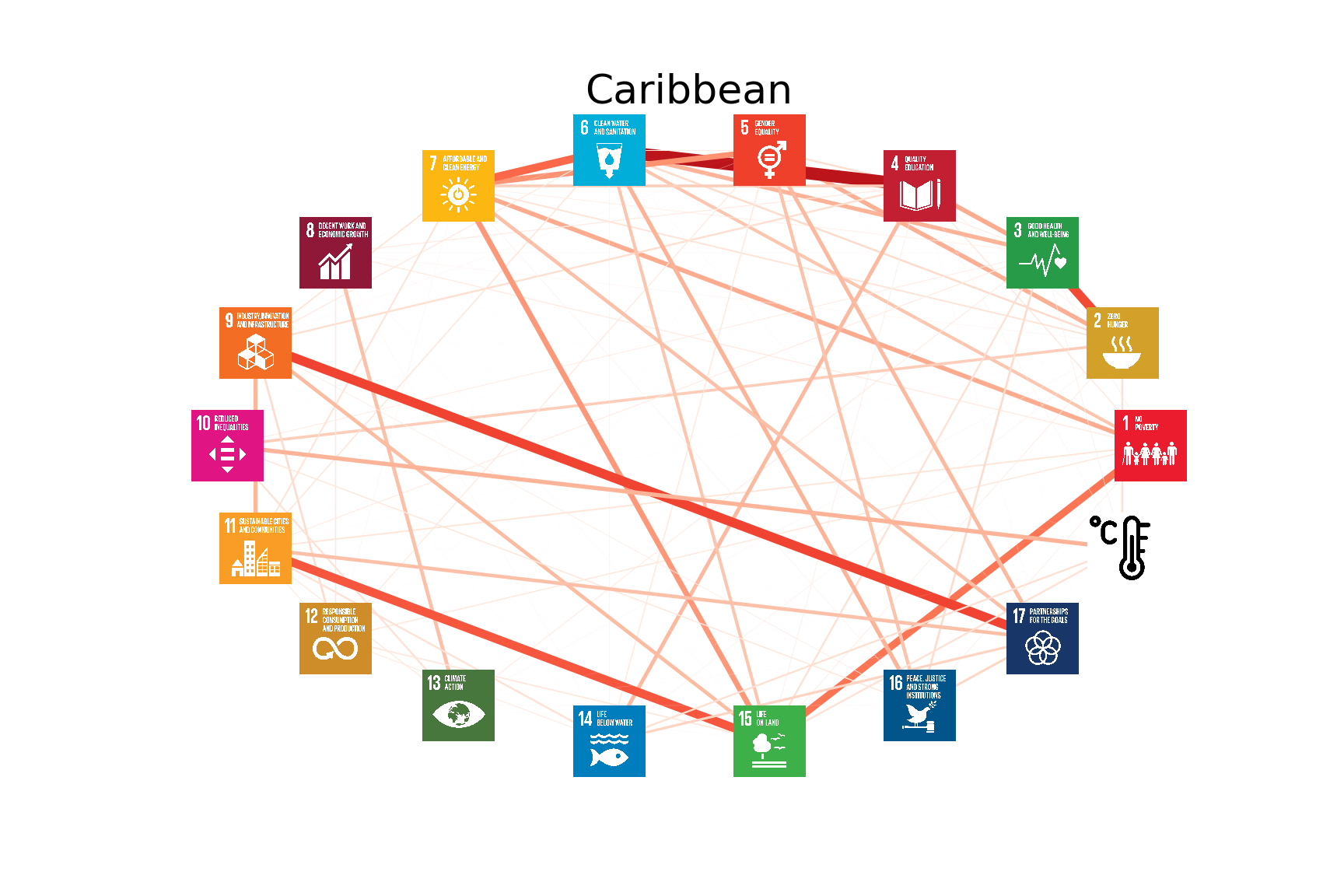}
\end{minipage}
\begin{minipage}{.47\textwidth}
  \centering
  \includegraphics[width=\linewidth]{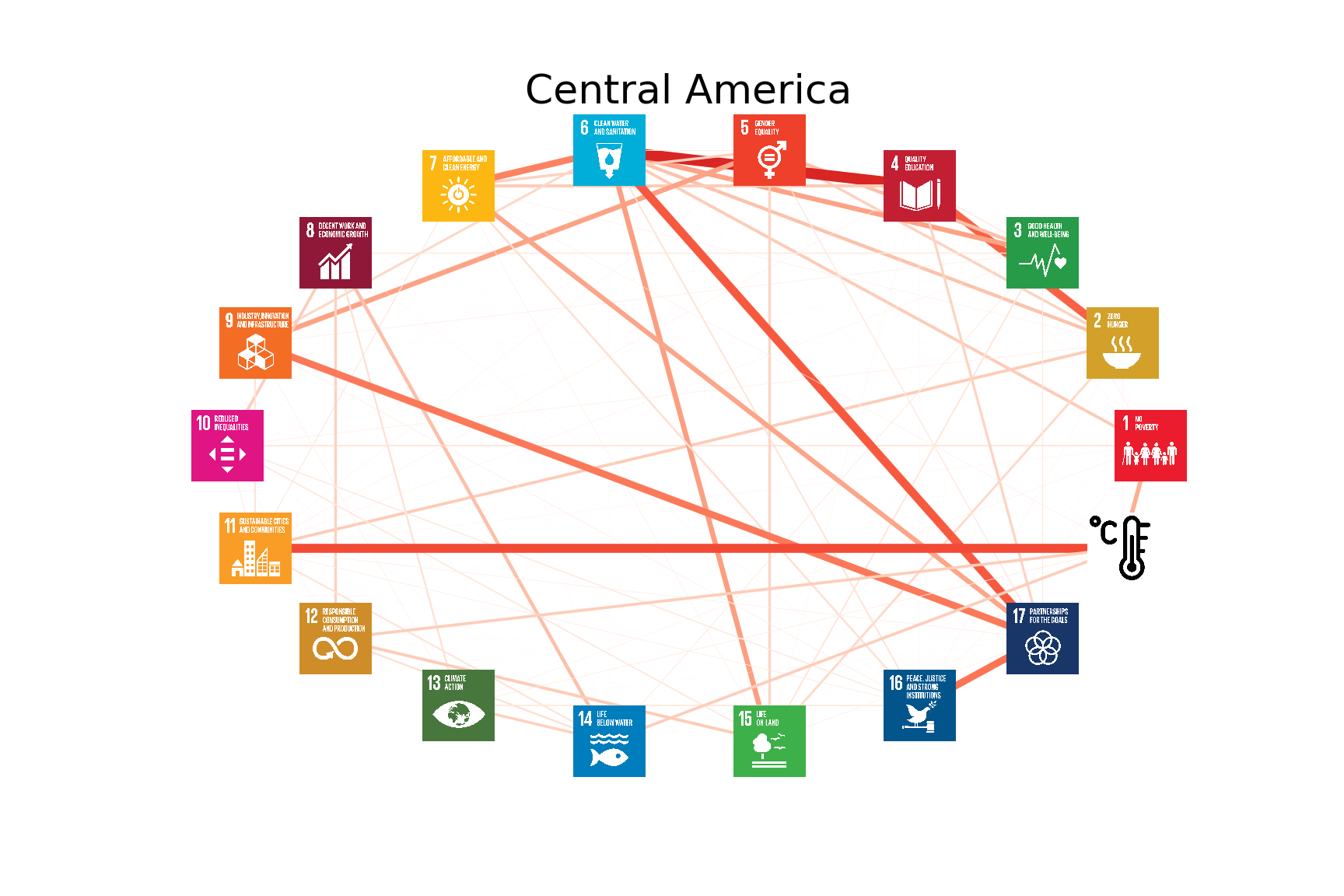}
\end{minipage}
\begin{minipage}{.04\textwidth}
  \centering
  \includegraphics[width=\linewidth]{networks/index.png}
\end{minipage}
%\end{figure}

%\begin{figure}[h]
%\centering
\begin{minipage}{.47\textwidth}
  \centering
  \includegraphics[width=\linewidth]{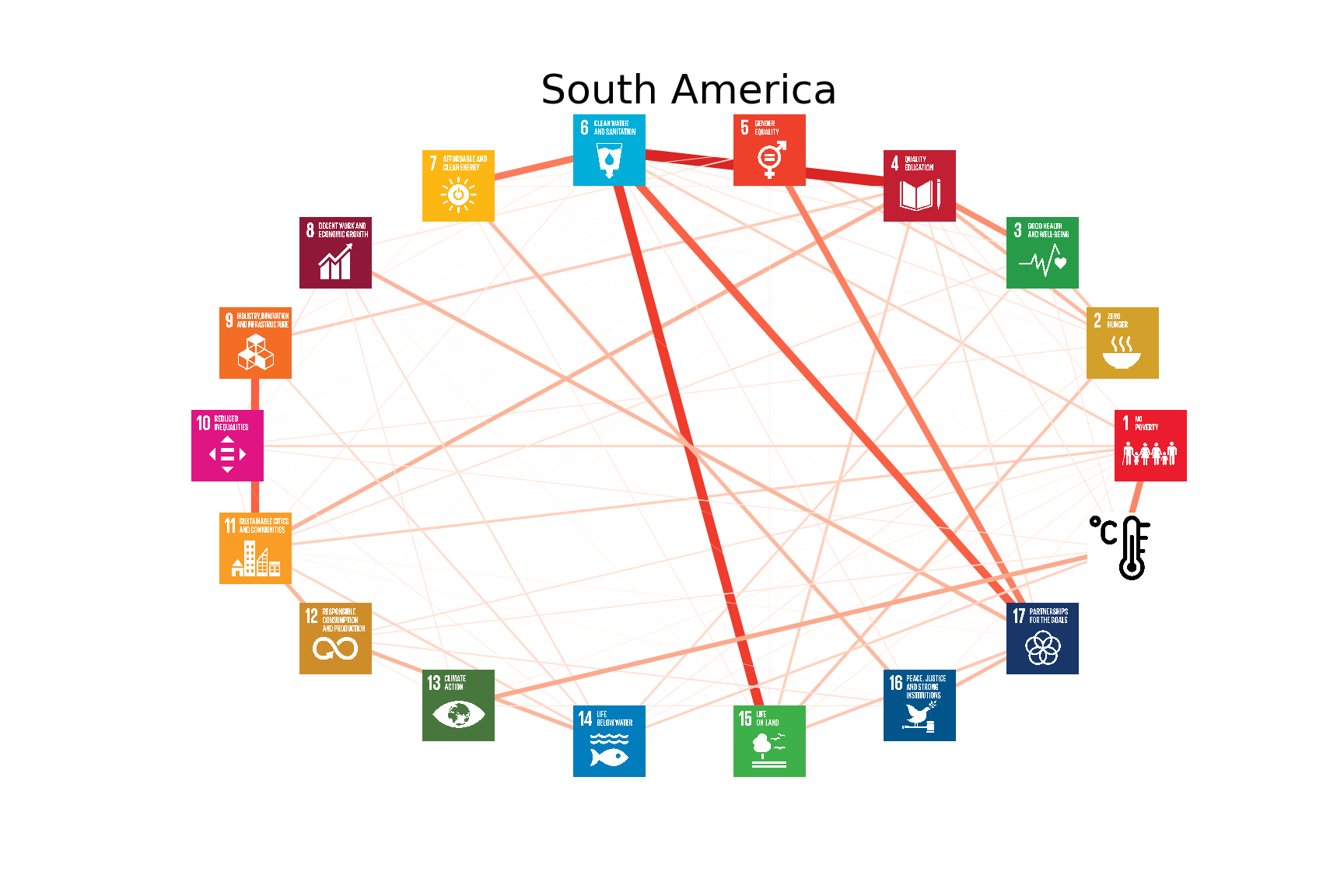}
\end{minipage}
\begin{minipage}{.47\textwidth}
  \centering
  \includegraphics[width=\linewidth]{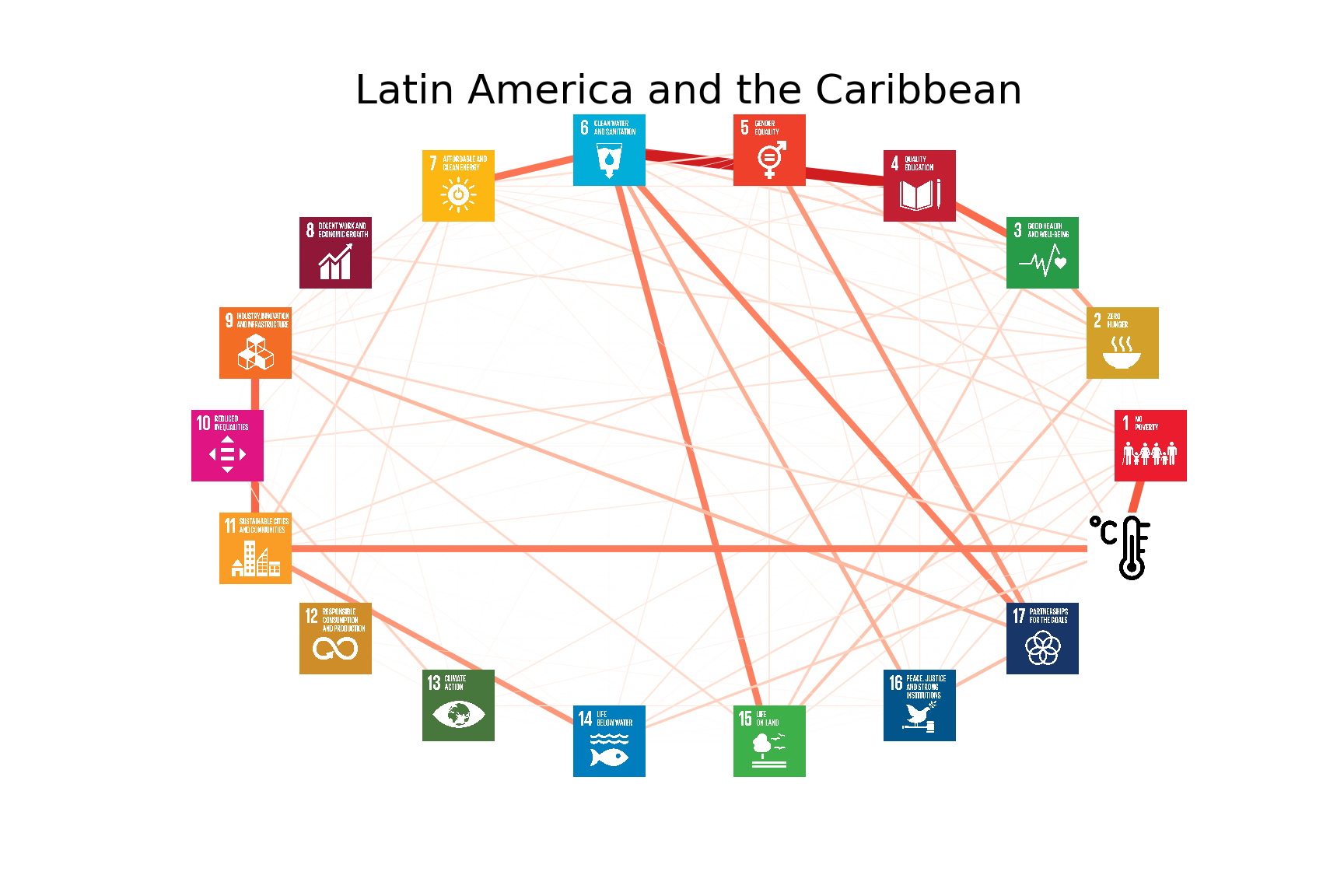}
\end{minipage}
\begin{minipage}{.04\textwidth}
  \centering
  \includegraphics[width=\linewidth]{networks/index.png}
\end{minipage}
%\end{figure}

%\begin{figure}[h]
%\centering
\begin{minipage}{.47\textwidth}
  \centering
  \includegraphics[width=\linewidth]{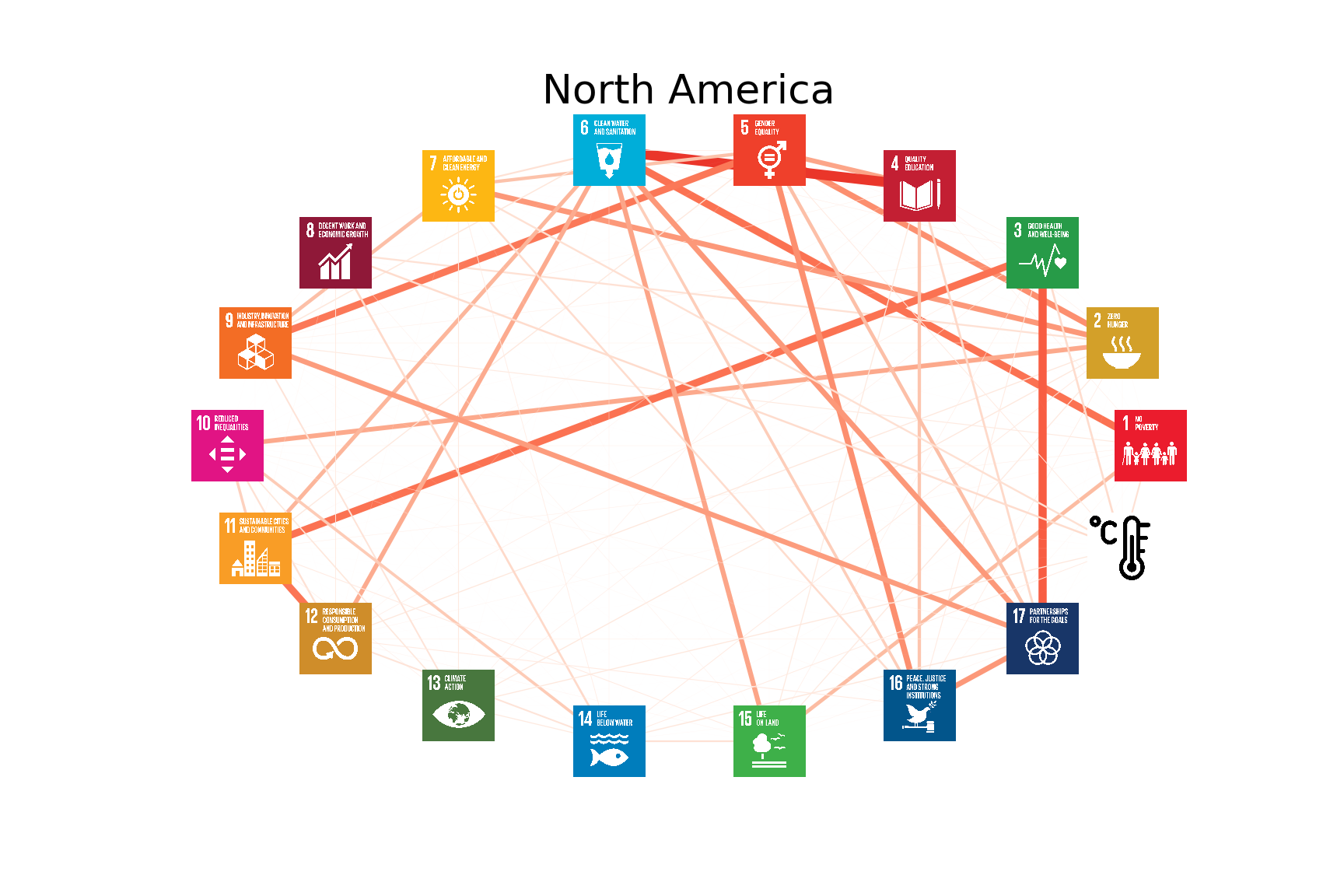}
\end{minipage}
\begin{minipage}{.47\textwidth}
  \centering
  \includegraphics[width=\linewidth]{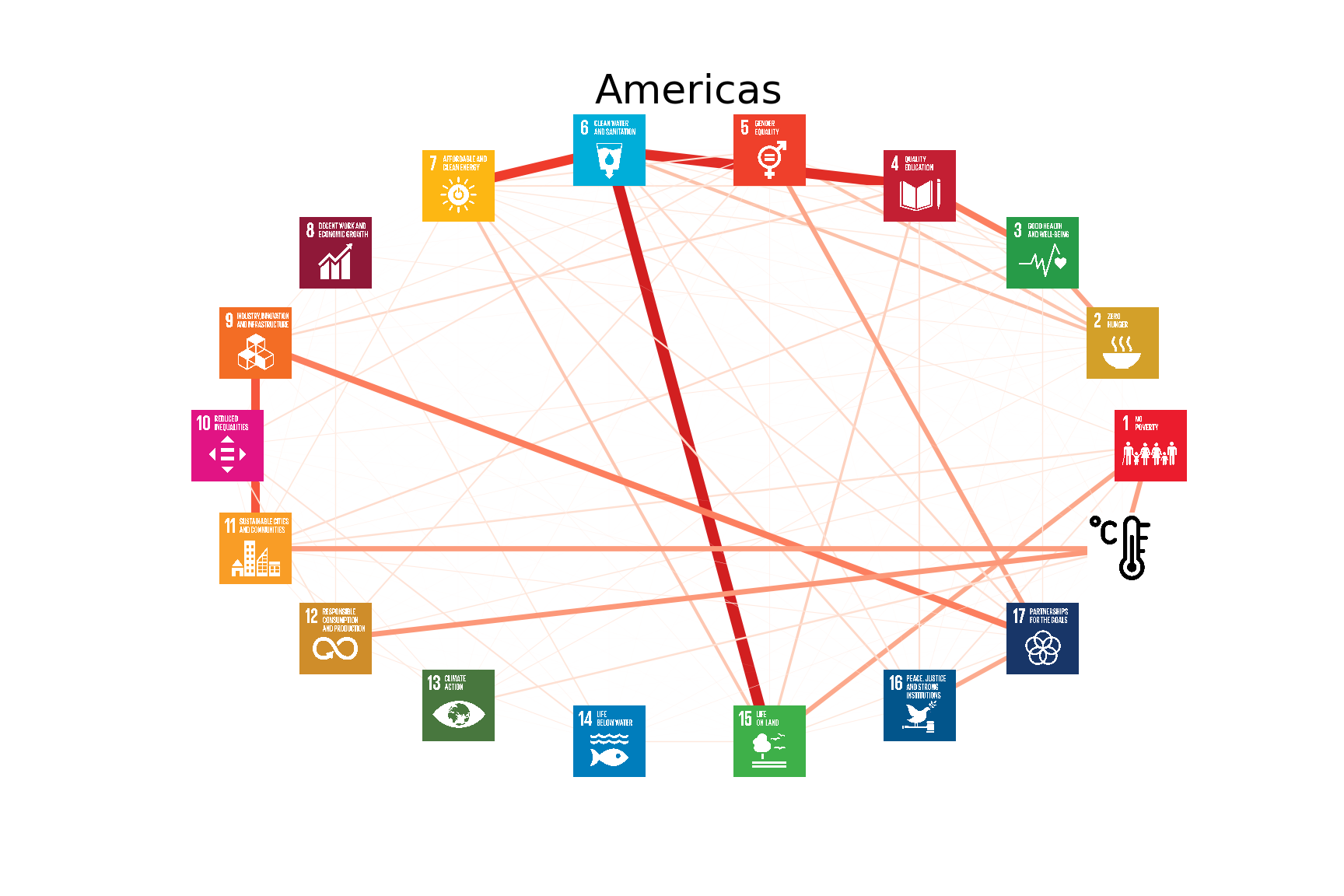}
\end{minipage}
\begin{minipage}{.04\textwidth}
  \centering
  \includegraphics[width=\linewidth]{networks/index.png}
\end{minipage}
%\end{figure}

%\begin{figure}[h]
%\centering
\begin{minipage}{.47\textwidth}
  \centering
  \includegraphics[width=\linewidth]{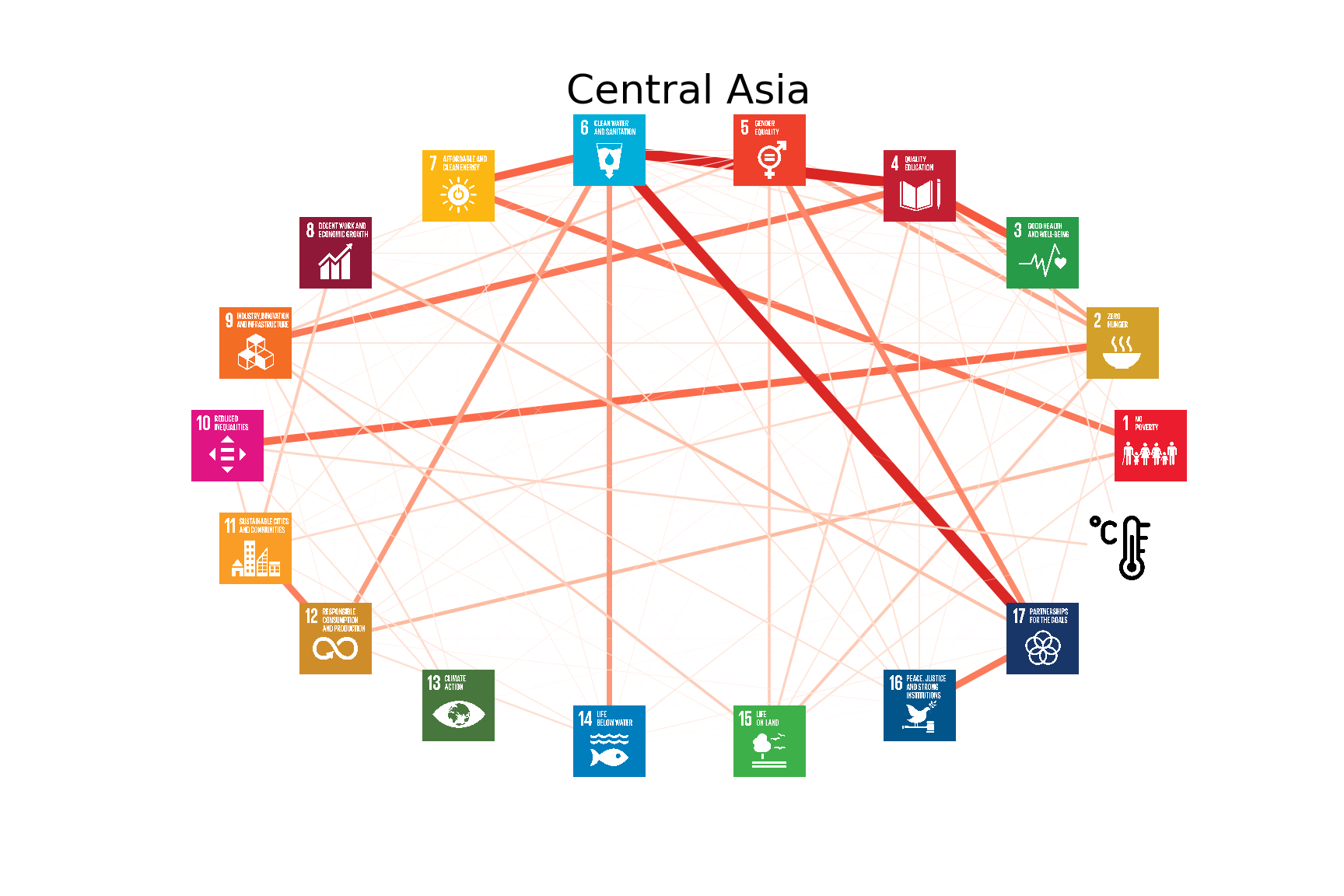}
\end{minipage}
\begin{minipage}{.47\textwidth}
  \centering
  \includegraphics[width=\linewidth]{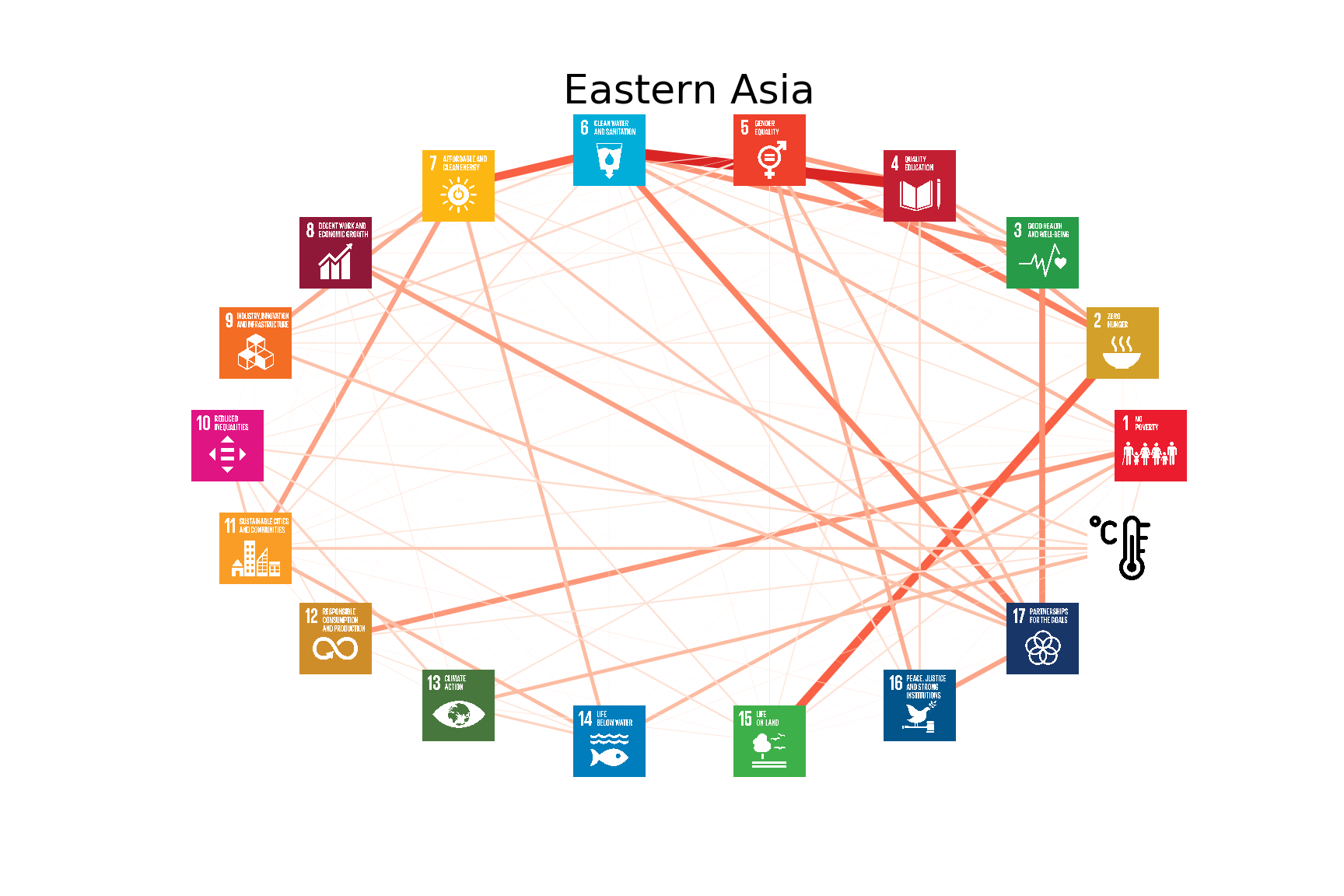}
\end{minipage}
\begin{minipage}{.04\textwidth}
  \centering
  \includegraphics[width=\linewidth]{networks/index.png}
\end{minipage}
%\end{figure}

%\begin{figure}[h]
%\centering
\begin{minipage}{.47\textwidth}
  \centering
  \includegraphics[width=\linewidth]{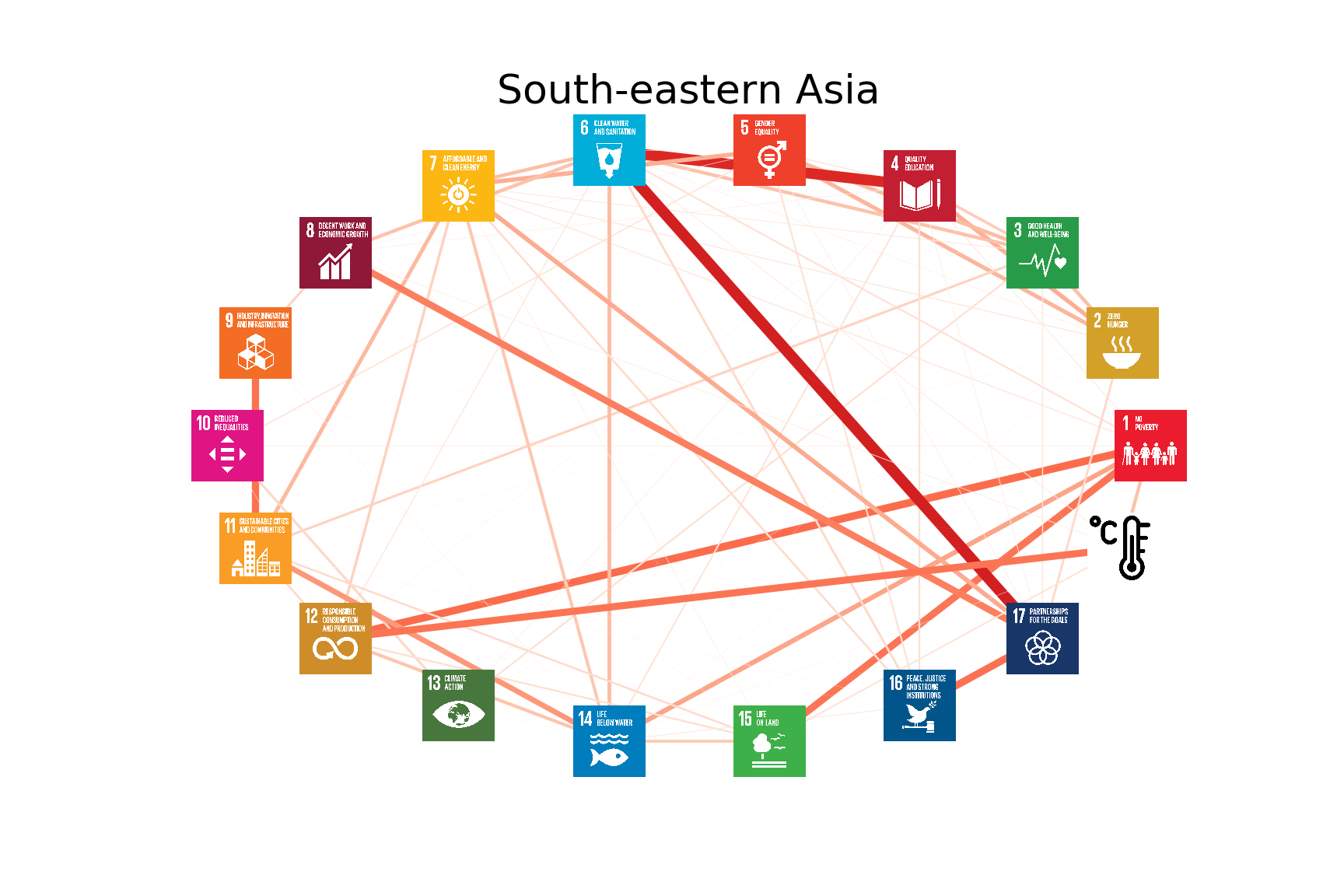}
\end{minipage}
\begin{minipage}{.47\textwidth}
  \centering
  \includegraphics[width=\linewidth]{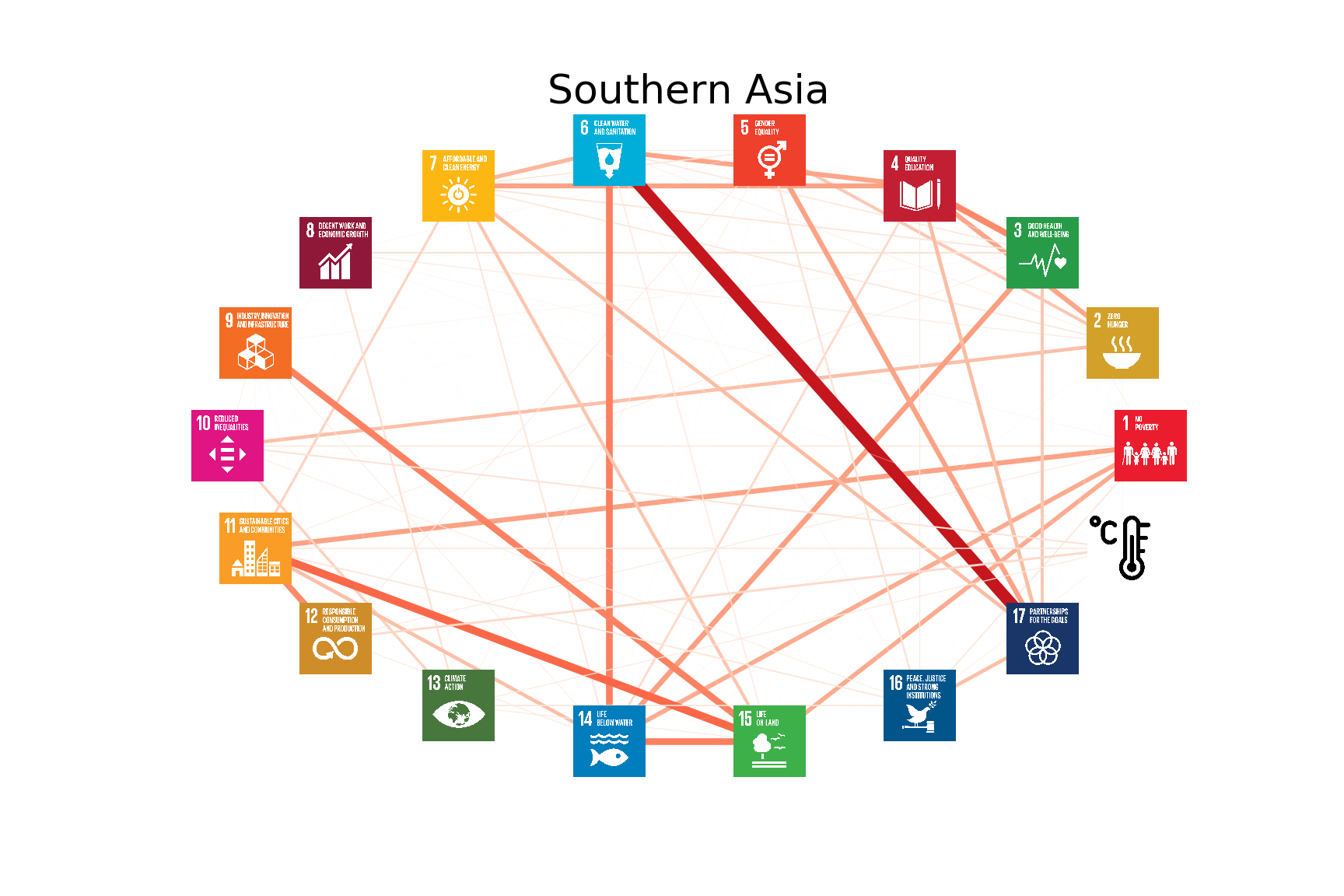}
\end{minipage}
\begin{minipage}{.04\textwidth}
  \centering
  \includegraphics[width=\linewidth]{networks/index.png}
\end{minipage}
%\end{figure}

%\begin{figure}[h]
%\centering
\begin{minipage}{.47\textwidth}
  \centering
  \includegraphics[width=\linewidth]{networks/Western_Asia_circular_network_logos.png}
\end{minipage}
\begin{minipage}{.47\textwidth}
  \centering
  \includegraphics[width=\linewidth]{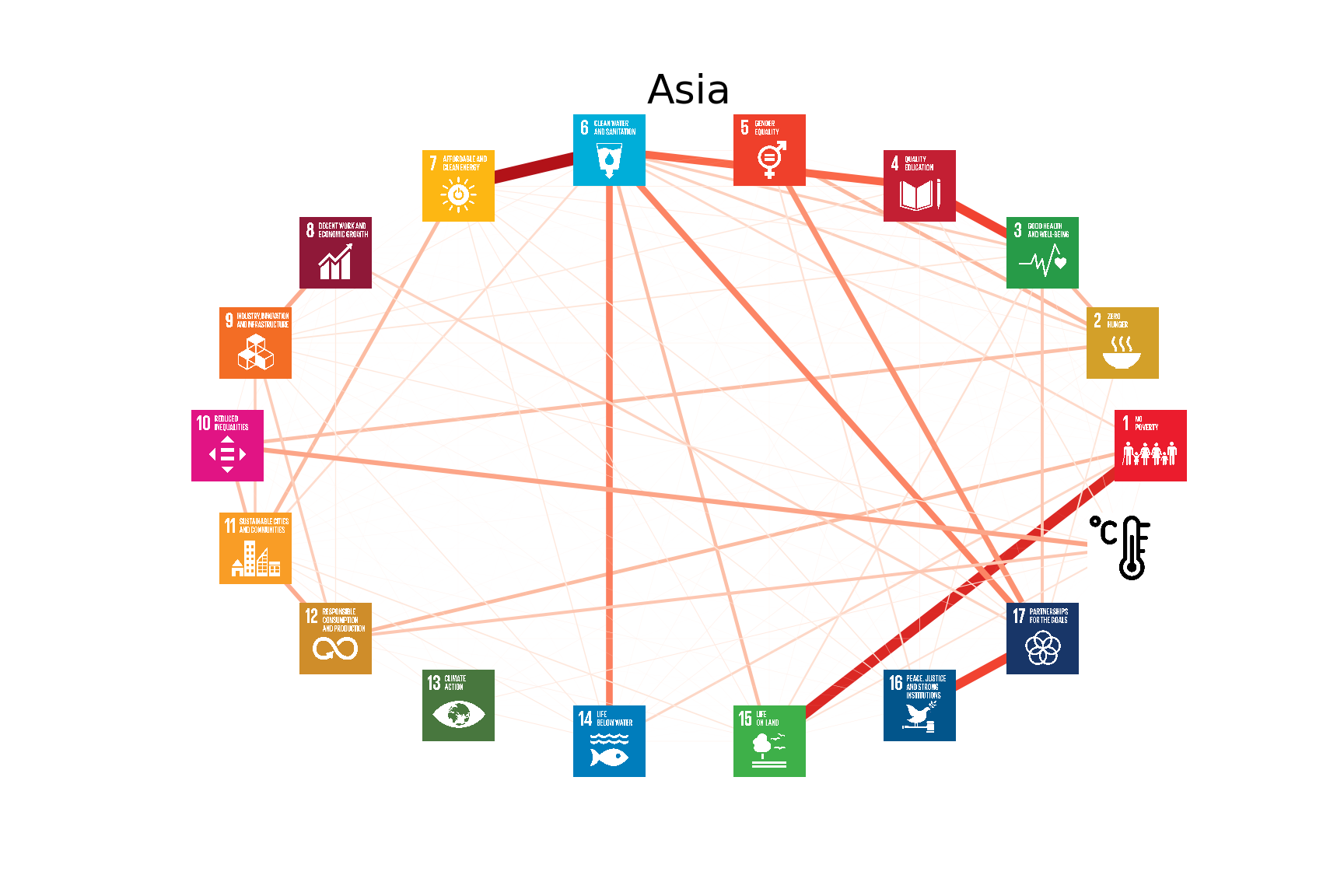}
\end{minipage}
\begin{minipage}{.04\textwidth}
  \centering
  \includegraphics[width=\linewidth]{networks/index.png}
\end{minipage}
%\end{figure}

%\begin{figure}[h]
%\centering
\begin{minipage}{.47\textwidth}
  \centering
  \includegraphics[width=\linewidth]{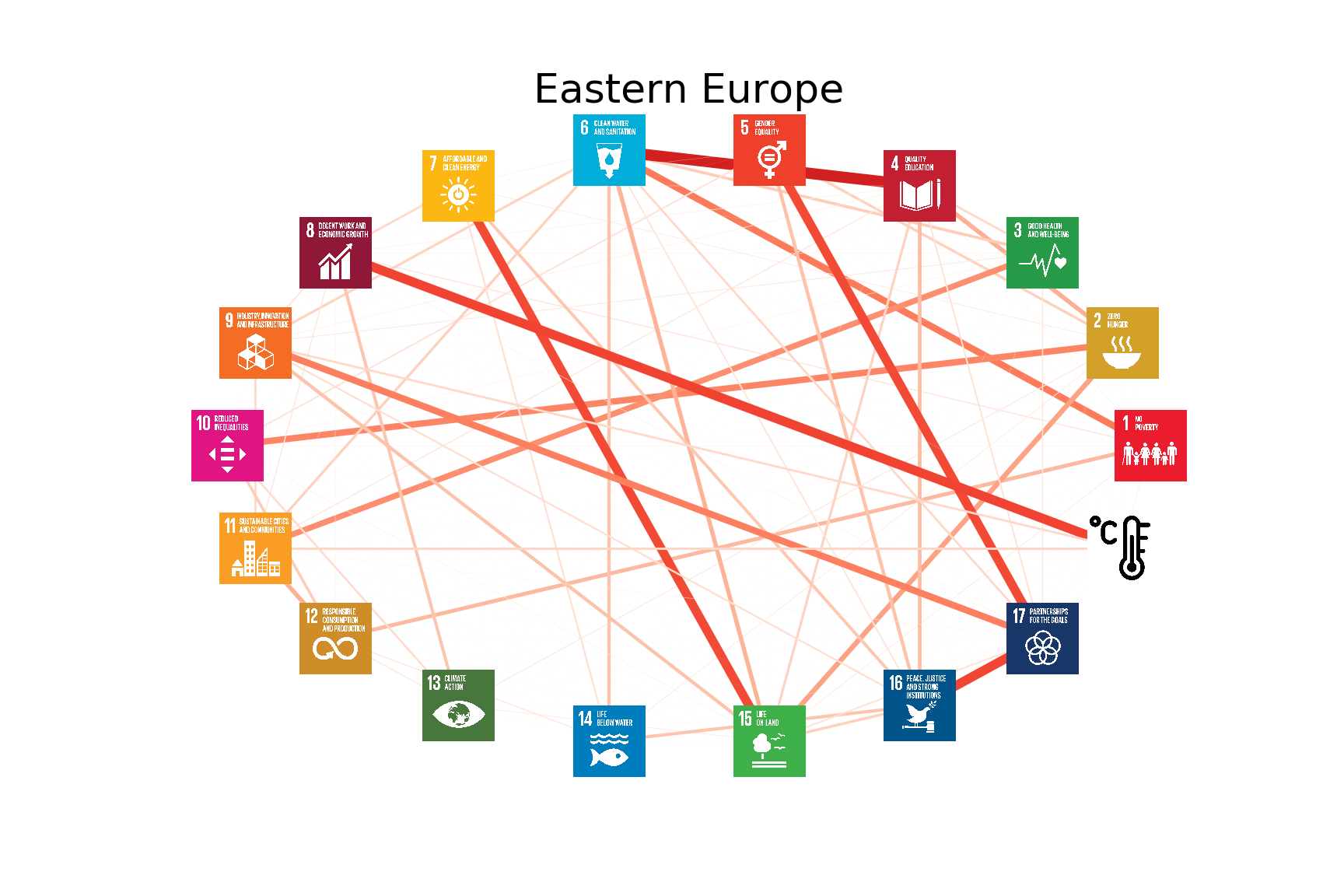}
\end{minipage}
\begin{minipage}{.47\textwidth}
  \centering
  \includegraphics[width=\linewidth]{networks/Northern_Europe_circular_network_logos.png}
\end{minipage}
\begin{minipage}{.04\textwidth}
  \centering
  \includegraphics[width=\linewidth]{networks/index.png}
\end{minipage}
%\end{figure}

%\begin{figure}[h]
%\centering
\begin{minipage}{.47\textwidth}
  \centering
  \includegraphics[width=\linewidth]{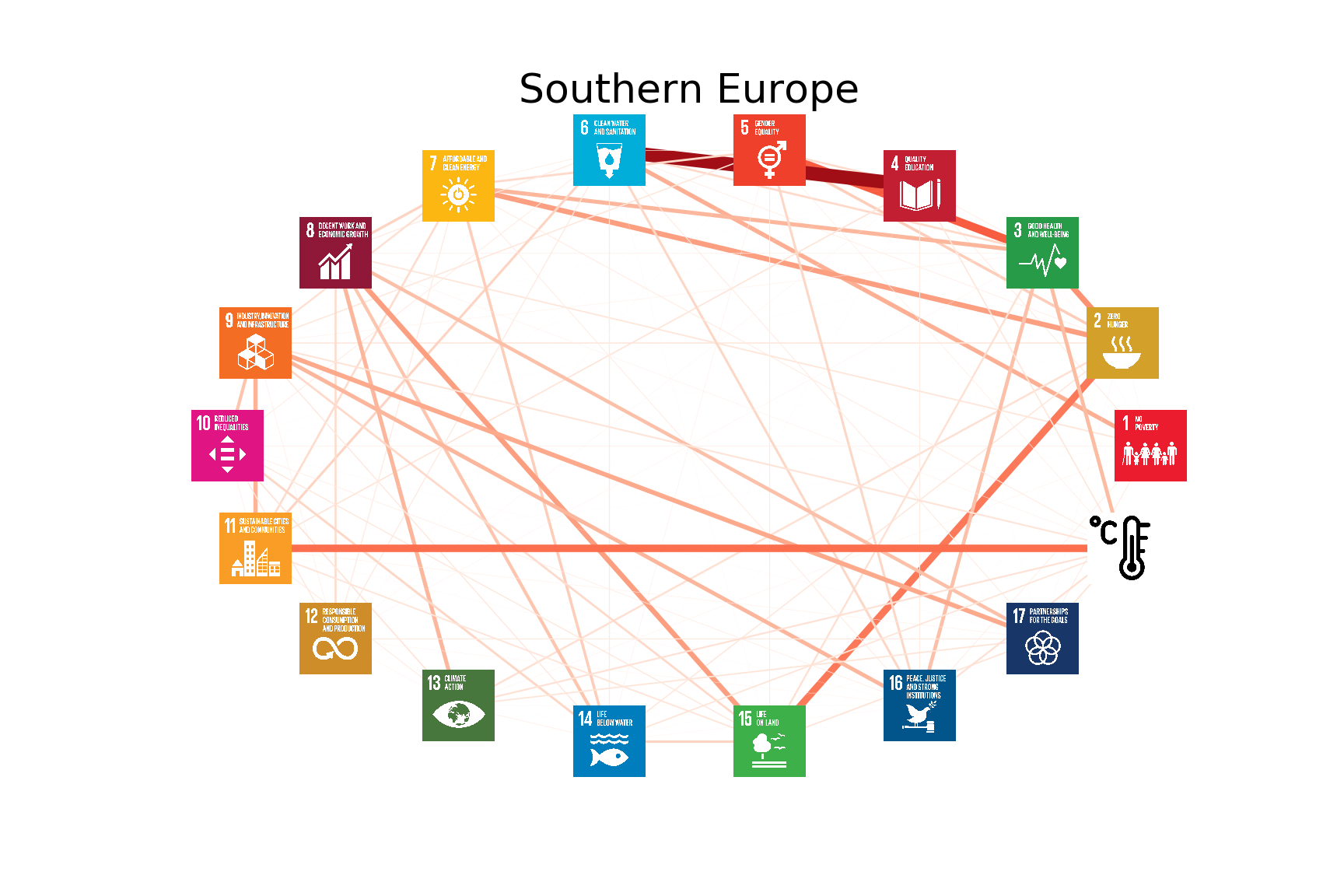}
\end{minipage}
\begin{minipage}{.47\textwidth}
  \centering
  \includegraphics[width=\linewidth]{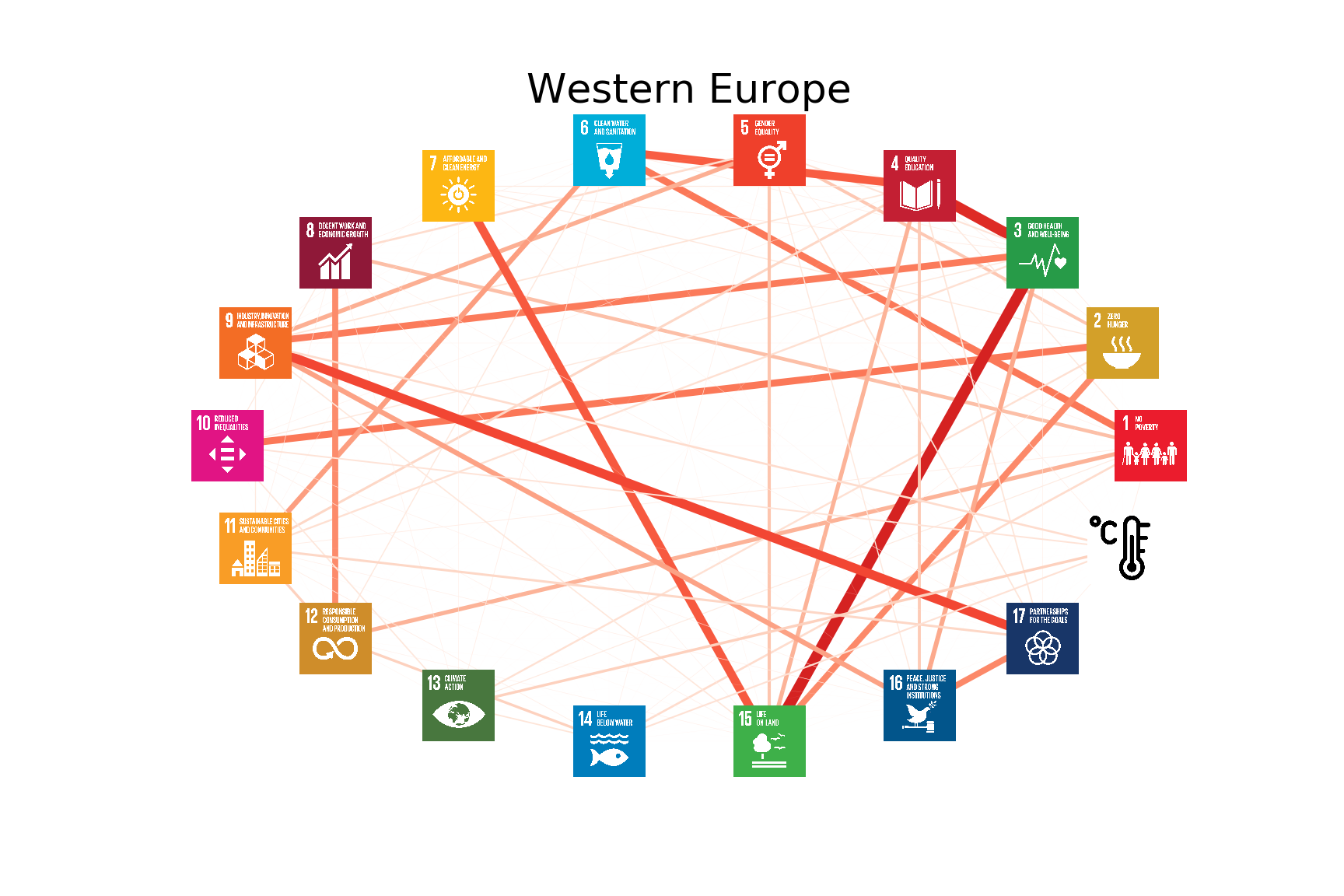}
\end{minipage}
\begin{minipage}{.04\textwidth}
  \centering
  \includegraphics[width=\linewidth]{networks/index.png}
\end{minipage}
%\end{figure}

%\begin{figure}[h]
%\centering
\begin{minipage}{.47\textwidth}
  \centering
  \includegraphics[width=\linewidth]{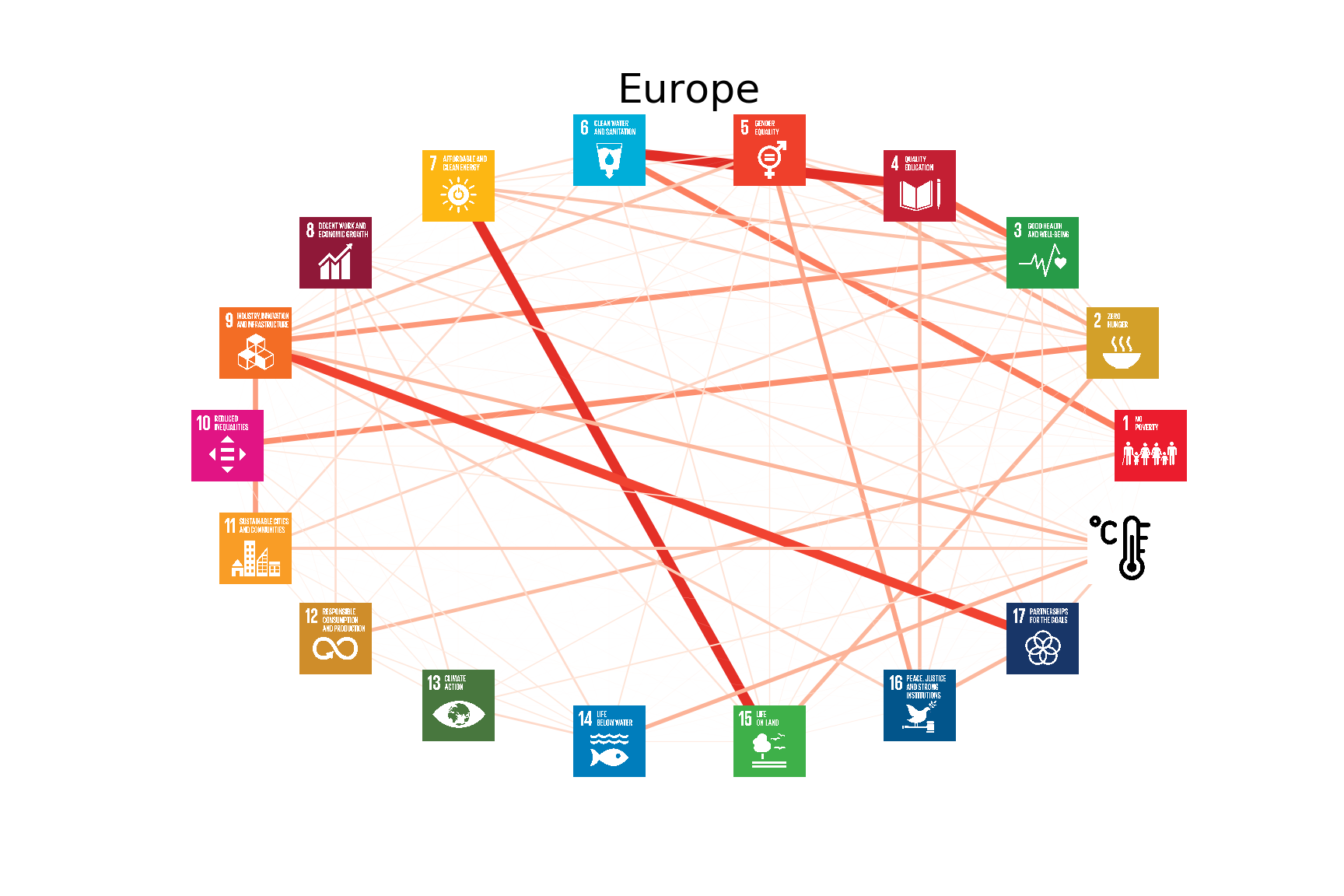}
\end{minipage}
\begin{minipage}{.47\textwidth}
  \centering
  \includegraphics[width=\linewidth]{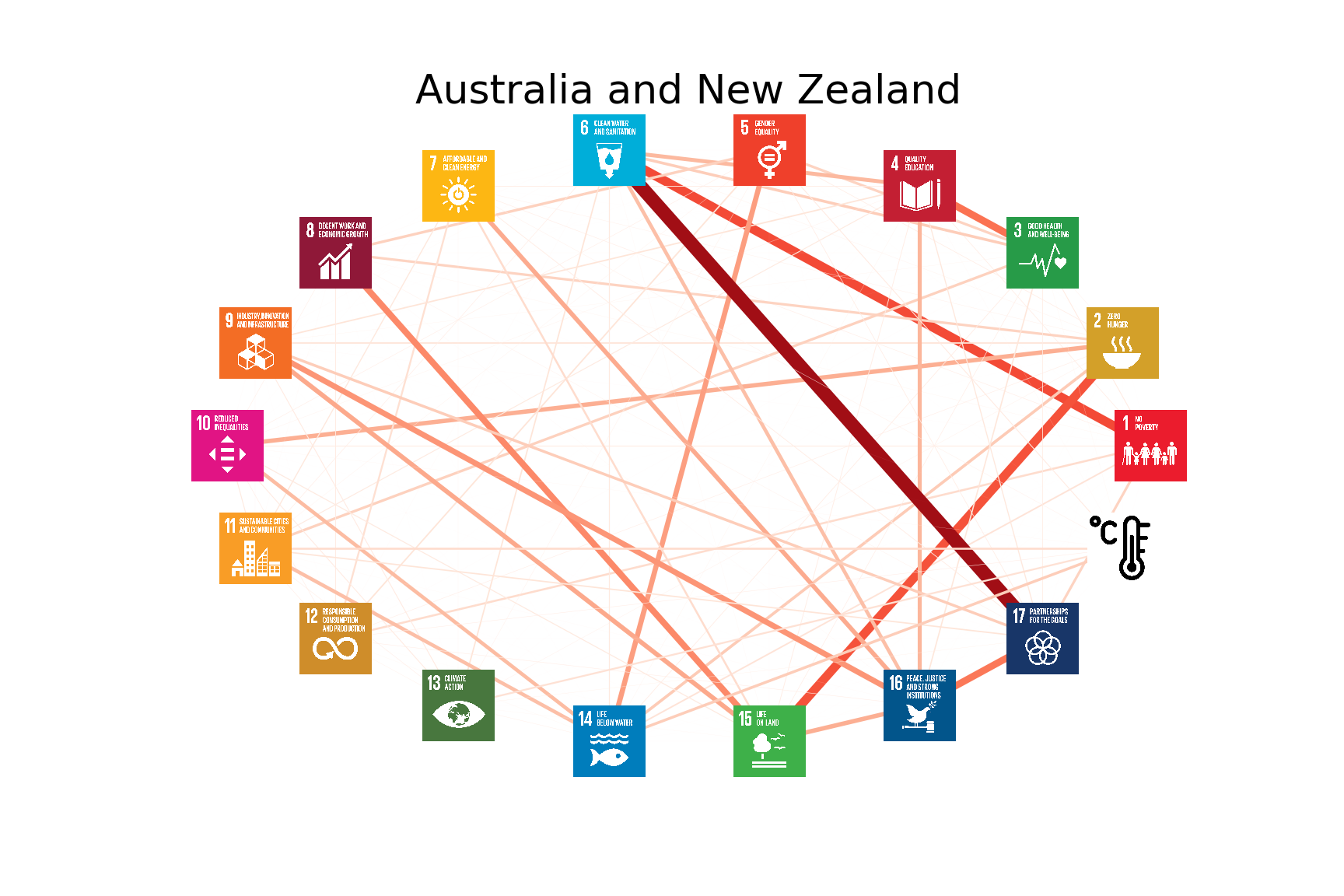}
\end{minipage}
\begin{minipage}{.04\textwidth}
  \centering
  \includegraphics[width=\linewidth]{networks/index.png}
\end{minipage}
%\end{figure}

%\begin{figure}[h]
%\centering
\begin{minipage}{.47\textwidth}
  \centering
  \includegraphics[width=\linewidth]{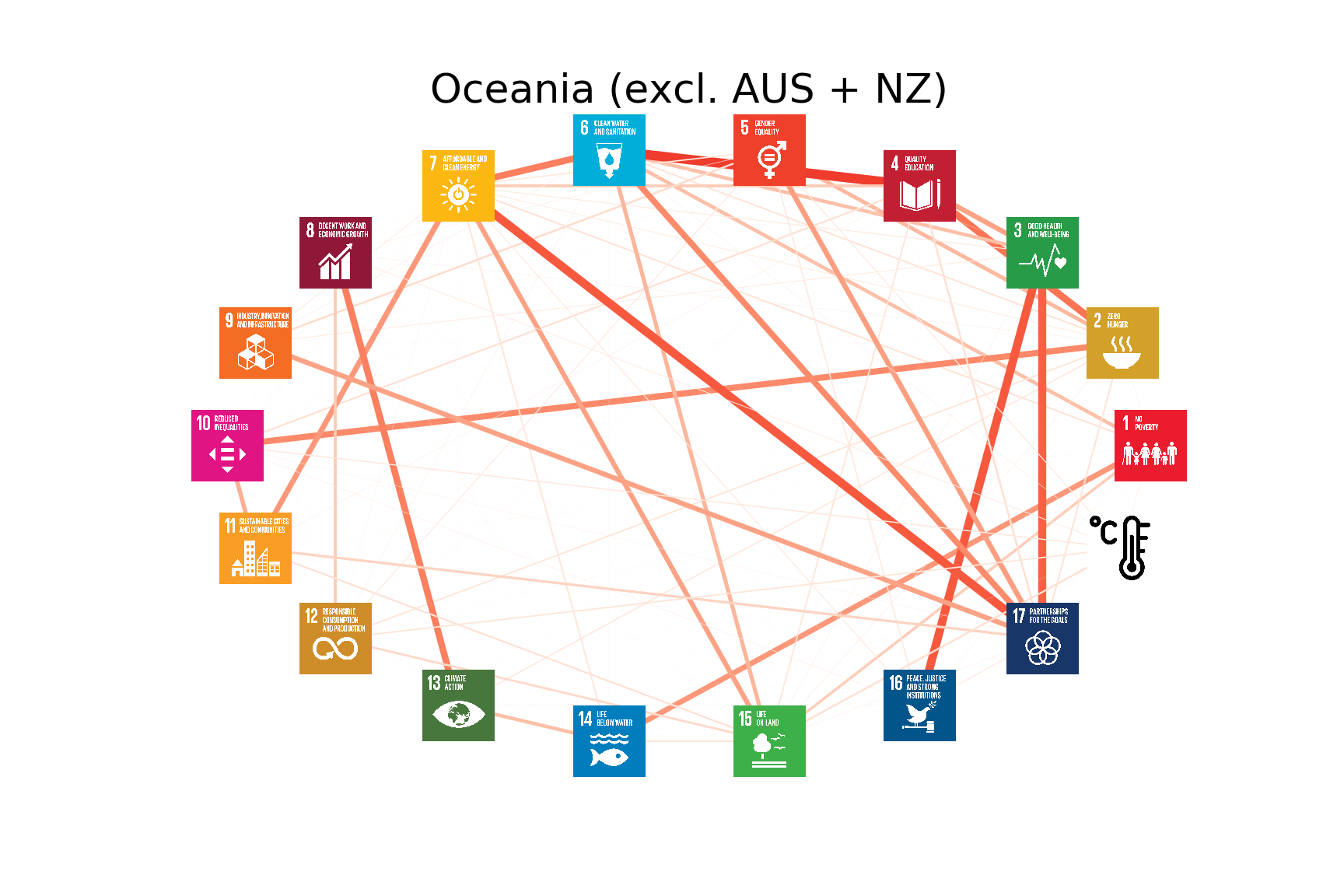}
\end{minipage}
\begin{minipage}{.47\textwidth}
  \centering
  \includegraphics[width=\linewidth]{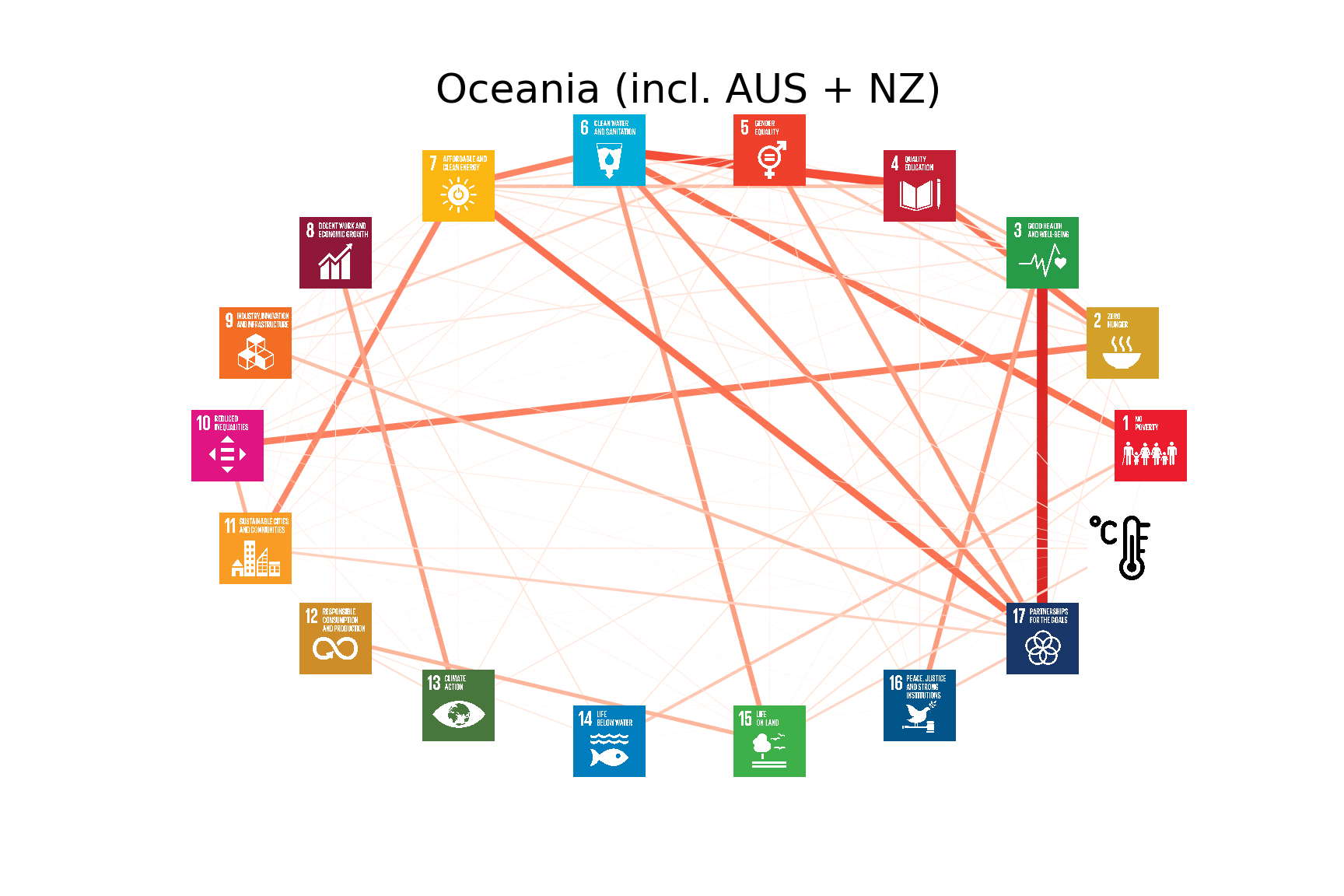}
\end{minipage}
\begin{minipage}{.04\textwidth}
  \centering
  \includegraphics[width=\linewidth]{networks/index.png}
\end{minipage}
%\end{figure}

%\begin{figure}[h]
\centering
\begin{minipage}{.47\textwidth}
  \centering
  \includegraphics[width=\linewidth]{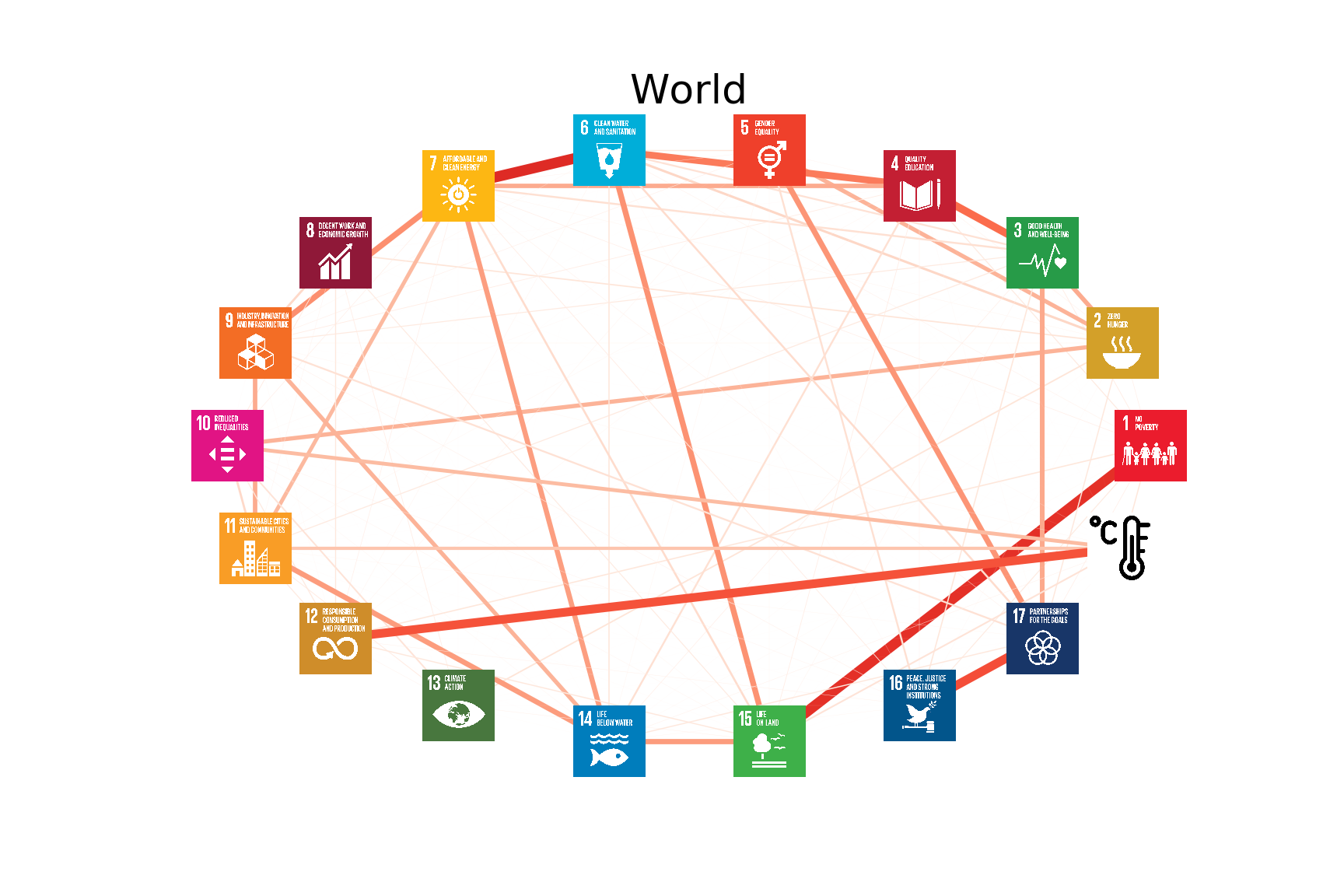}
\end{minipage}
\begin{minipage}{.04\textwidth}
  \centering
  \includegraphics[width=\linewidth]{networks/index.png}
\end{minipage}
%\end{figure}

% GROUPS

%\begin{figure}[h]
%\centering
\begin{minipage}{.47\textwidth}
  \centering
  \includegraphics[width=\linewidth]{networks/Global_North_circular_network_logos.png}
\end{minipage}
\begin{minipage}{.47\textwidth}
  \centering
  \includegraphics[width=\linewidth]{networks/Global_South_circular_network_logos.png}
\end{minipage}
\begin{minipage}{.04\textwidth}
  \centering
  \includegraphics[width=\linewidth]{networks/index.png}
\end{minipage}
%\end{figure}

%\begin{figure}[h]
%\centering
\begin{minipage}{.47\textwidth}
  \centering
  \includegraphics[width=\linewidth]{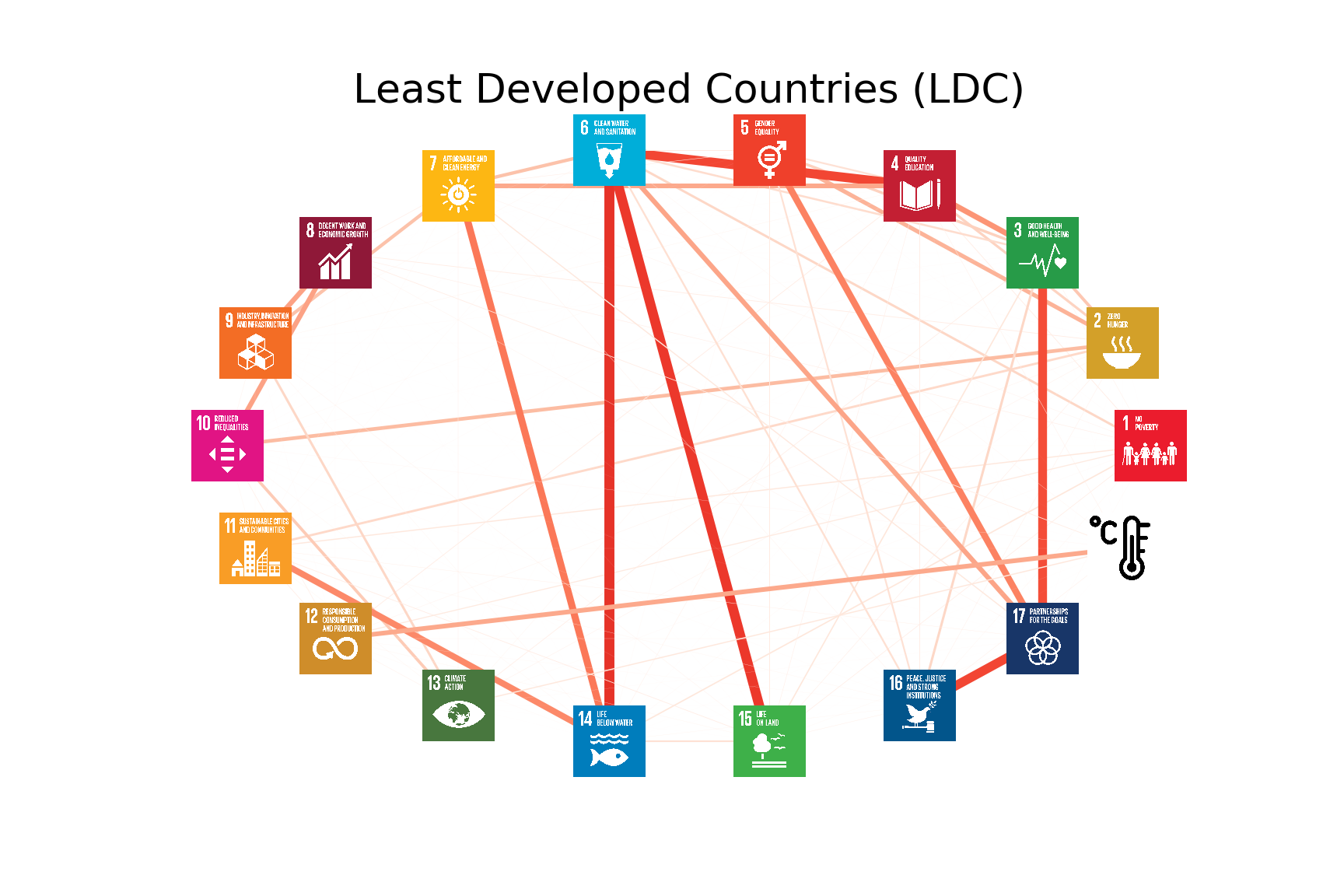}
\end{minipage}
\begin{minipage}{.47\textwidth}
  \centering
  \includegraphics[width=\linewidth]{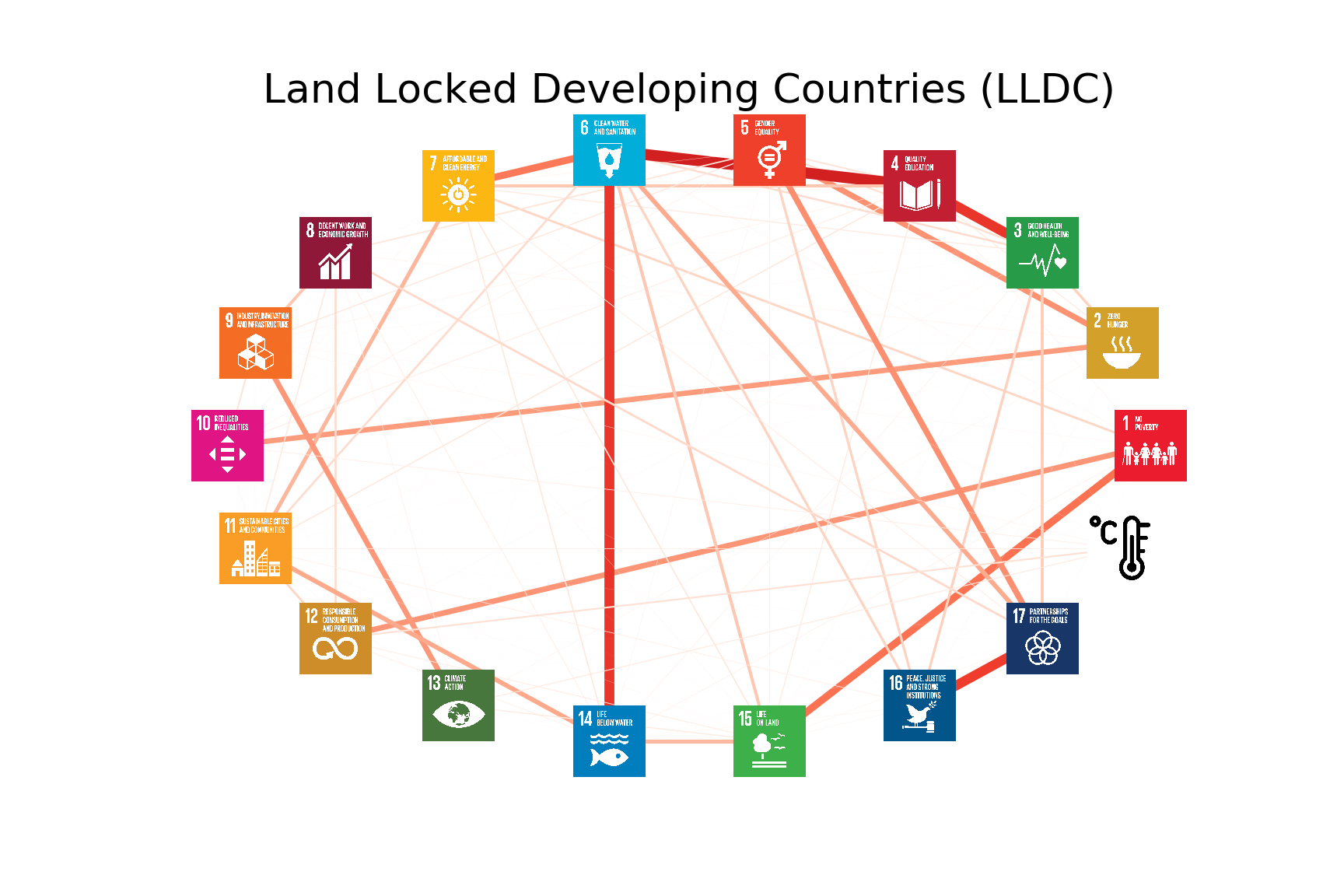}
\end{minipage}
\begin{minipage}{.04\textwidth}
  \centering
  \includegraphics[width=\linewidth]{networks/index.png}
\end{minipage}
%\end{figure}

%\begin{figure}[h]
%\centering
\begin{minipage}{.47\textwidth}
  \centering
  \includegraphics[width=\linewidth]{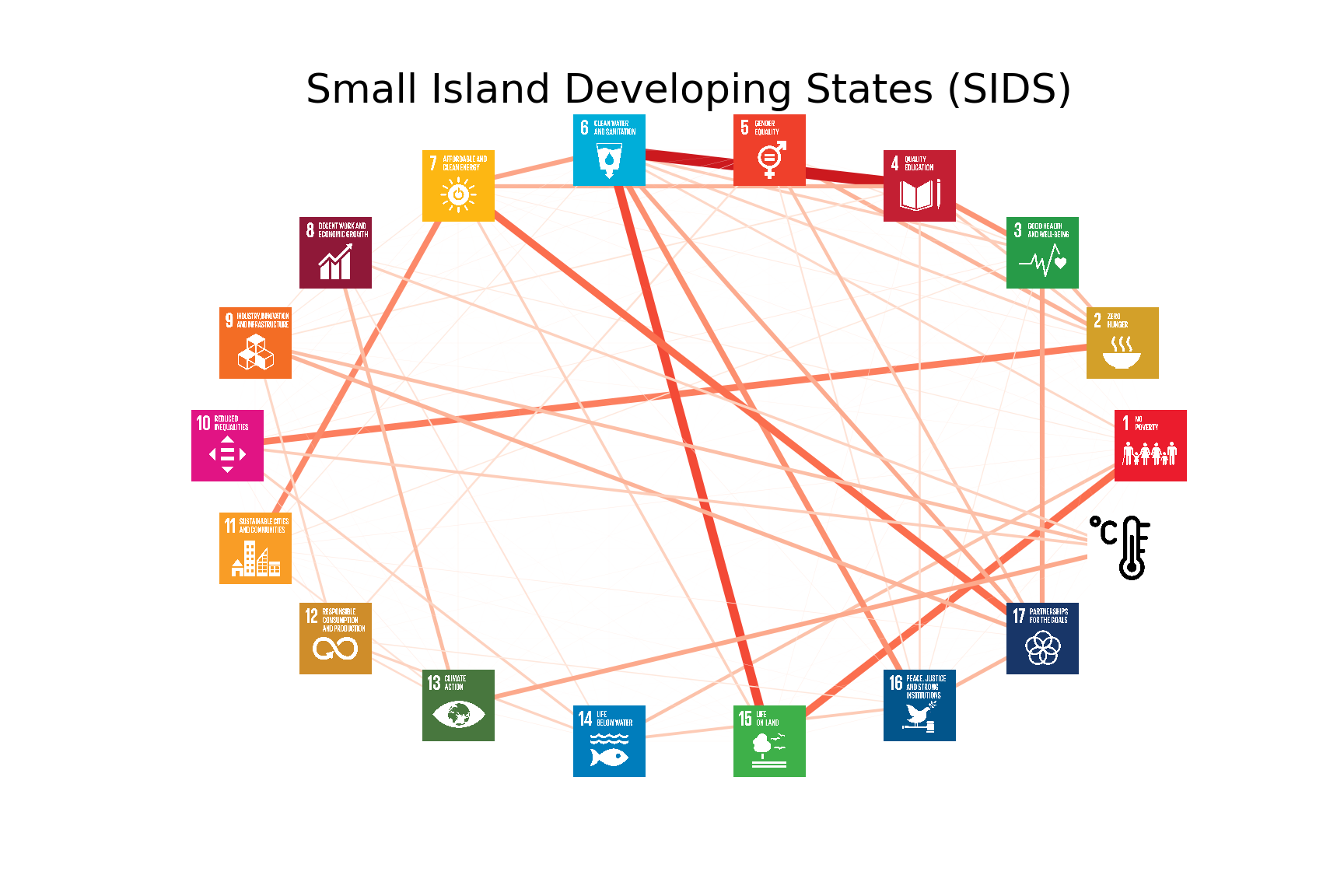}
\end{minipage}
\begin{minipage}{.47\textwidth}
  \centering
  \includegraphics[width=\linewidth]{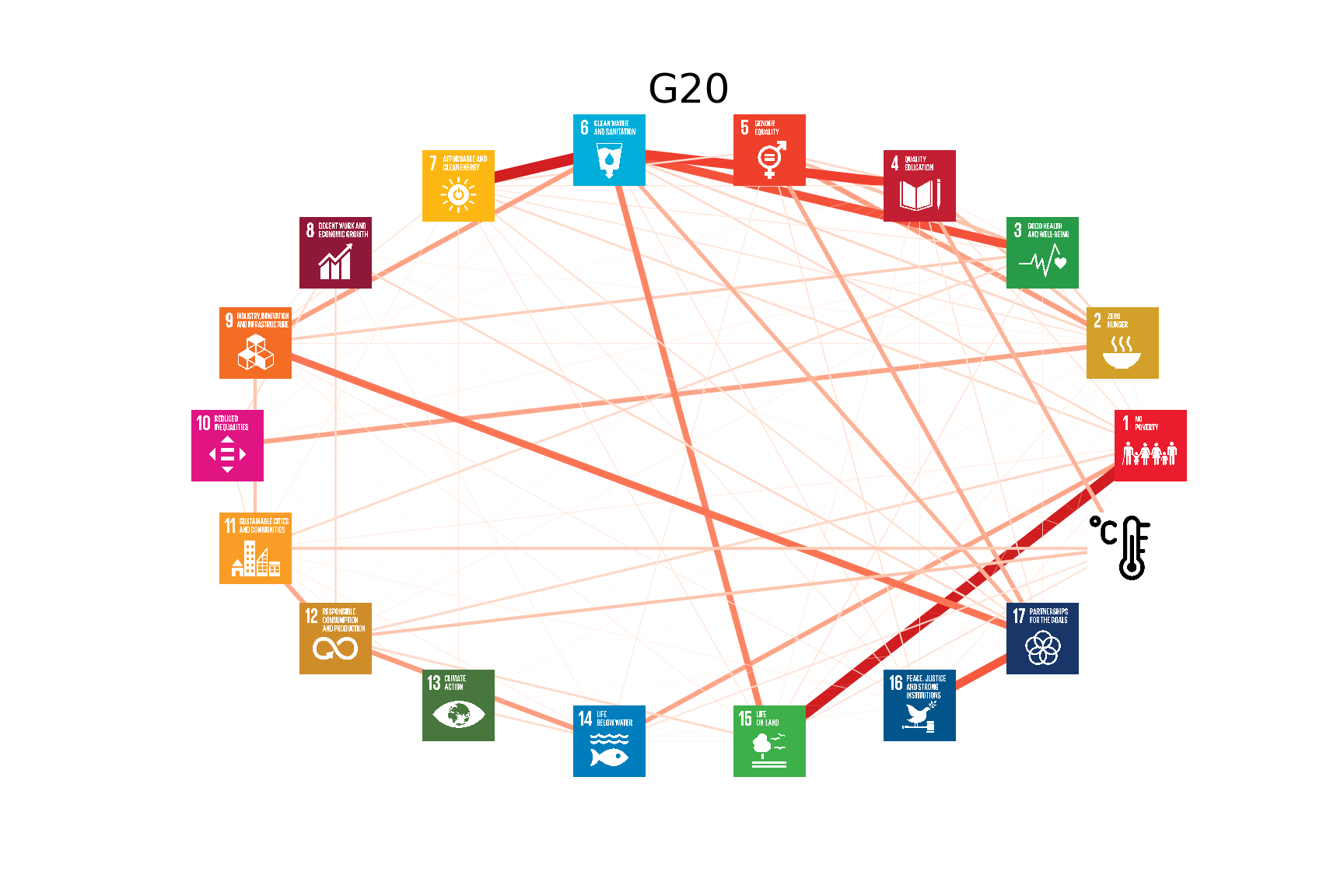}
\end{minipage}
\begin{minipage}{.04\textwidth}
  \centering
  \includegraphics[width=\linewidth]{networks/index.png}
\end{minipage}
%\end{figure}

%\begin{figure}[h]
%\centering
\begin{minipage}{.47\textwidth}
  \centering
  \includegraphics[width=\linewidth]{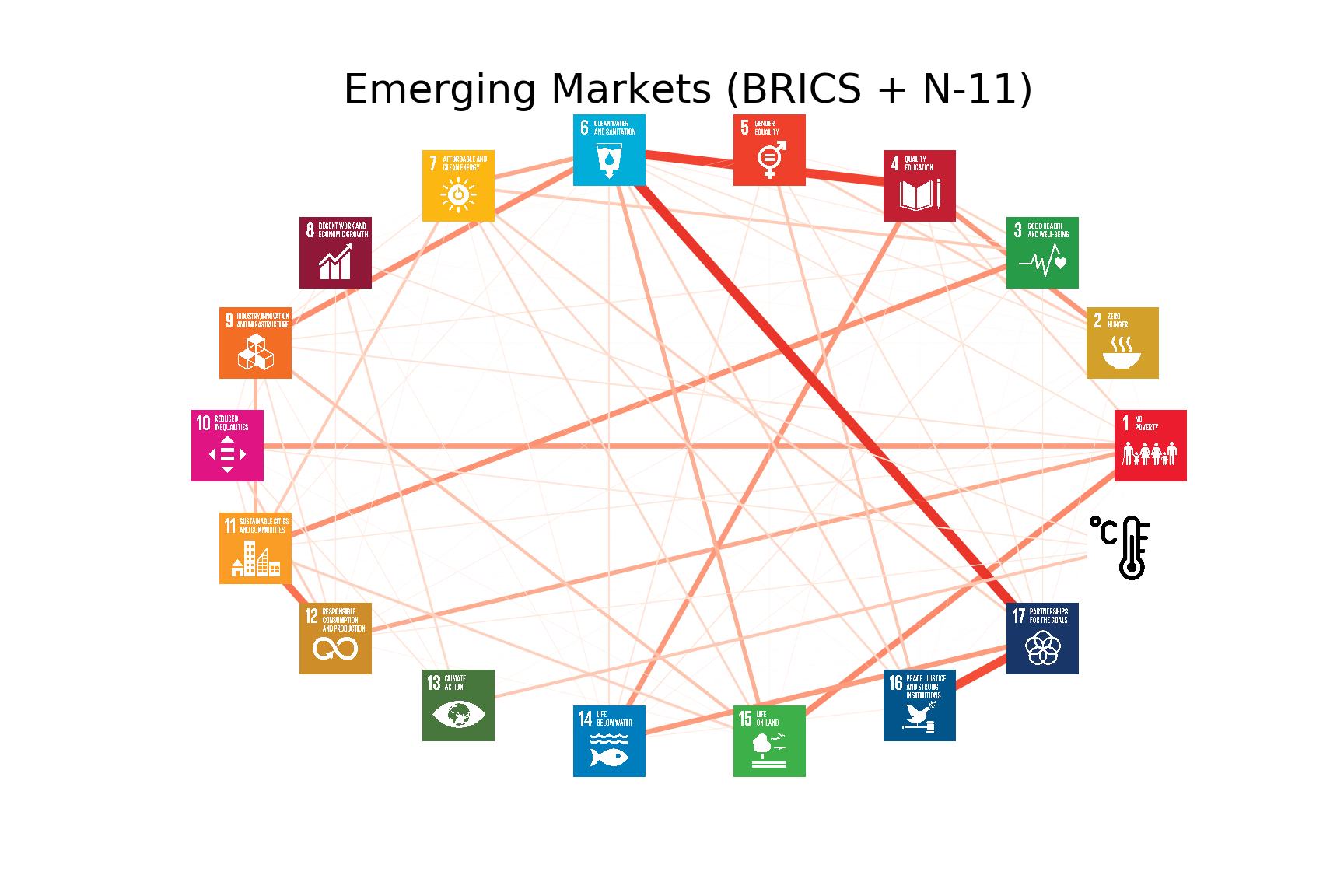}
\end{minipage}
\begin{minipage}{.47\textwidth}
  \centering
  \includegraphics[width=\linewidth]{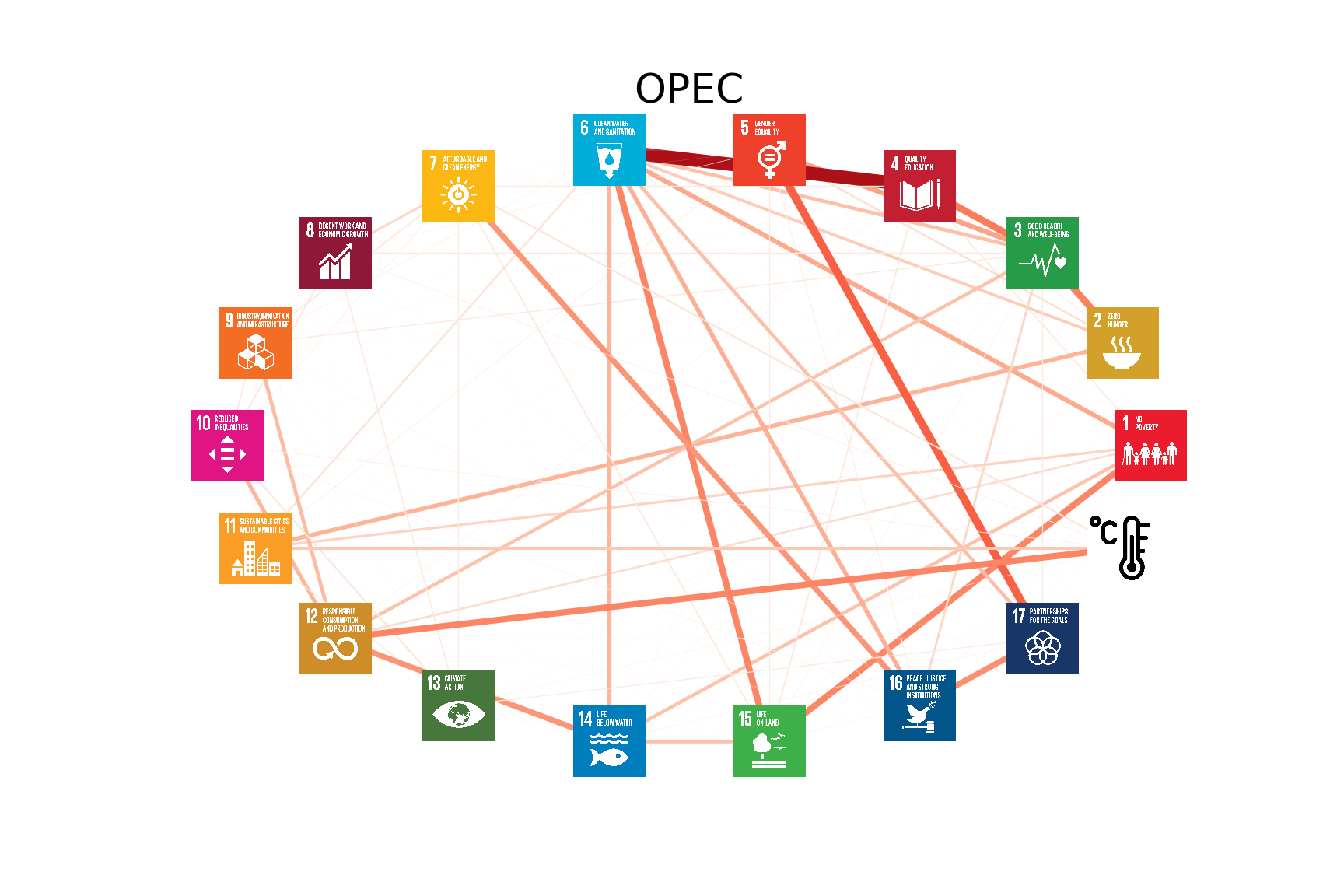}
\end{minipage}
\begin{minipage}{.04\textwidth}
  \centering
  \includegraphics[width=\linewidth]{networks/index.png}
\end{minipage}
%\end{figure}

%\begin{figure}[h]
%\centering
\begin{minipage}{.47\textwidth}
  \centering
  \includegraphics[width=\linewidth]{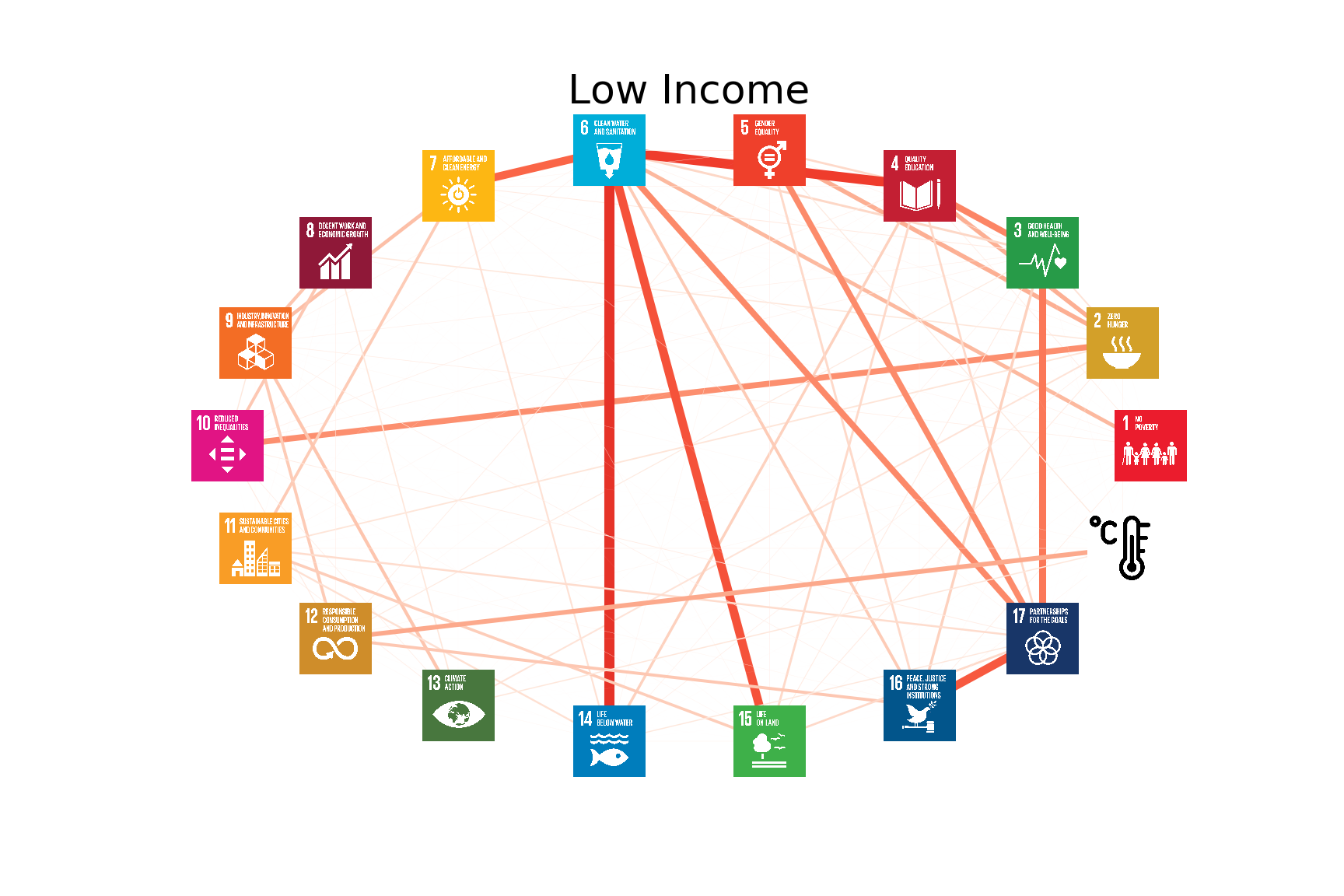}
\end{minipage}
\begin{minipage}{.47\textwidth}
  \centering
  \includegraphics[width=\linewidth]{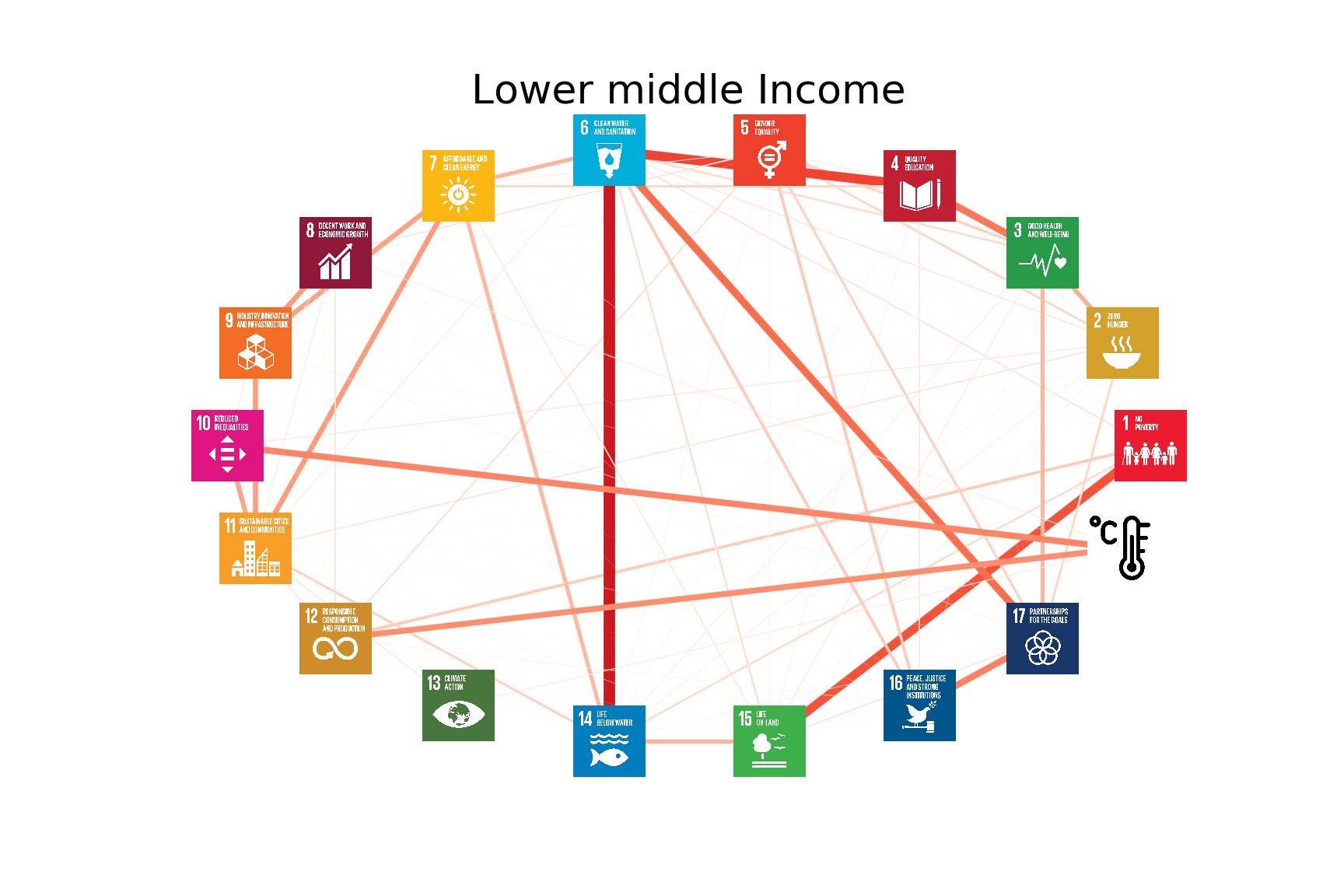}
\end{minipage}
\begin{minipage}{.04\textwidth}
  \centering
  \includegraphics[width=\linewidth]{networks/index.png}
\end{minipage}
%\end{figure}

%\begin{figure}[h]
%\centering
\begin{minipage}{.47\textwidth}
  \centering
  \includegraphics[width=\linewidth]{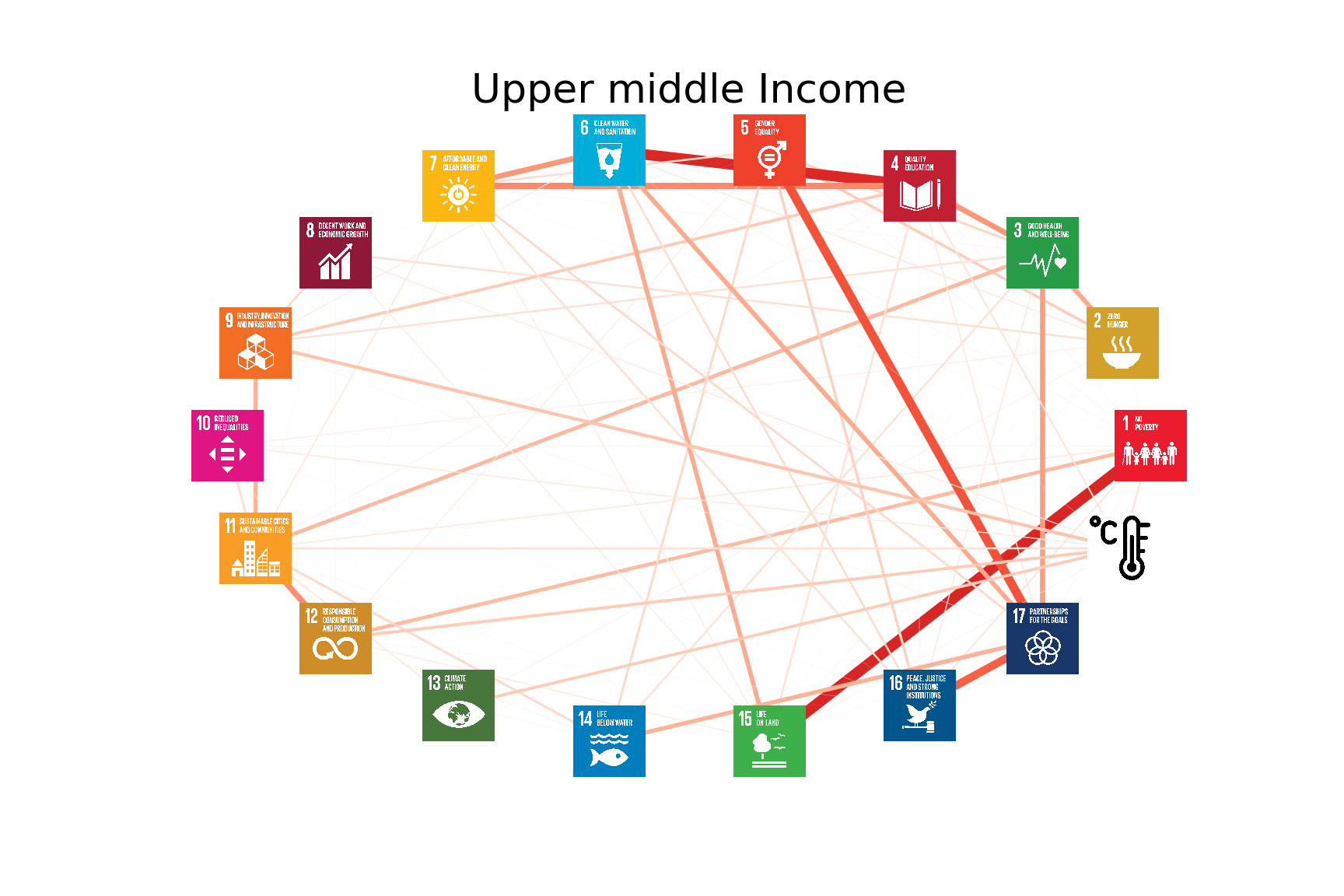}
\end{minipage}
\begin{minipage}{.47\textwidth}
  \centering
  \includegraphics[width=\linewidth]{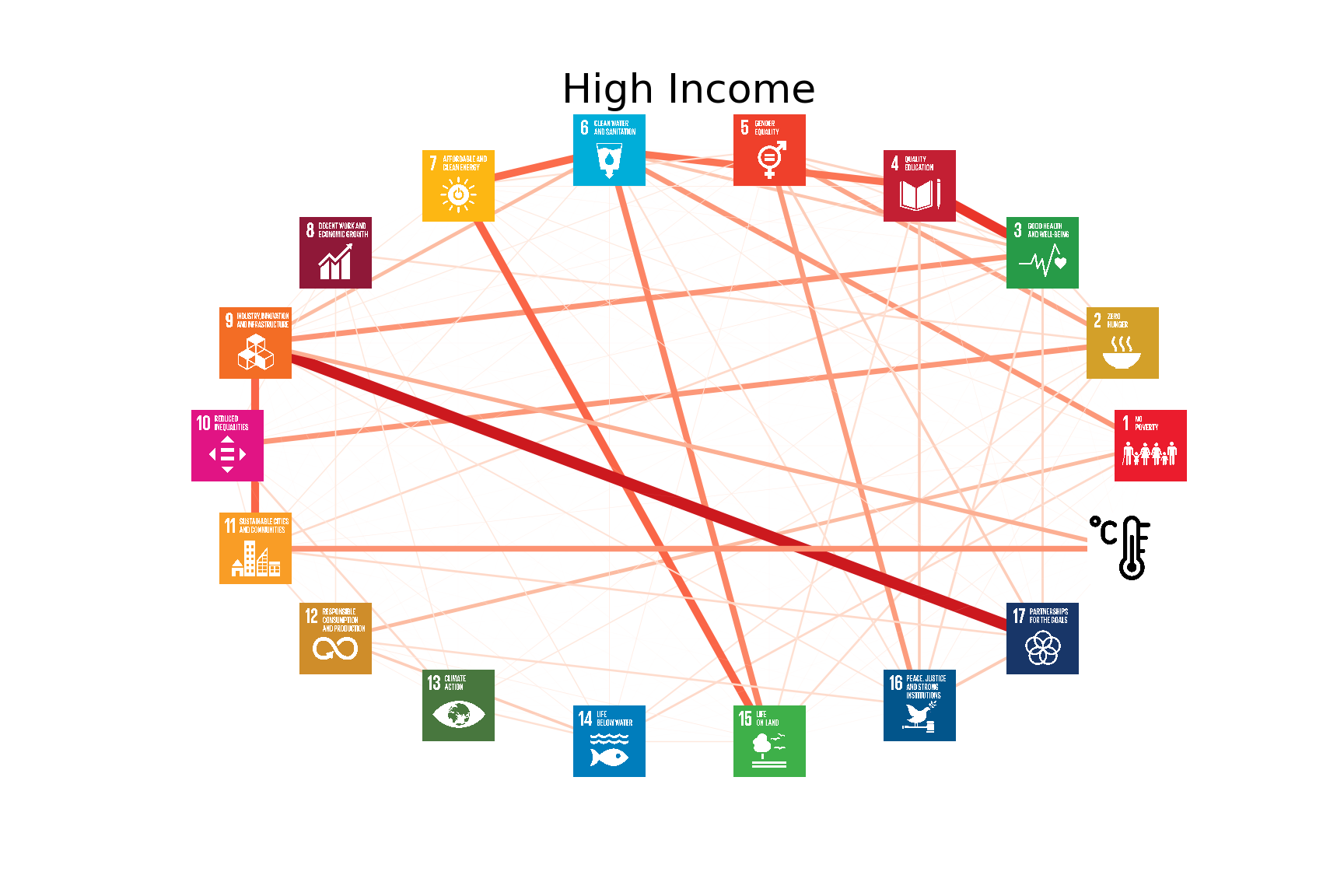}
\end{minipage}
\begin{minipage}{.04\textwidth}
  \centering
  \includegraphics[width=\linewidth]{networks/index.png}
\end{minipage}
%\end{figure}

\newpage

%\begin{landscape}
\subsection{Eigenvector centralities} \label{app_central}

%\begin{figure}[!h]
%\centering
\begin{minipage}{.47\textwidth}
  \centering
  \includegraphics[width=\linewidth]{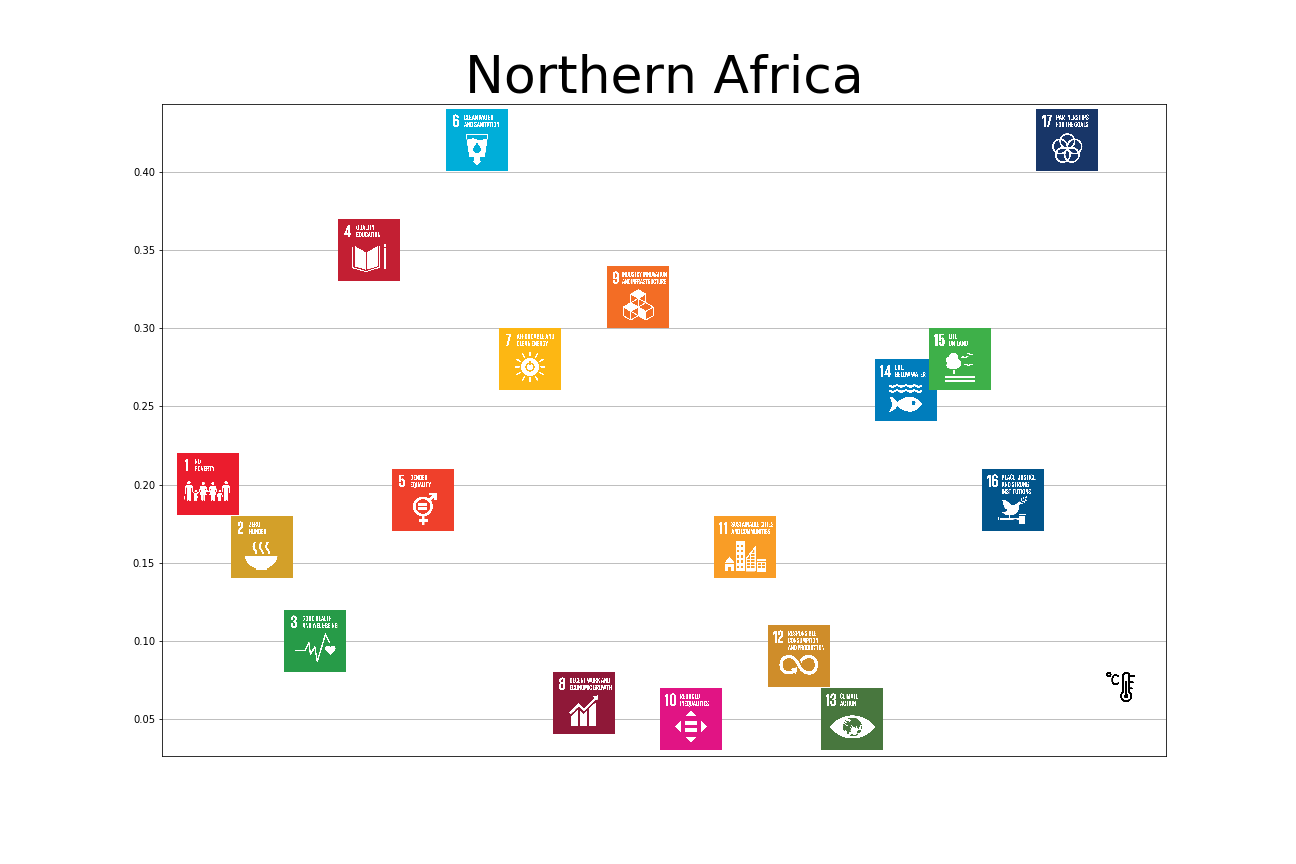}
\end{minipage}
\begin{minipage}{.47\textwidth}
  \centering
  \includegraphics[width=\linewidth]{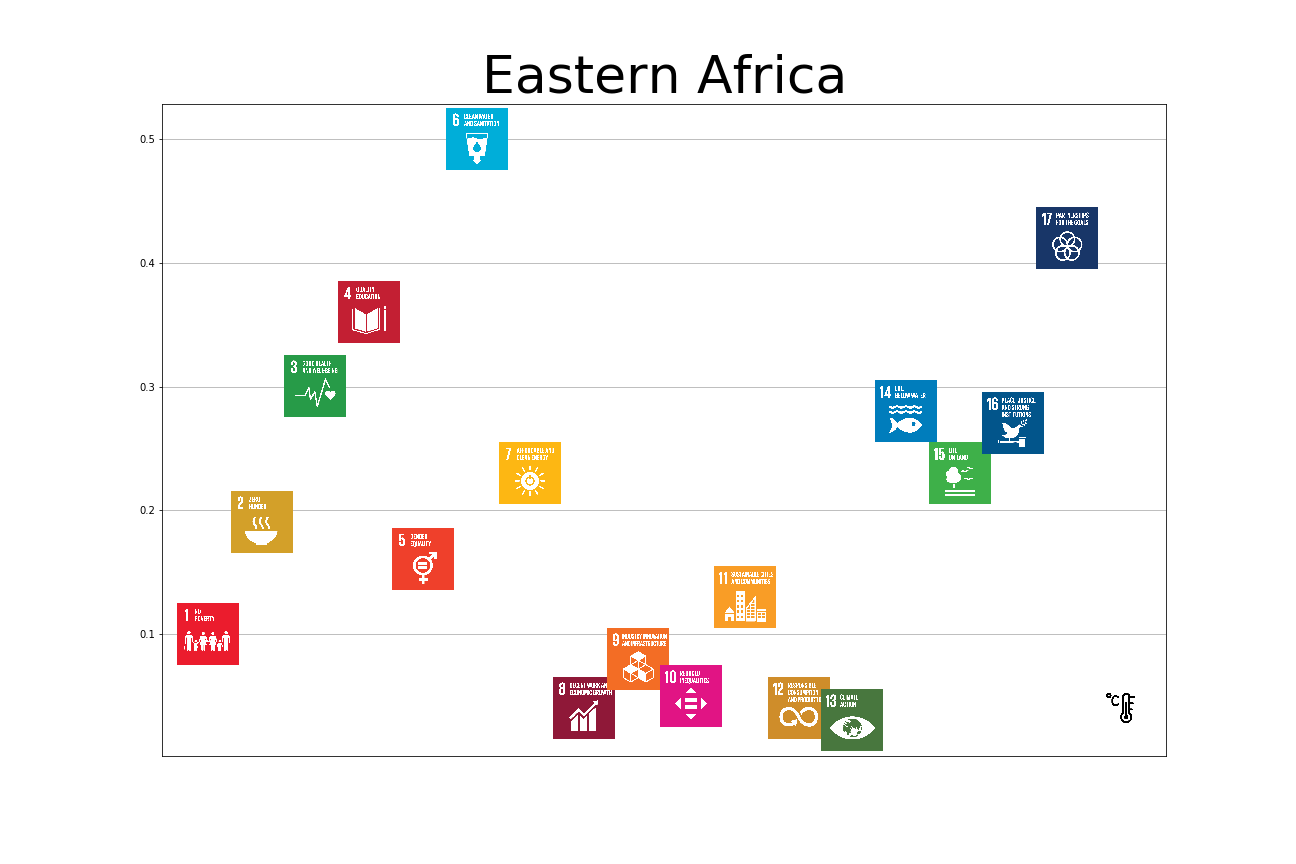}
\end{minipage}
%\end{figure}

%\begin{figure}[!h]
%\centering
\begin{minipage}{.47\textwidth}
  \centering
  \includegraphics[width=\linewidth]{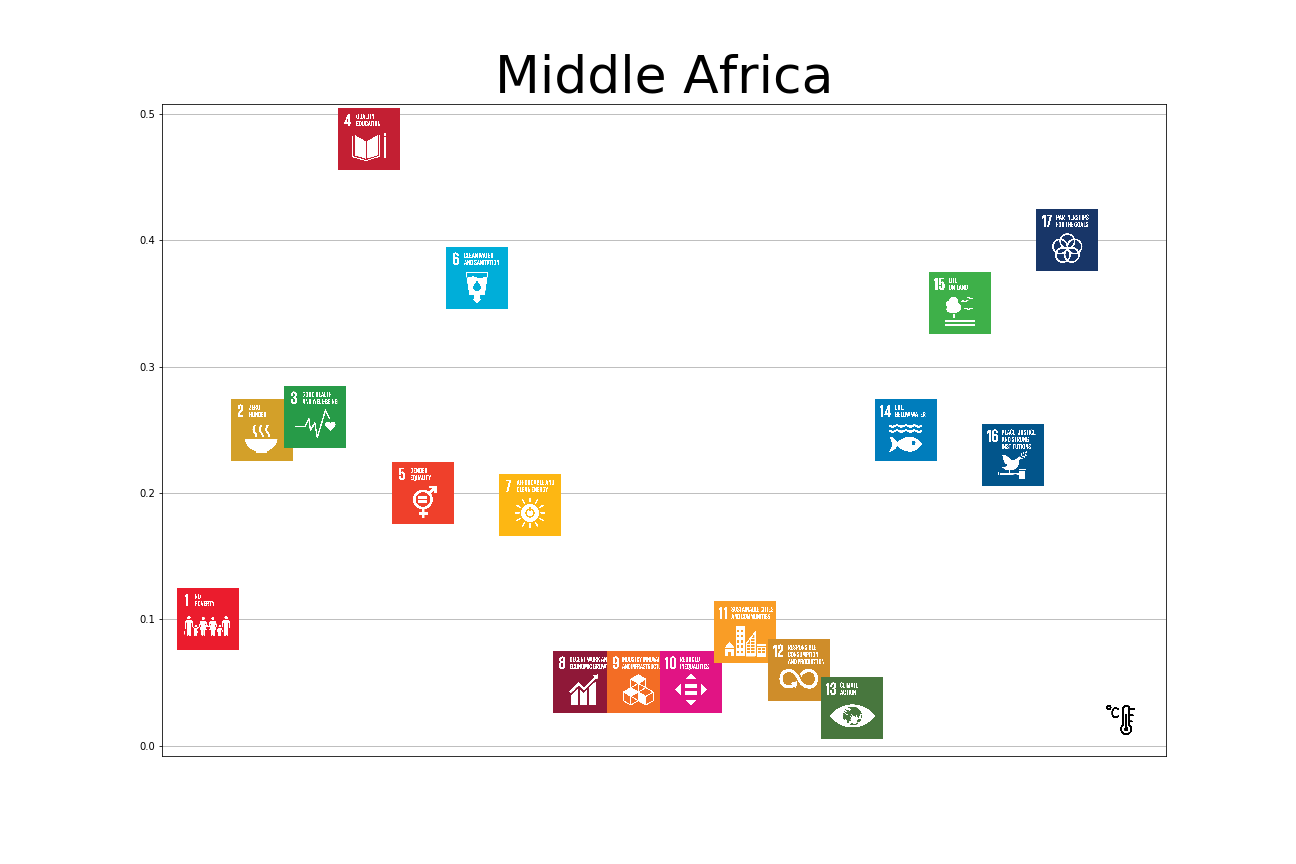}
\end{minipage}
\begin{minipage}{.47\textwidth}
  \centering
  \includegraphics[width=\linewidth]{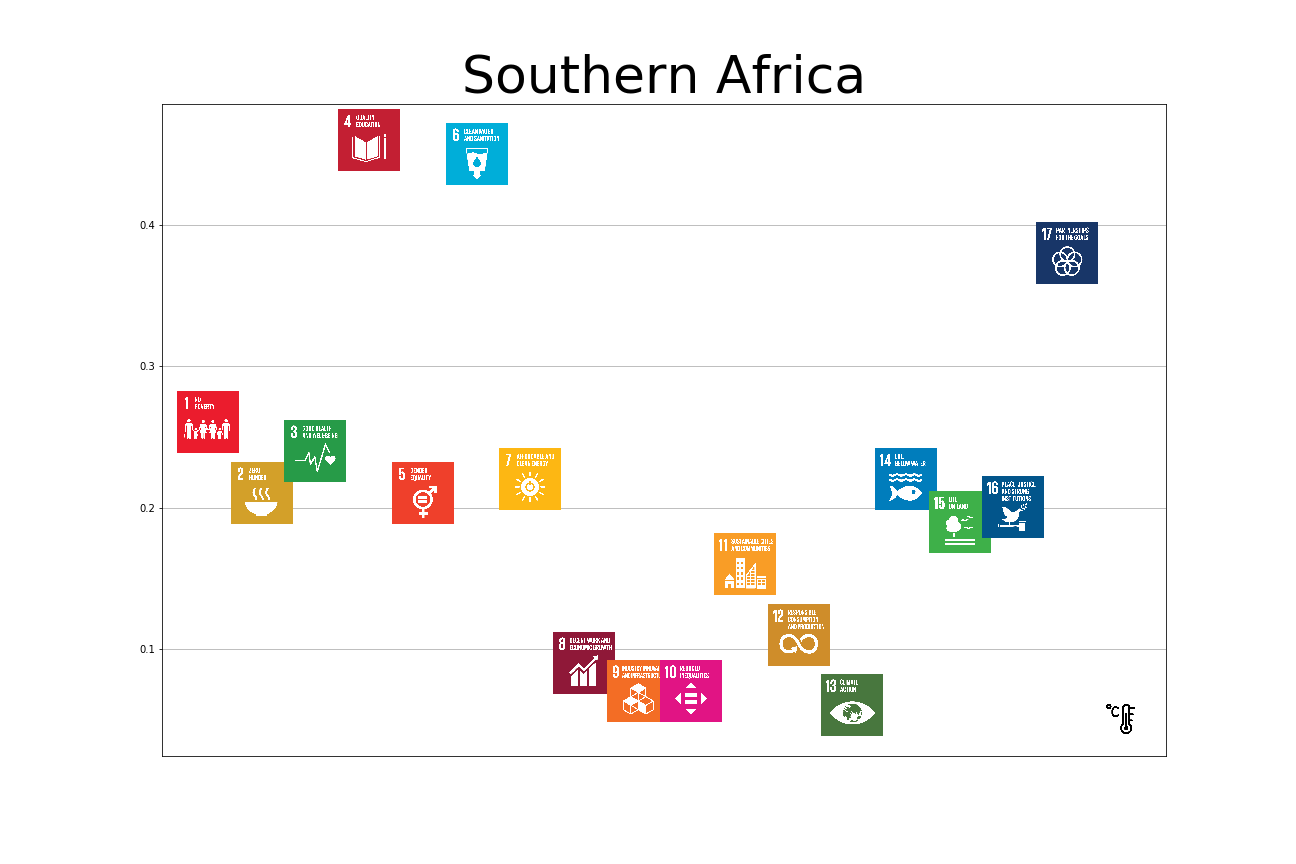}
\end{minipage}
%\end{figure}

%\begin{figure}[!h]
%\centering
\begin{minipage}{.47\textwidth}
  \centering
  \includegraphics[width=\linewidth]{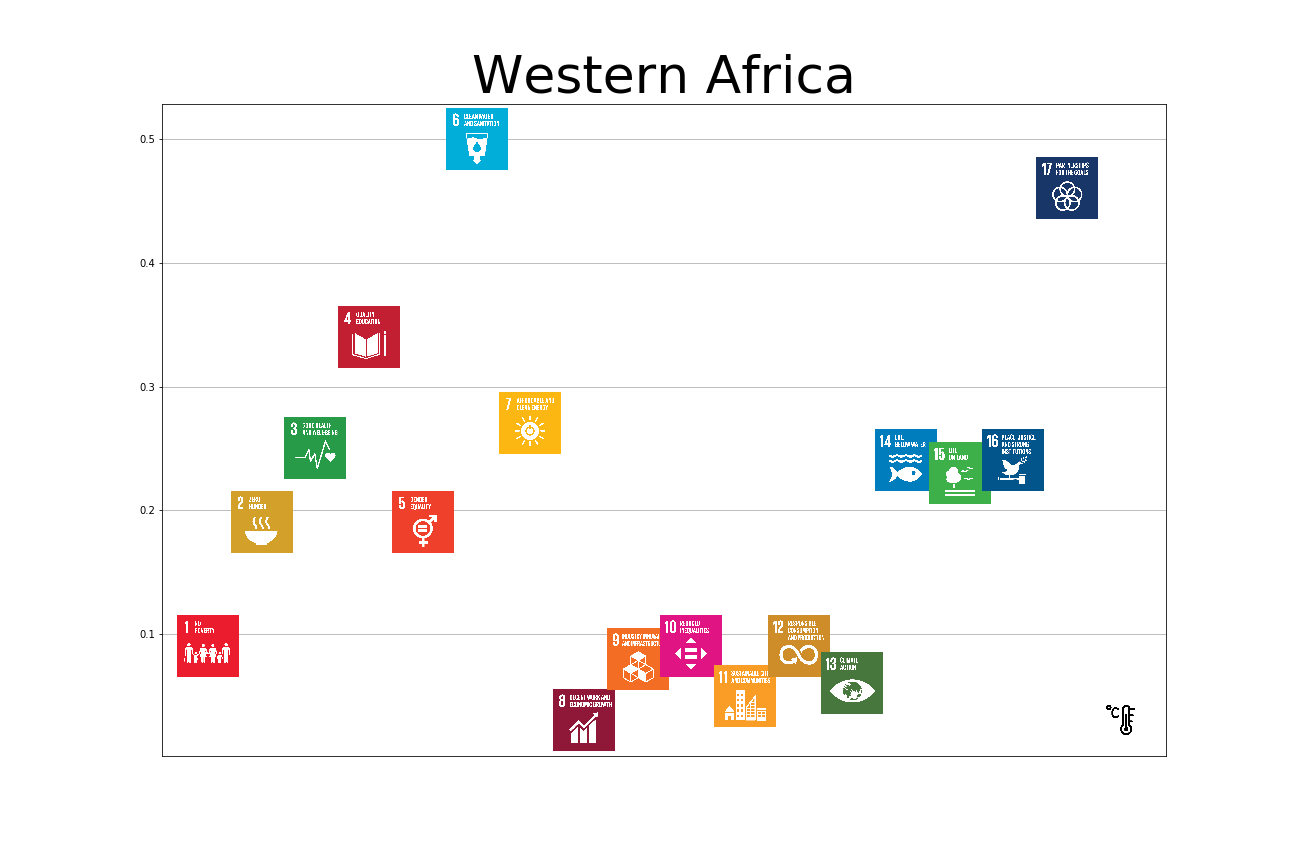}
\end{minipage}
\begin{minipage}{.47\textwidth}
  \centering
  \includegraphics[width=\linewidth]{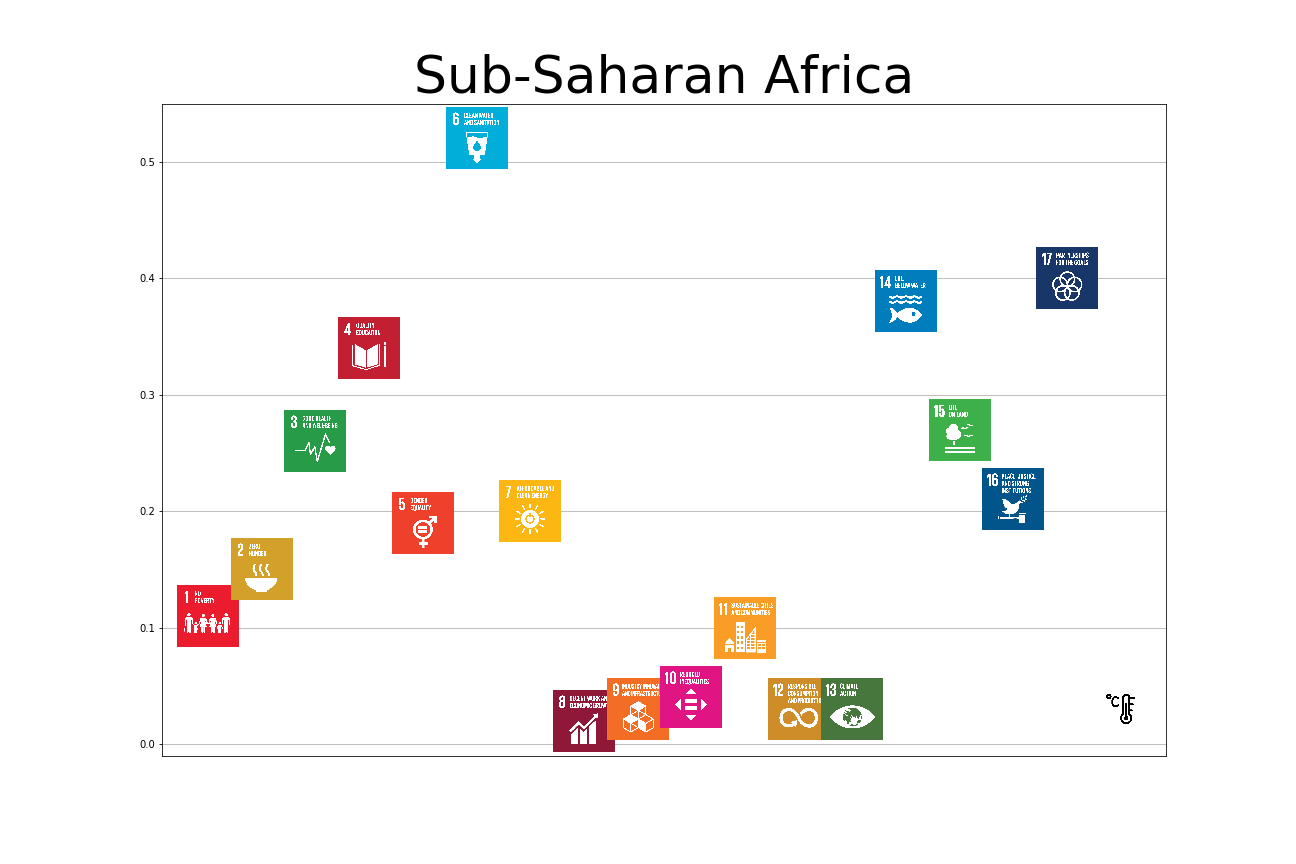}
\end{minipage}
%\end{figure}

%\begin{figure}[!h]
\centering
\begin{minipage}{.47\textwidth}
  \centering
  \includegraphics[width=\linewidth]{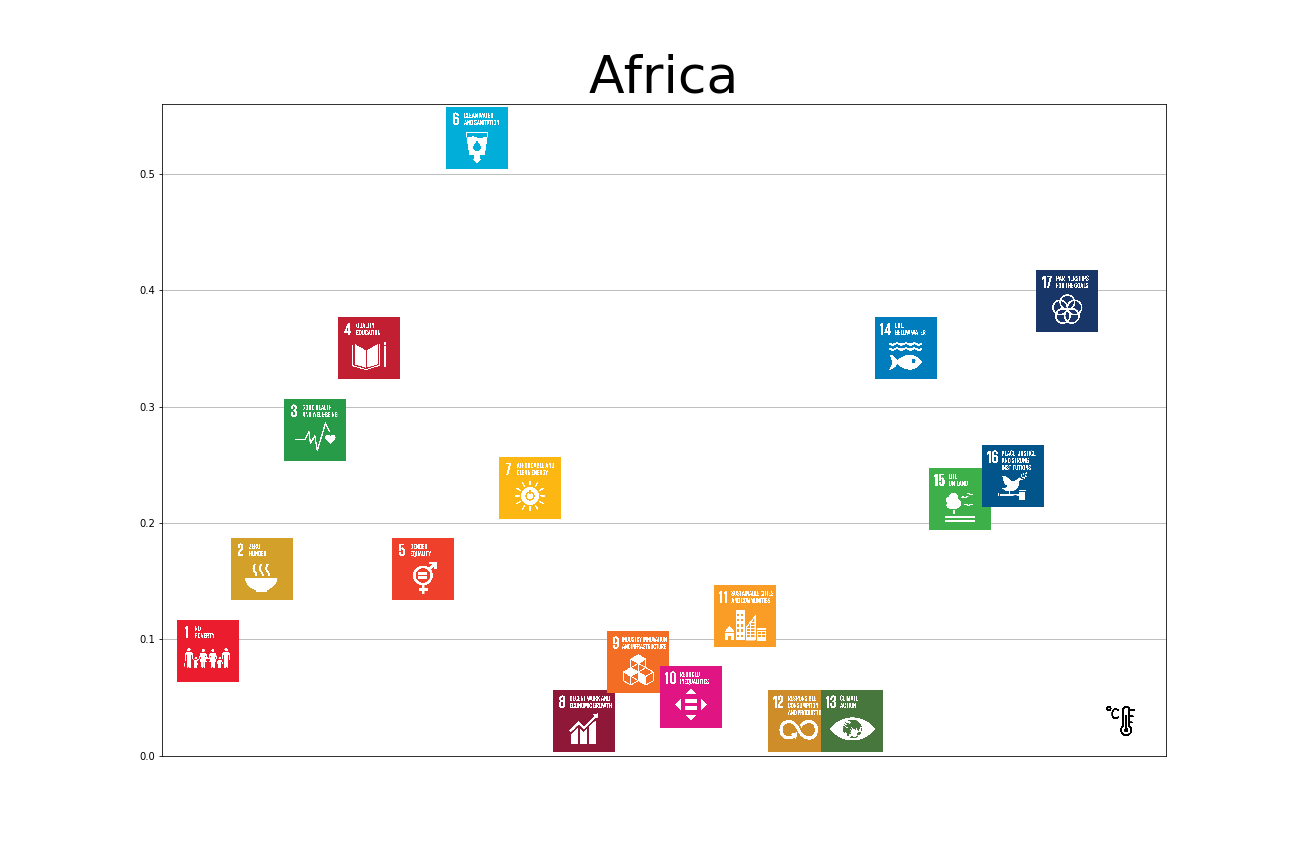}
\end{minipage}
%\end{figure}

%\begin{figure}[!h]
%\centering
\begin{minipage}{.47\textwidth}
  \centering
  \includegraphics[width=\linewidth]{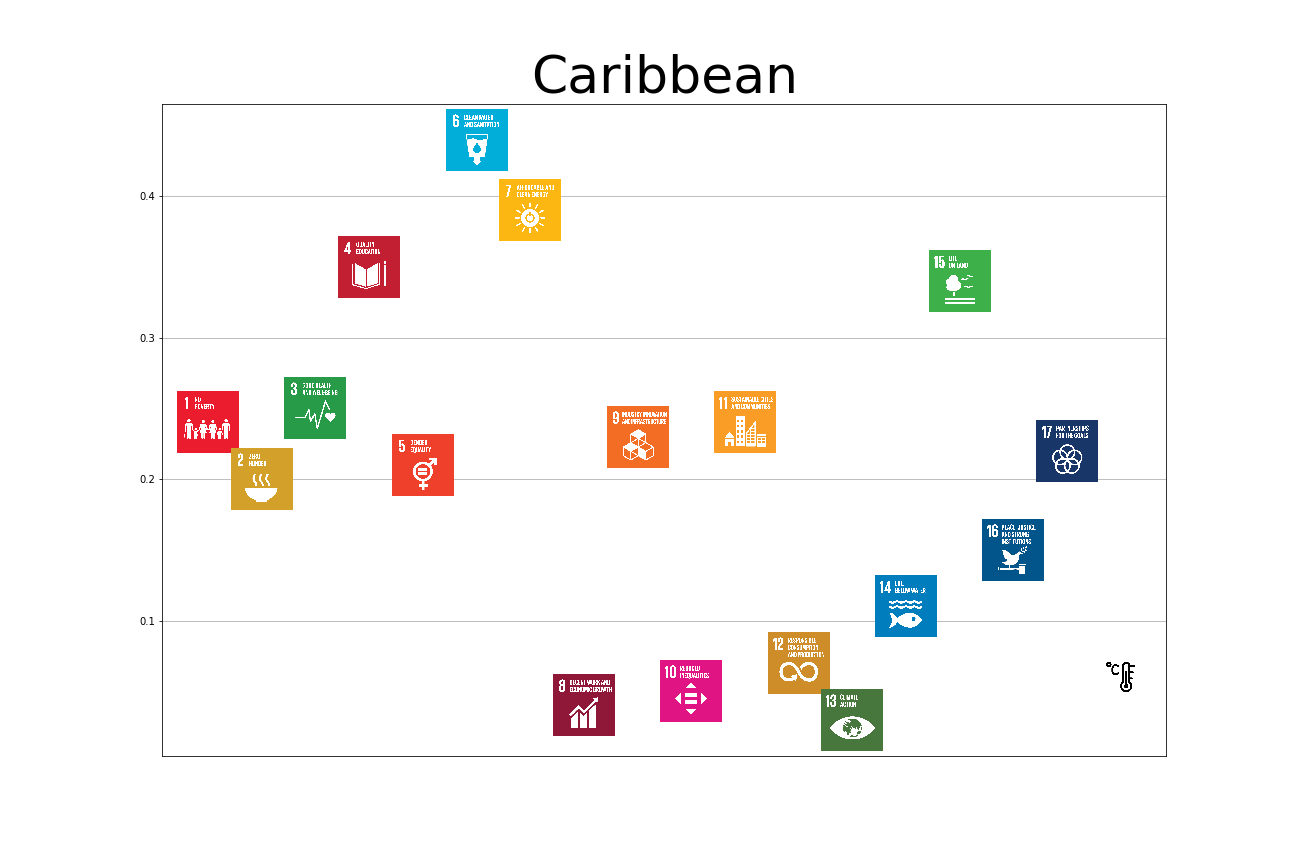}
\end{minipage}
\begin{minipage}{.47\textwidth}
  \centering
  \includegraphics[width=\linewidth]{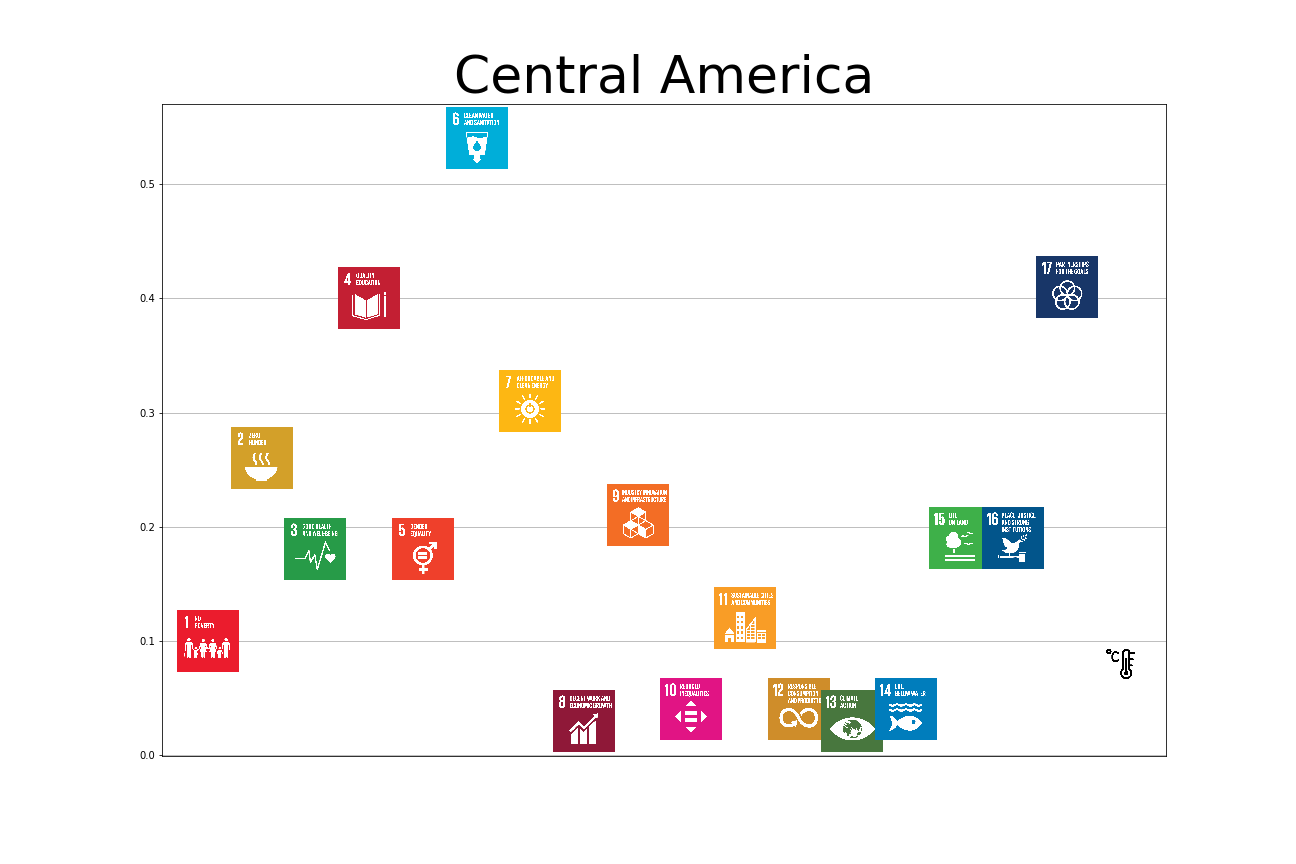}
\end{minipage}
%\end{figure}

%\begin{figure}[!h]
%\centering
\begin{minipage}{.47\textwidth}
  \centering
  \includegraphics[width=\linewidth]{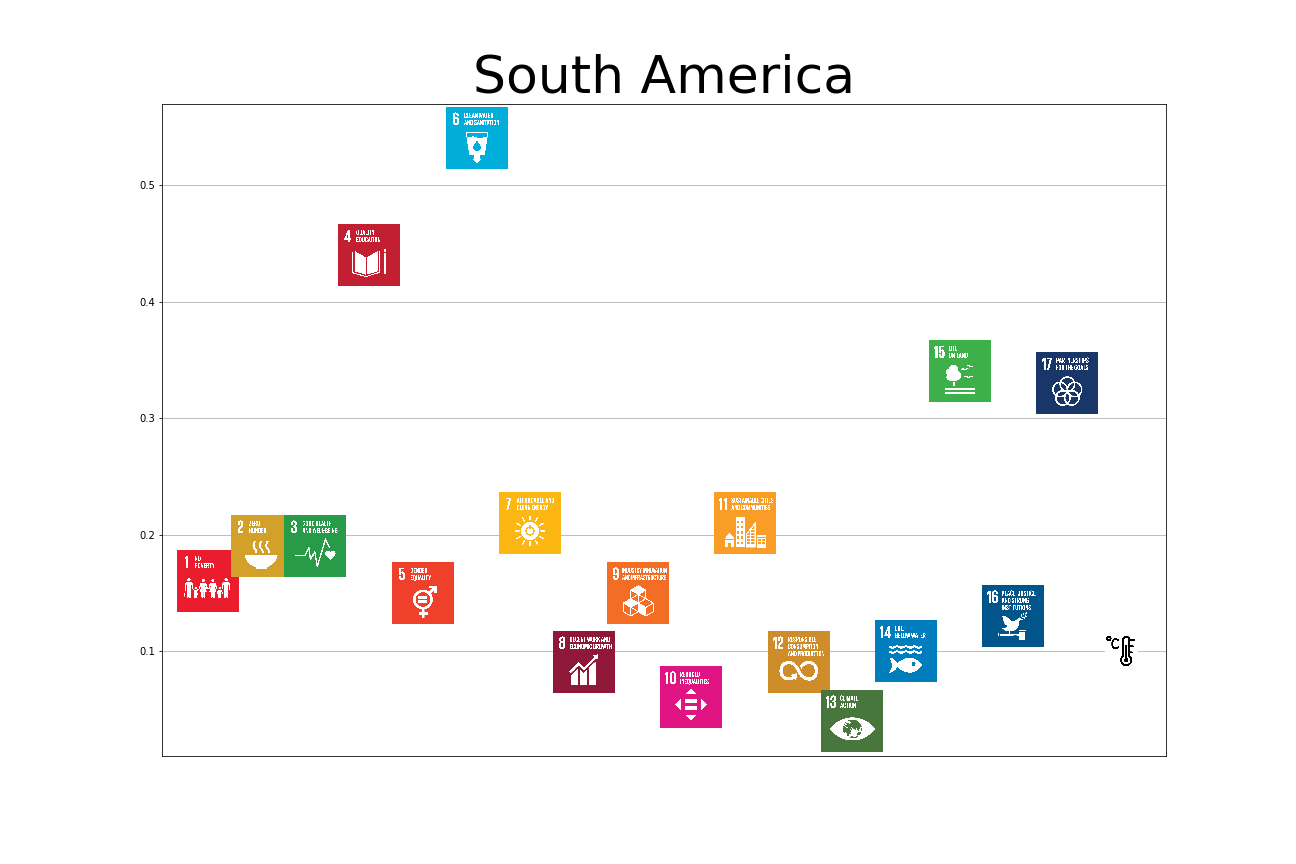}
\end{minipage}
\begin{minipage}{.47\textwidth}
  \centering
  \includegraphics[width=\linewidth]{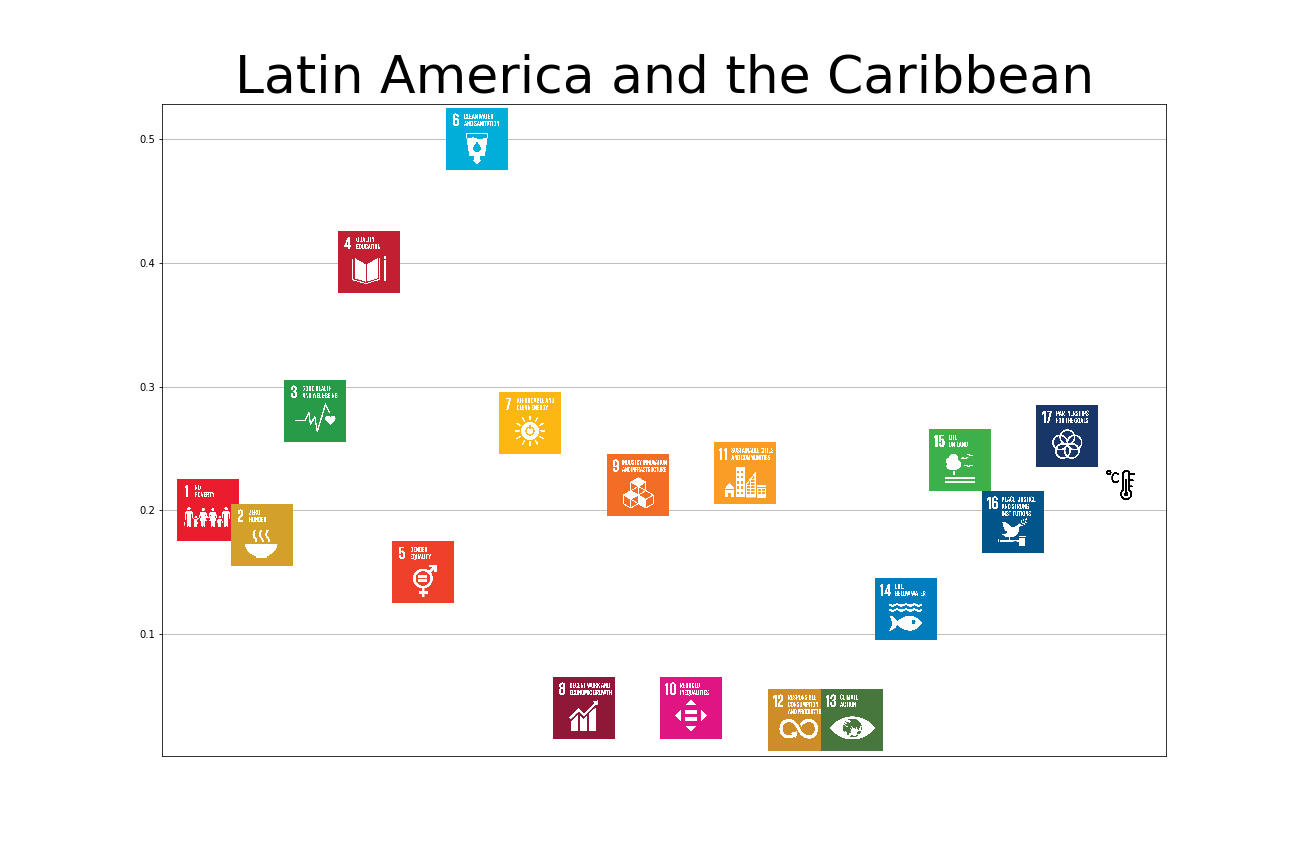}
\end{minipage}
%\end{figure}

%\begin{figure}[!h]
%\centering
\begin{minipage}{.47\textwidth}
  \centering
  \includegraphics[width=\linewidth]{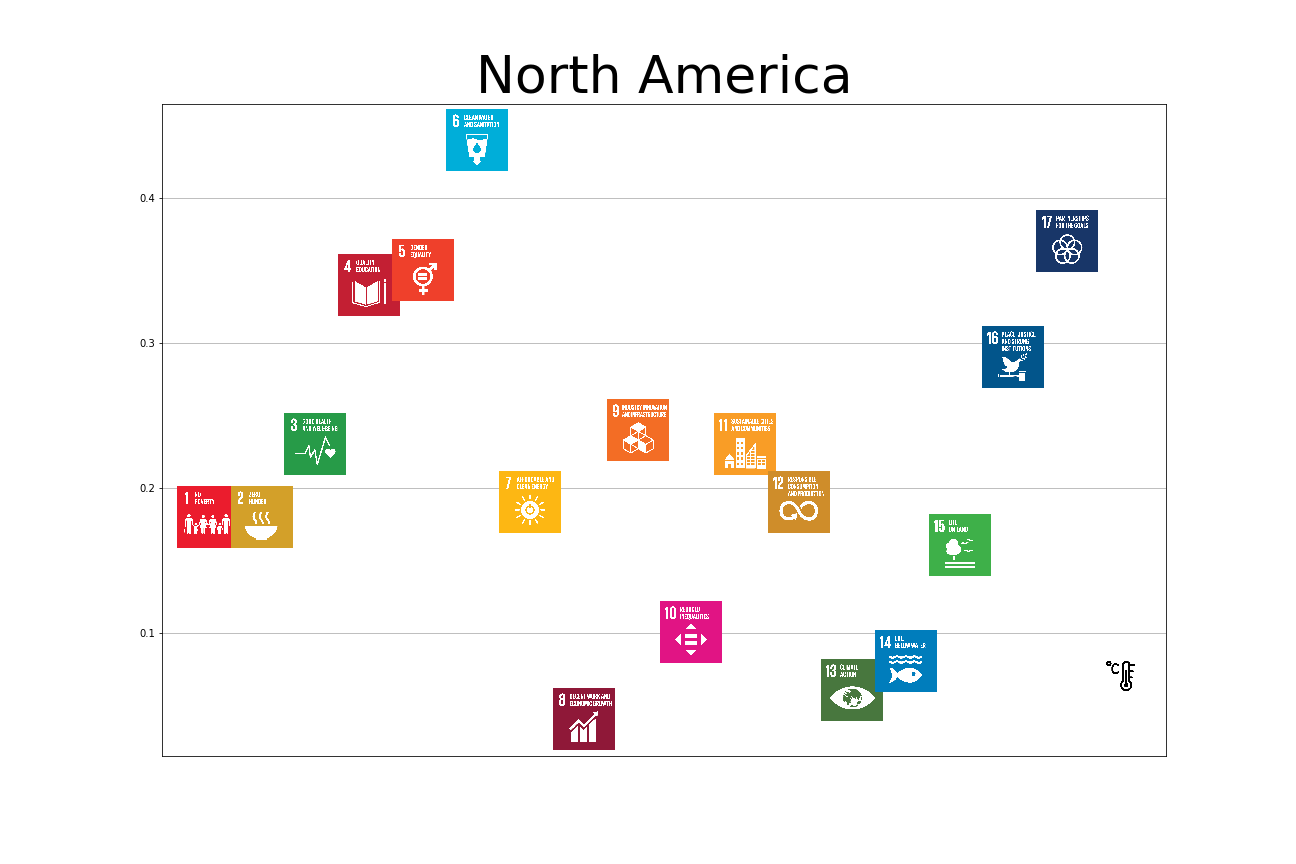}
\end{minipage}
\begin{minipage}{.47\textwidth}
  \centering
  \includegraphics[width=\linewidth]{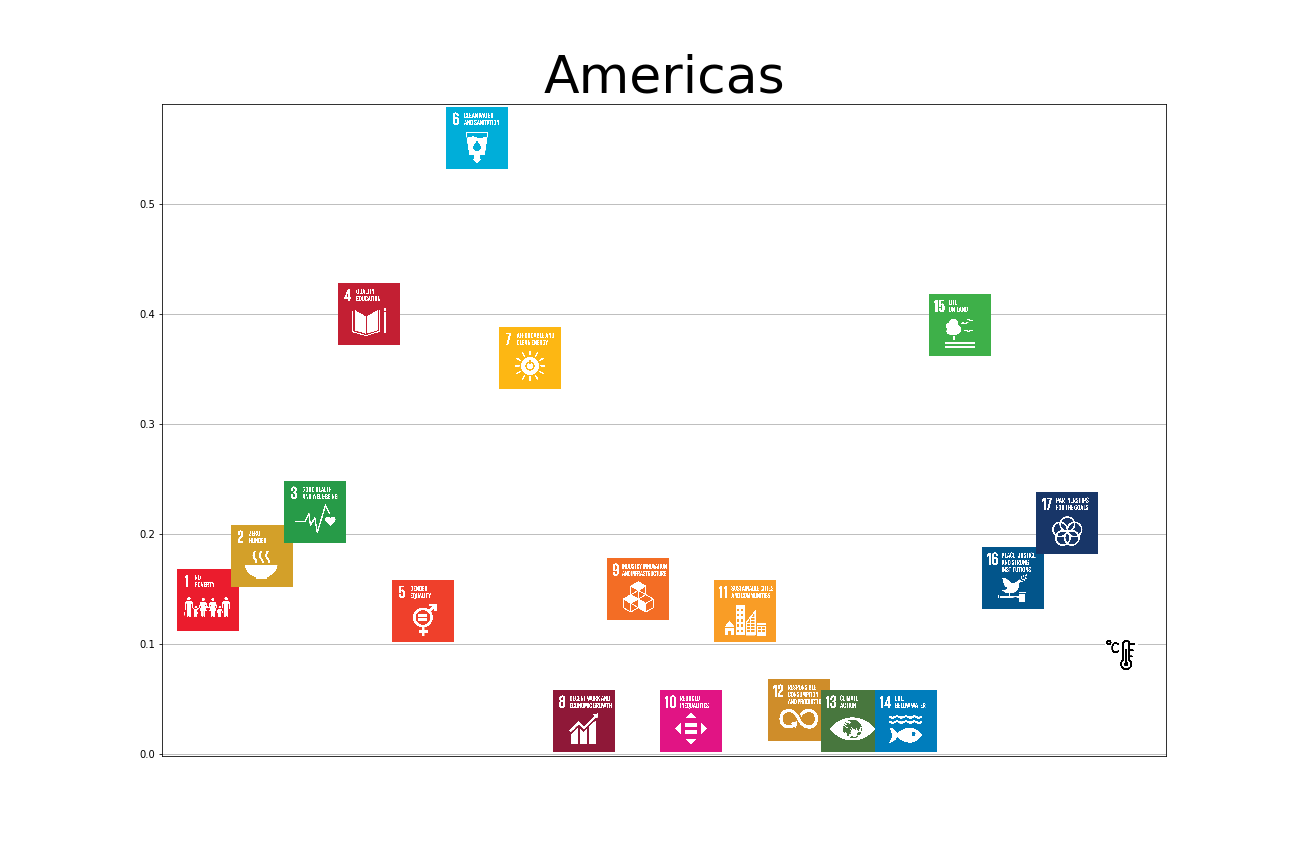}
\end{minipage}
%\end{figure}

%\begin{figure}[!h]
%\centering
\begin{minipage}{.47\textwidth}
  \centering
  \includegraphics[width=\linewidth]{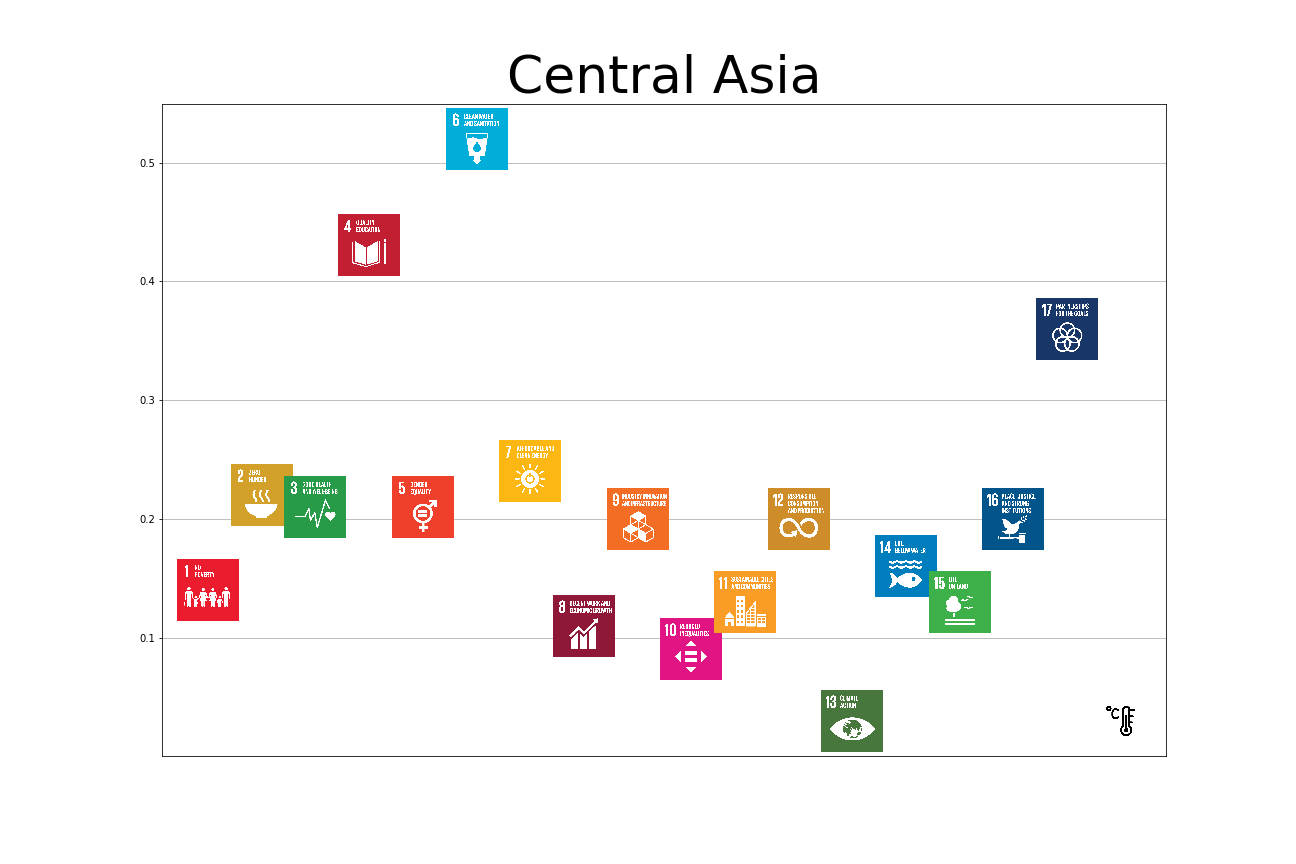}
\end{minipage}
\begin{minipage}{.47\textwidth}
  \centering
  \includegraphics[width=\linewidth]{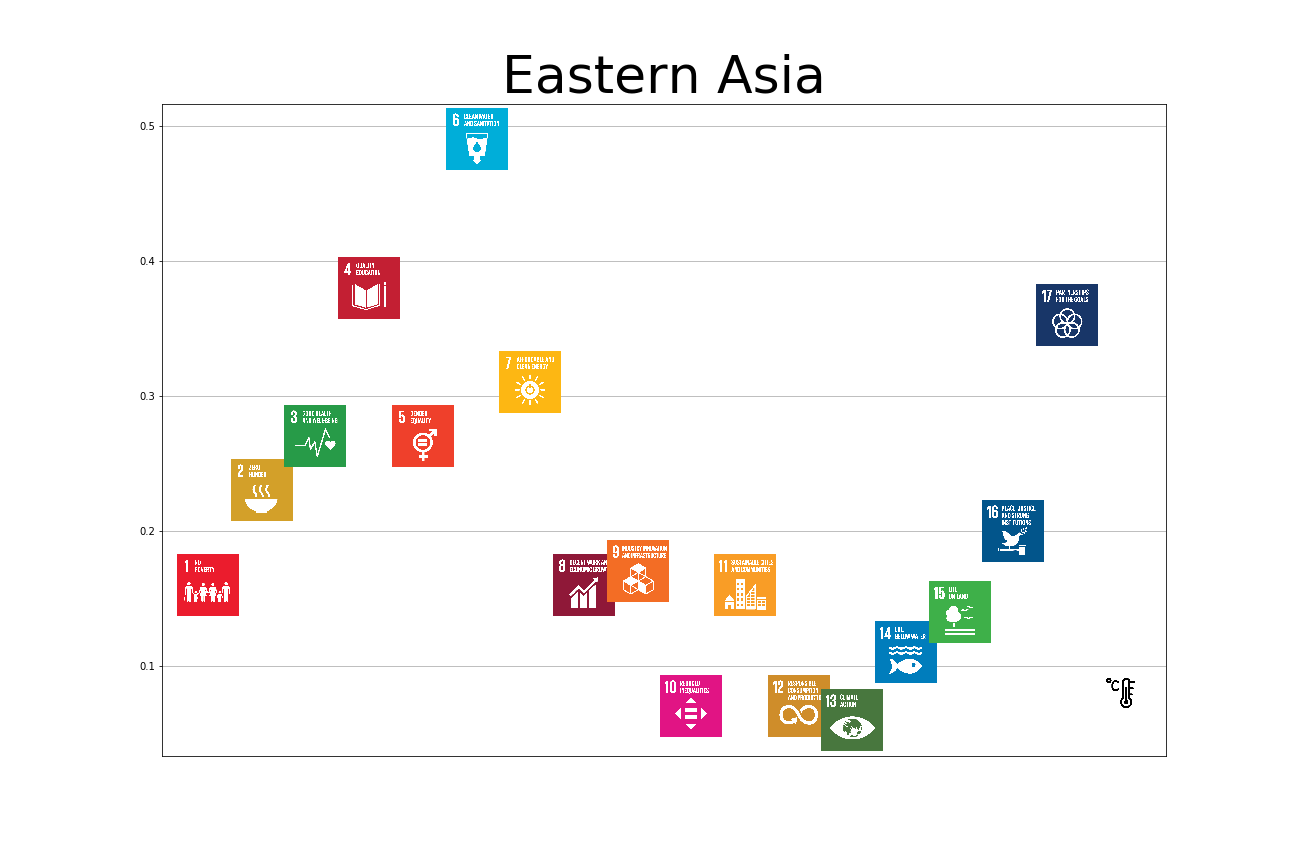}
\end{minipage}
%\end{figure}

%\begin{figure}[!h]
%\centering
\begin{minipage}{.47\textwidth}
  \centering
  \includegraphics[width=\linewidth]{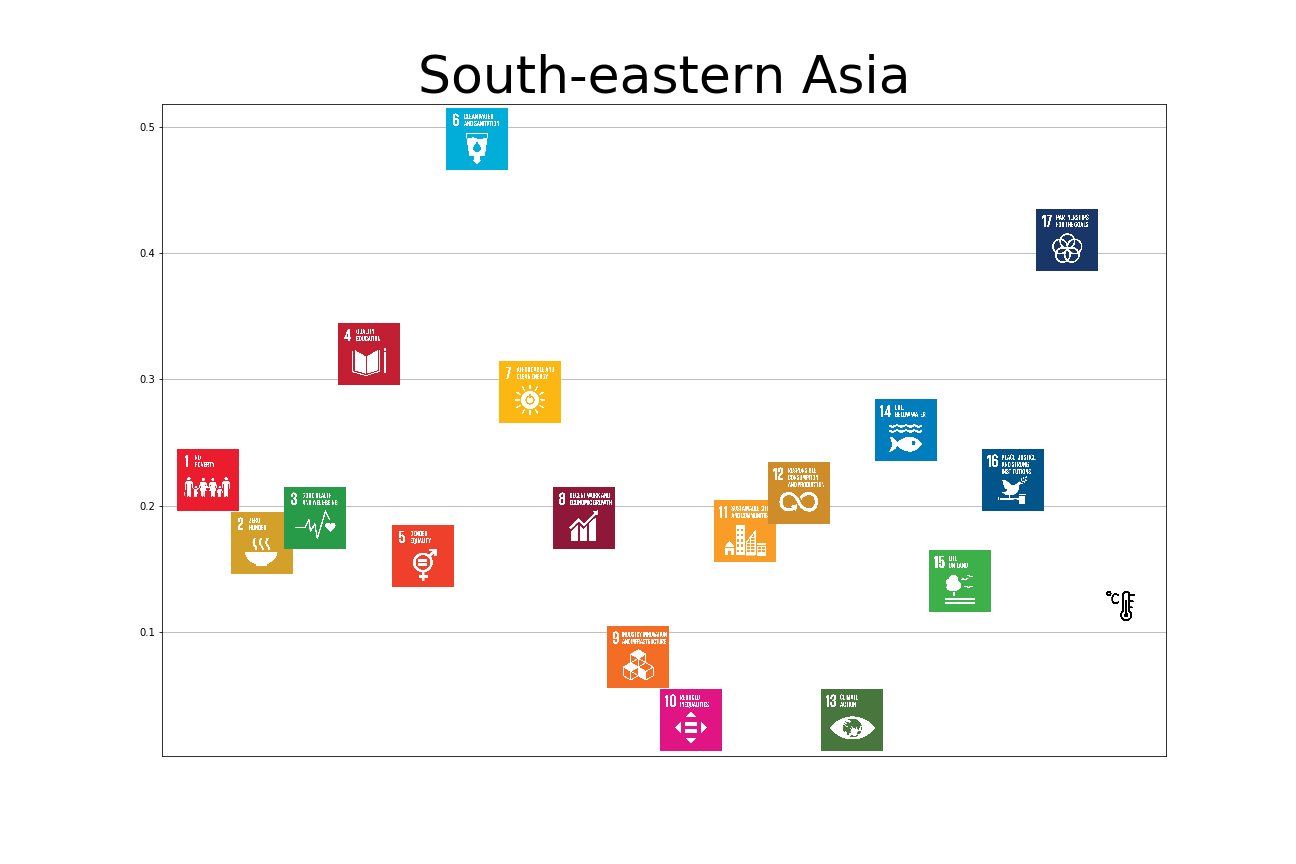}
\end{minipage}
\begin{minipage}{.47\textwidth}
  \centering
  \includegraphics[width=\linewidth]{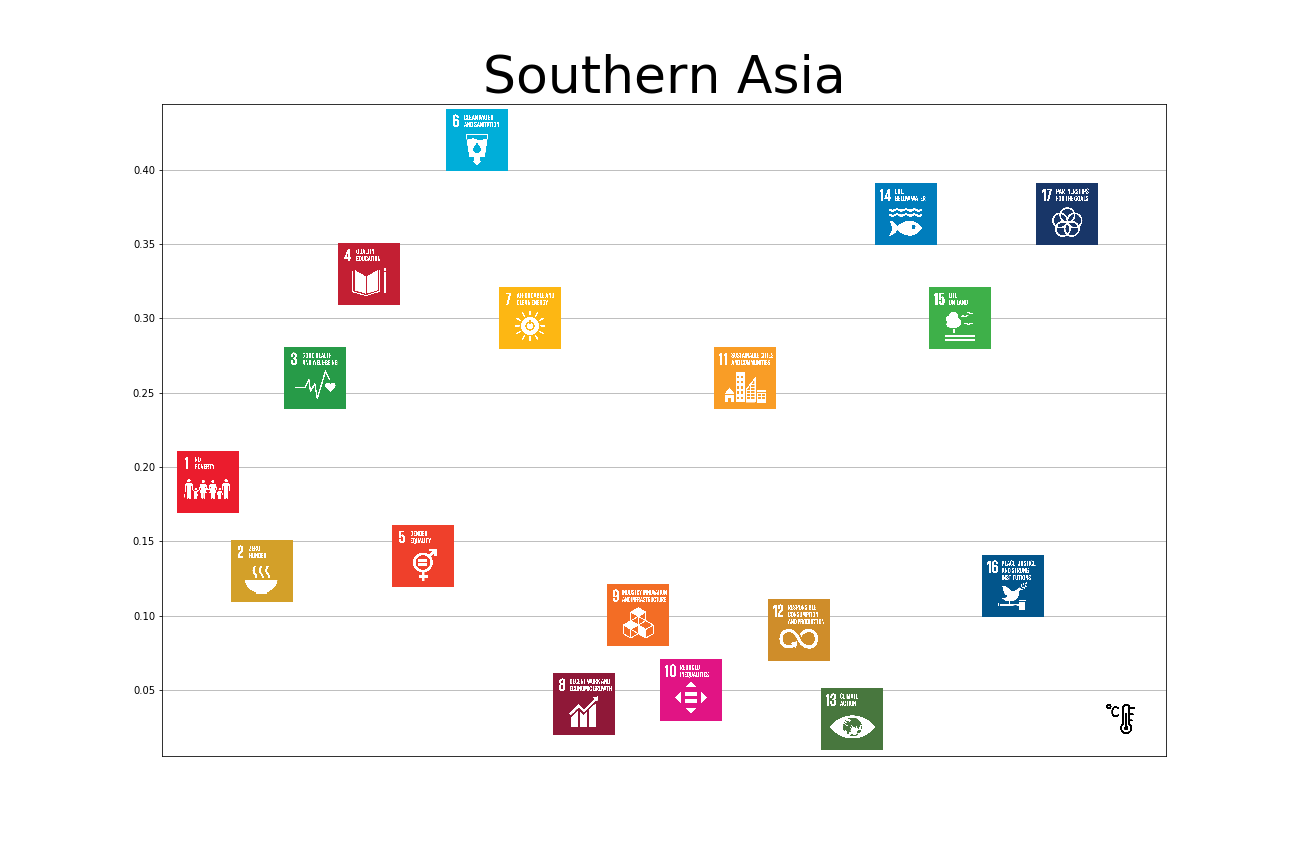}
\end{minipage}
%\end{figure}

%\begin{figure}[!h]
%\centering
\begin{minipage}{.47\textwidth}
  \centering
  \includegraphics[width=\linewidth]{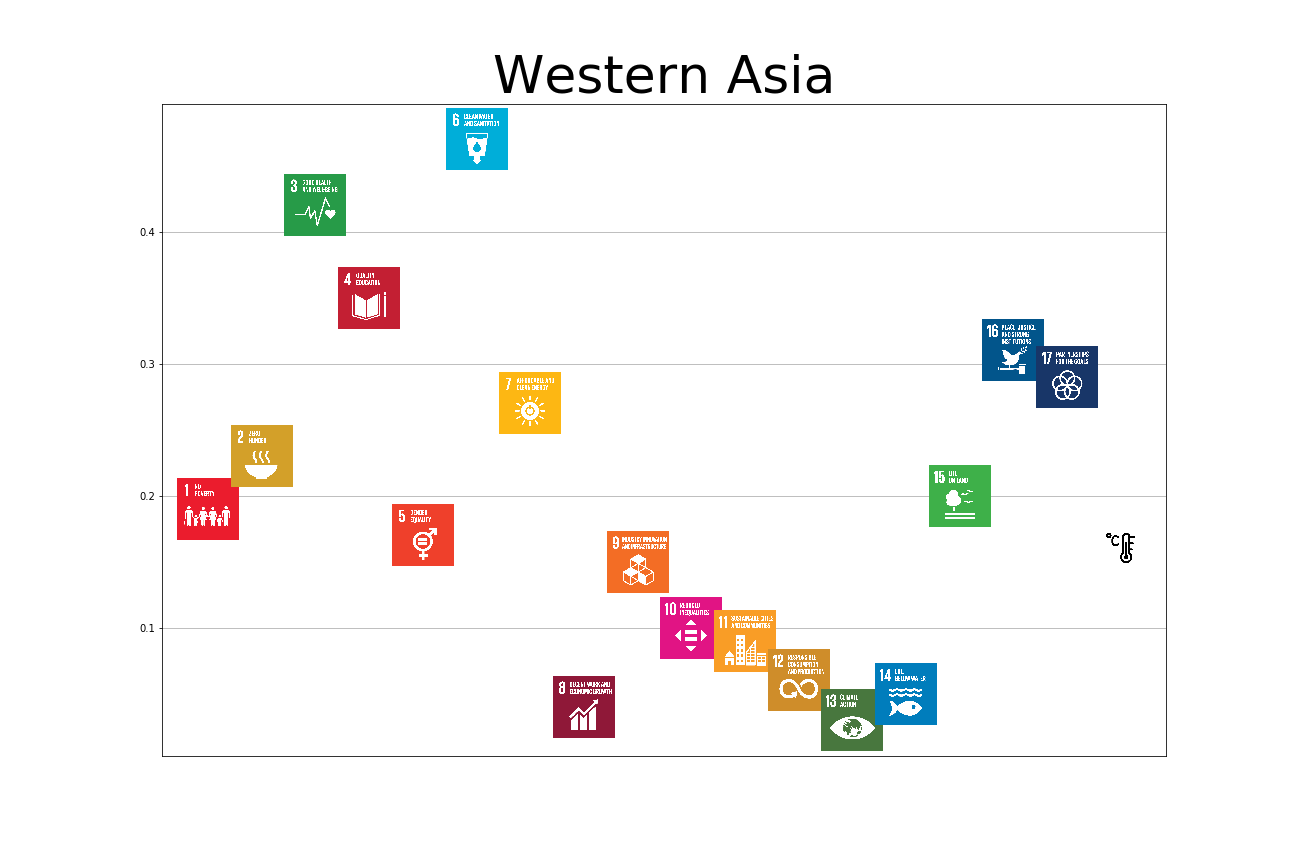}
\end{minipage}
\begin{minipage}{.47\textwidth}
  \centering
  \includegraphics[width=\linewidth]{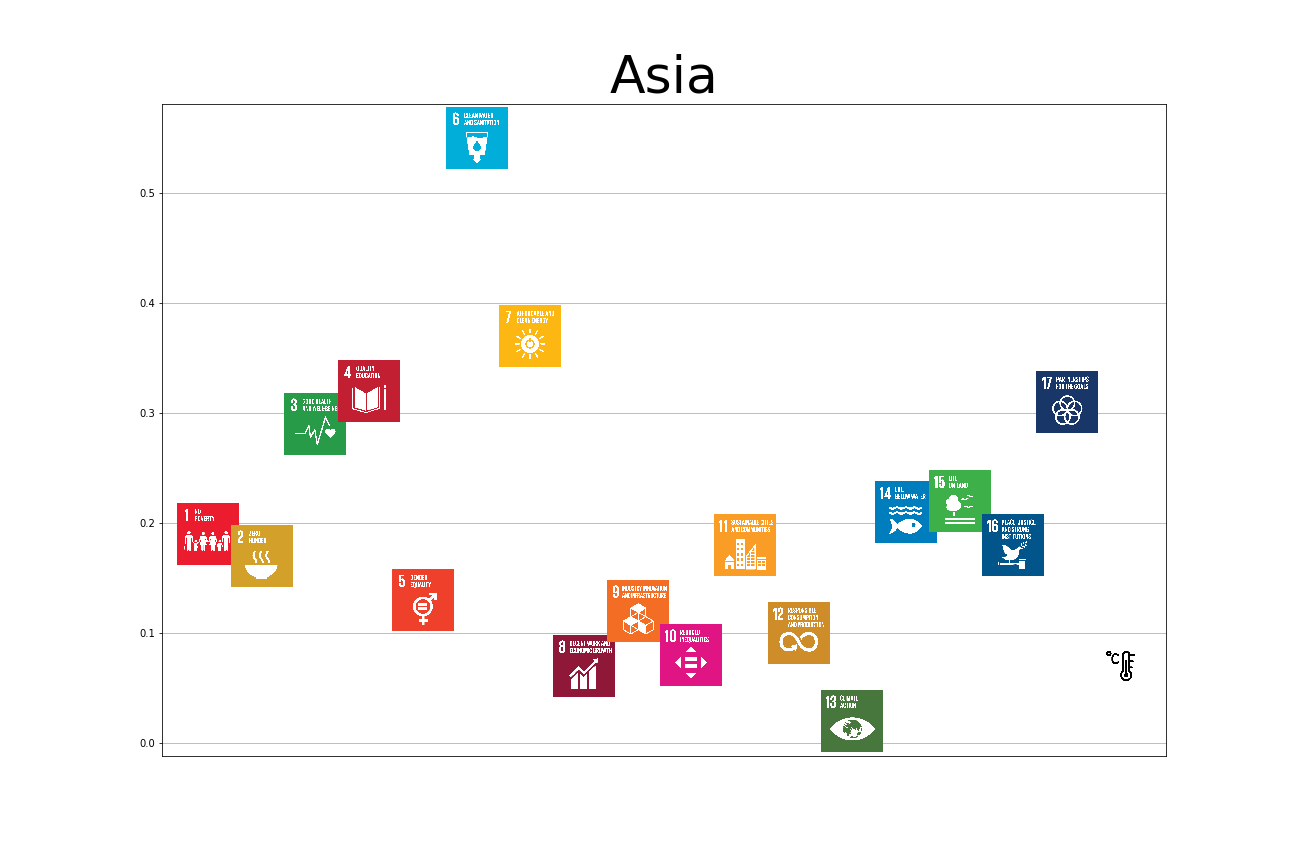}
\end{minipage}
%\end{figure}

%\begin{figure}[!h]
%\centering
\begin{minipage}{.47\textwidth}
  \centering
  \includegraphics[width=\linewidth]{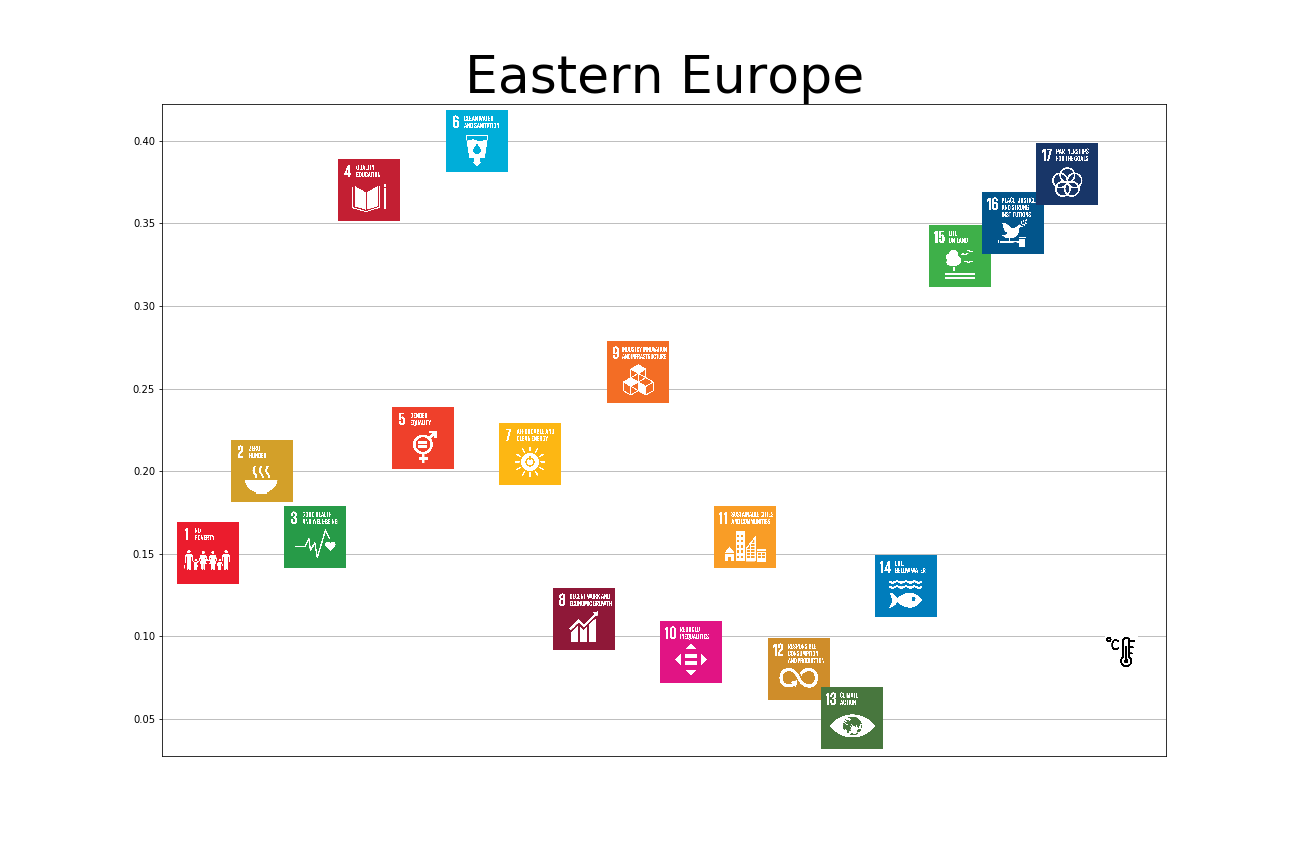}
\end{minipage}
\begin{minipage}{.47\textwidth}
  \centering
  \includegraphics[width=\linewidth]{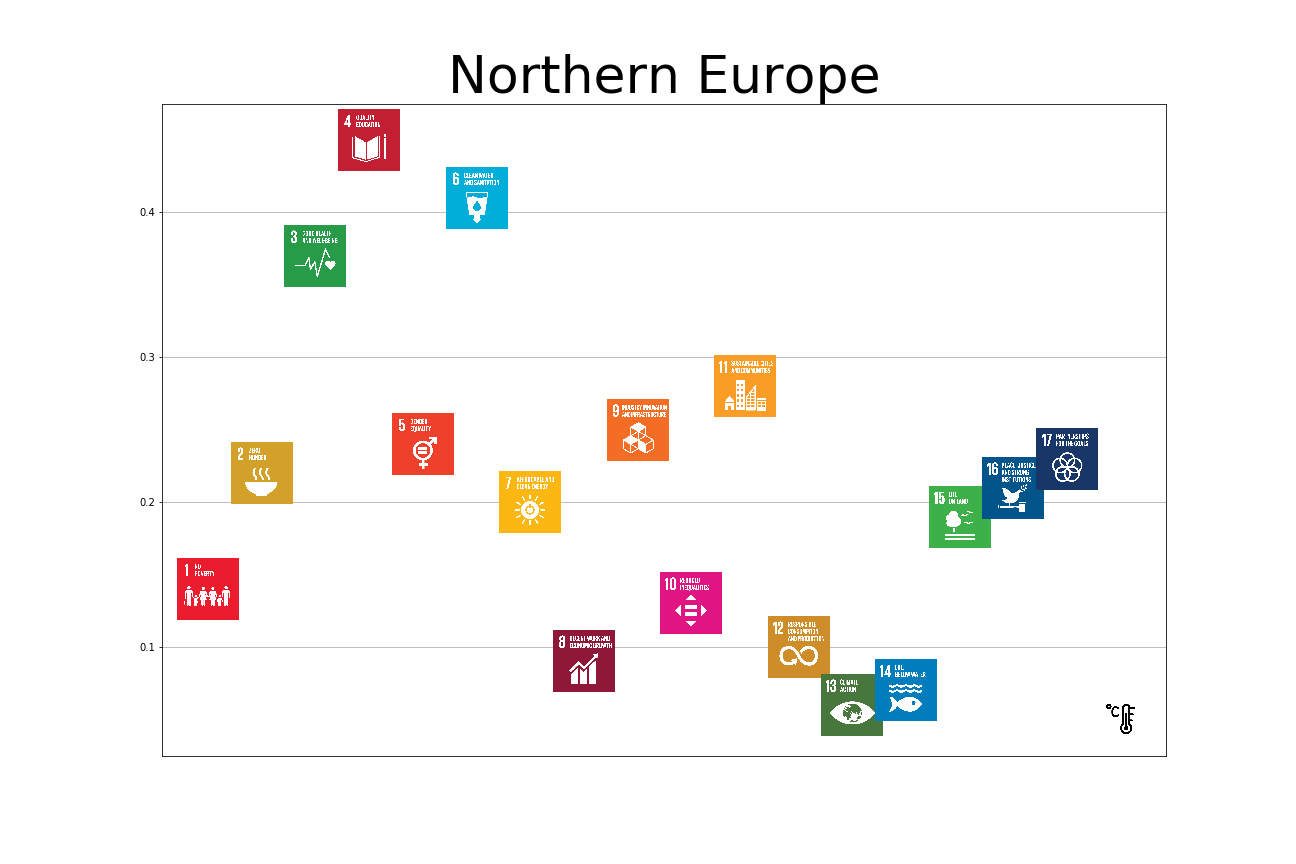}
\end{minipage}
%\end{figure}

%\begin{figure}[!h]
%\centering
\begin{minipage}{.47\textwidth}
  \centering
  \includegraphics[width=\linewidth]{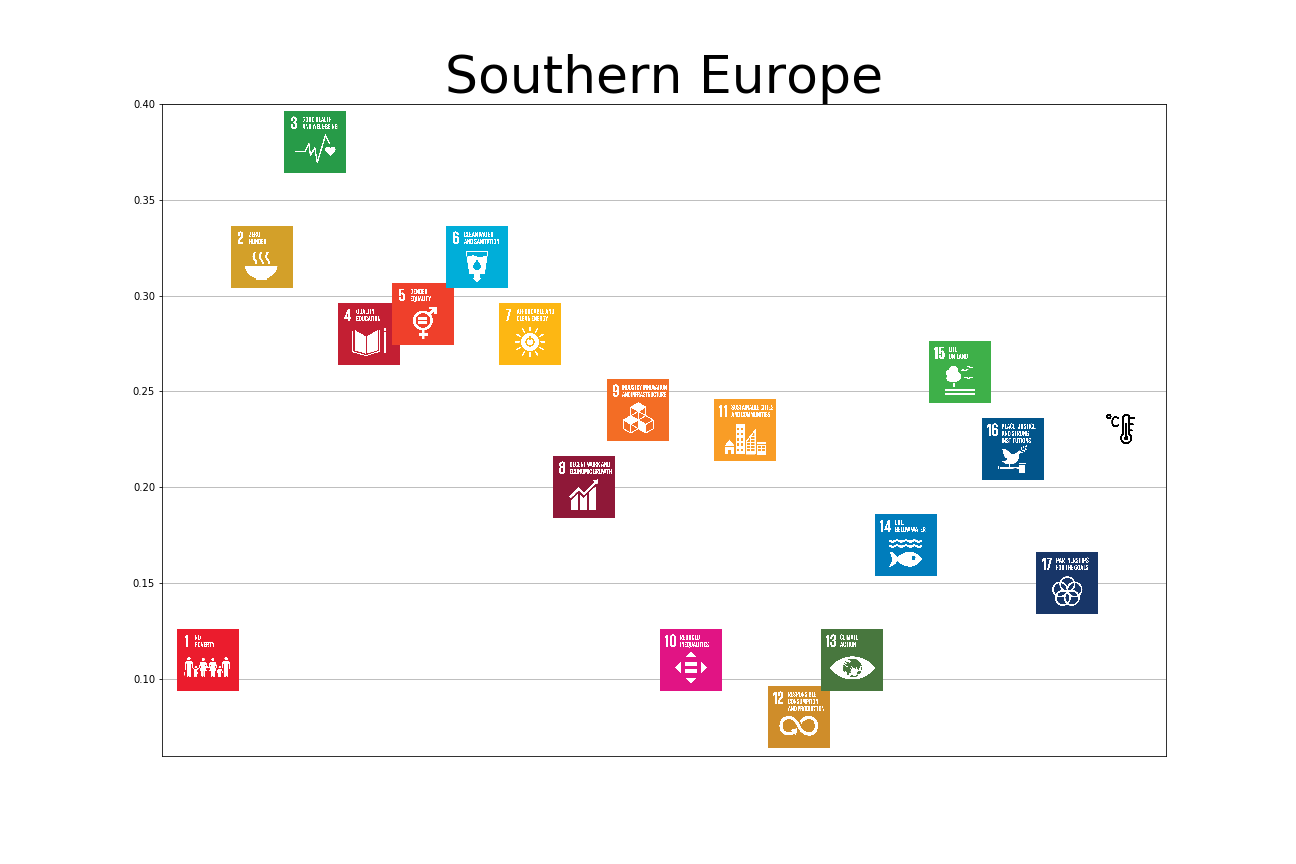}
\end{minipage}
\begin{minipage}{.47\textwidth}
  \centering
  \includegraphics[width=\linewidth]{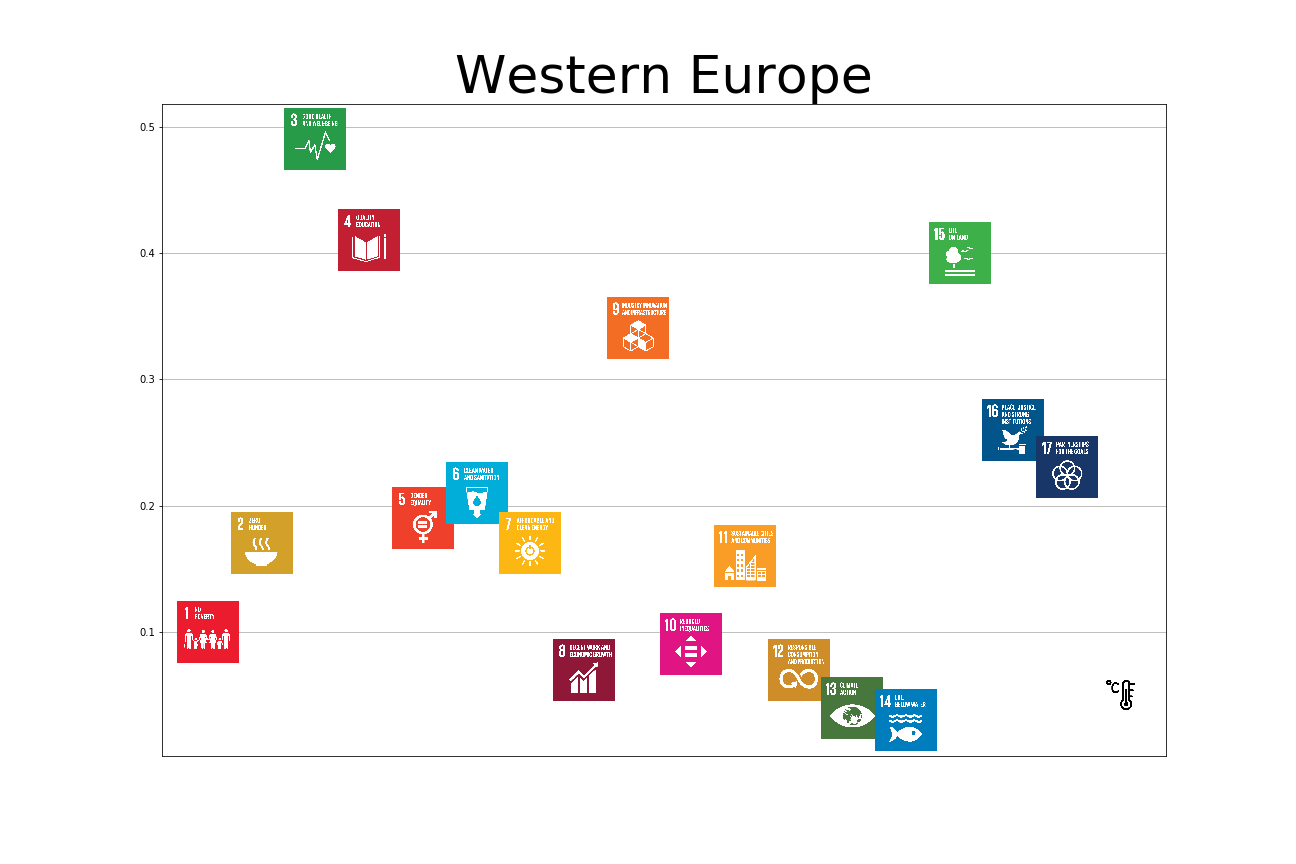}
\end{minipage}
%\end{figure}

%\begin{figure}[!h]
%\centering
\begin{minipage}{.47\textwidth}
  \centering
  \includegraphics[width=\linewidth]{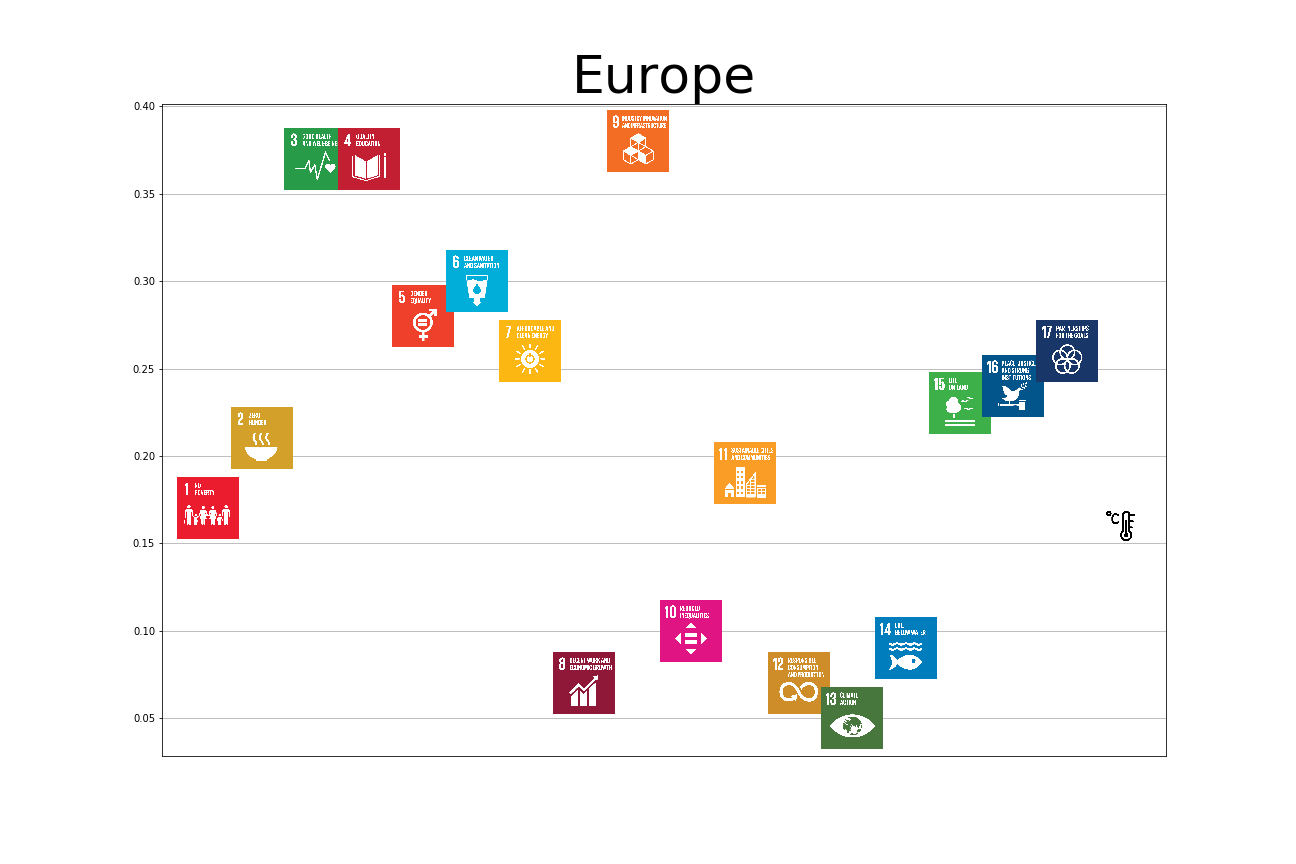}
\end{minipage}
\begin{minipage}{.47\textwidth}
  \centering
  \includegraphics[width=\linewidth]{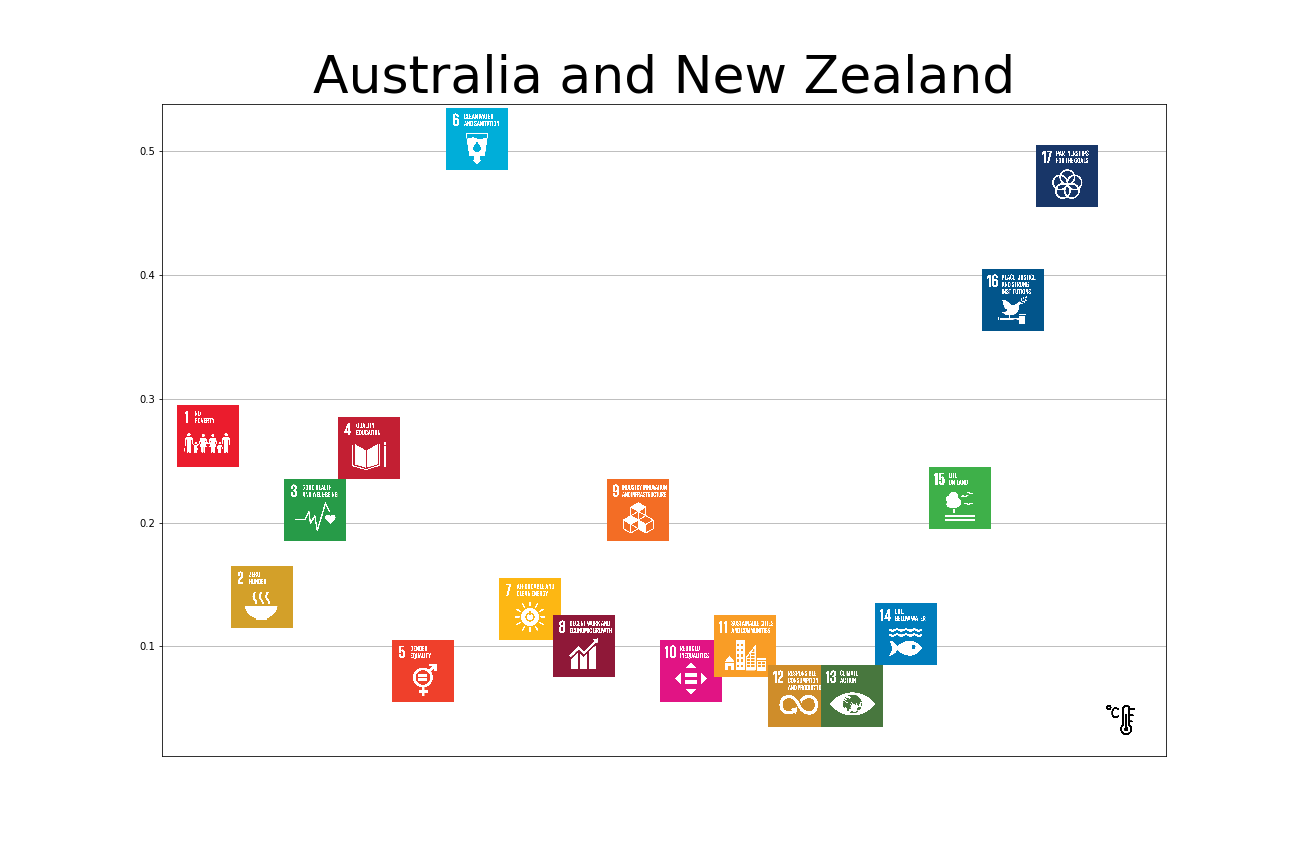}
\end{minipage}
%\end{figure}

%\begin{figure}[!h]
%\centering
\begin{minipage}{.47\textwidth}
  \centering
  \includegraphics[width=\linewidth]{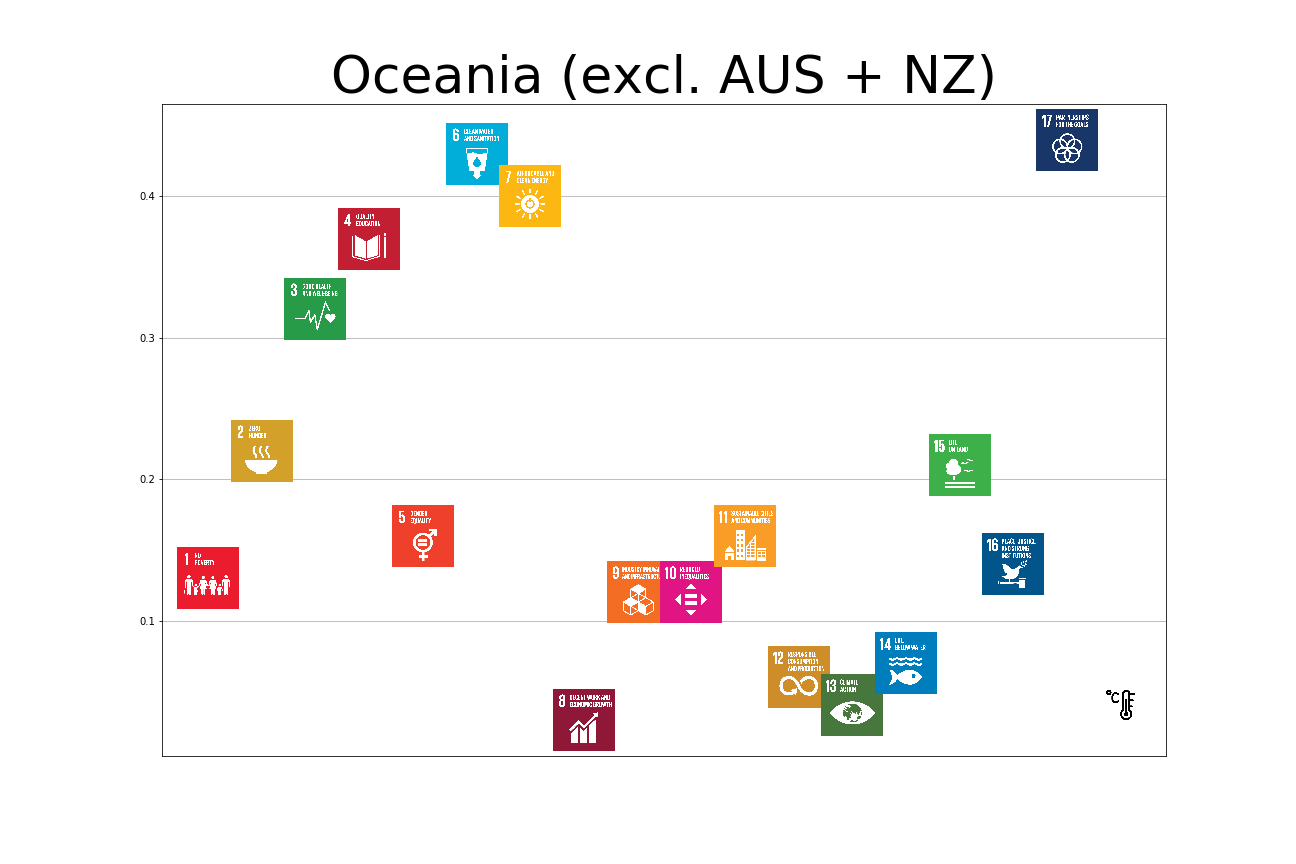}
\end{minipage}
\begin{minipage}{.47\textwidth}
  \centering
  \includegraphics[width=\linewidth]{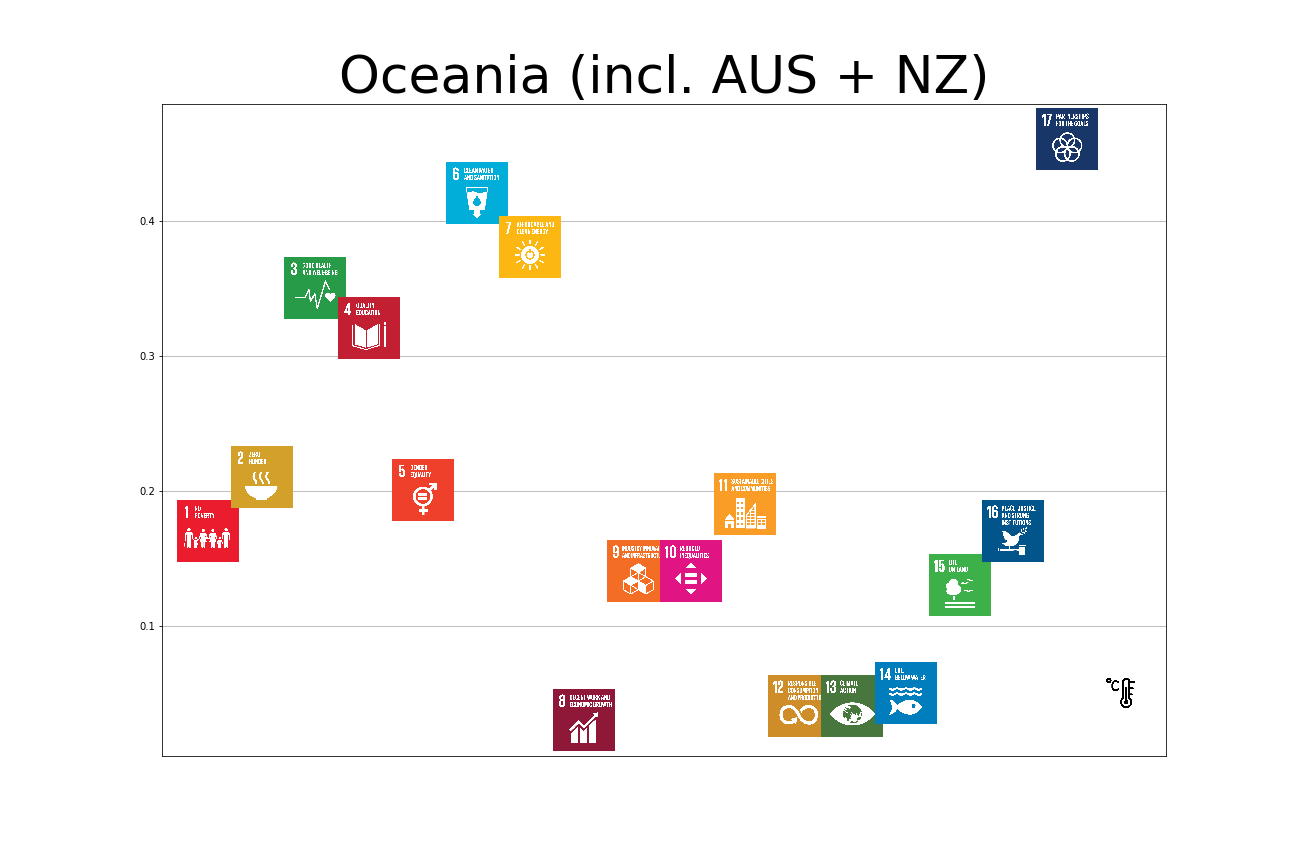}
\end{minipage}
%\end{figure}

%\begin{figure}[!h]
\centering
\begin{minipage}{.47\textwidth}
  \centering
  \includegraphics[width=\linewidth]{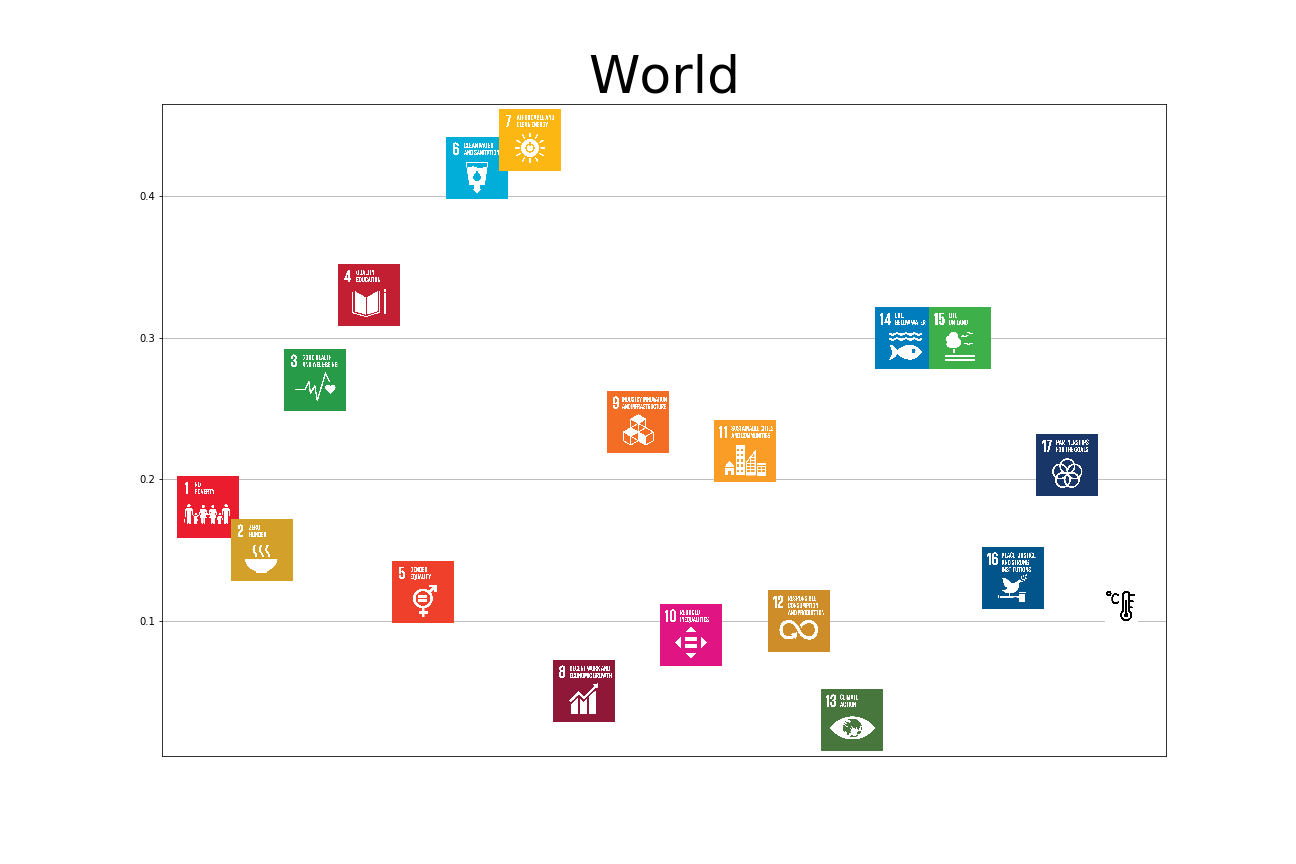}
\end{minipage}
%\end{figure}

% GROUPS

%\begin{figure}[!h]
%\centering
\begin{minipage}{.47\textwidth}
  \centering
  \includegraphics[width=\linewidth]{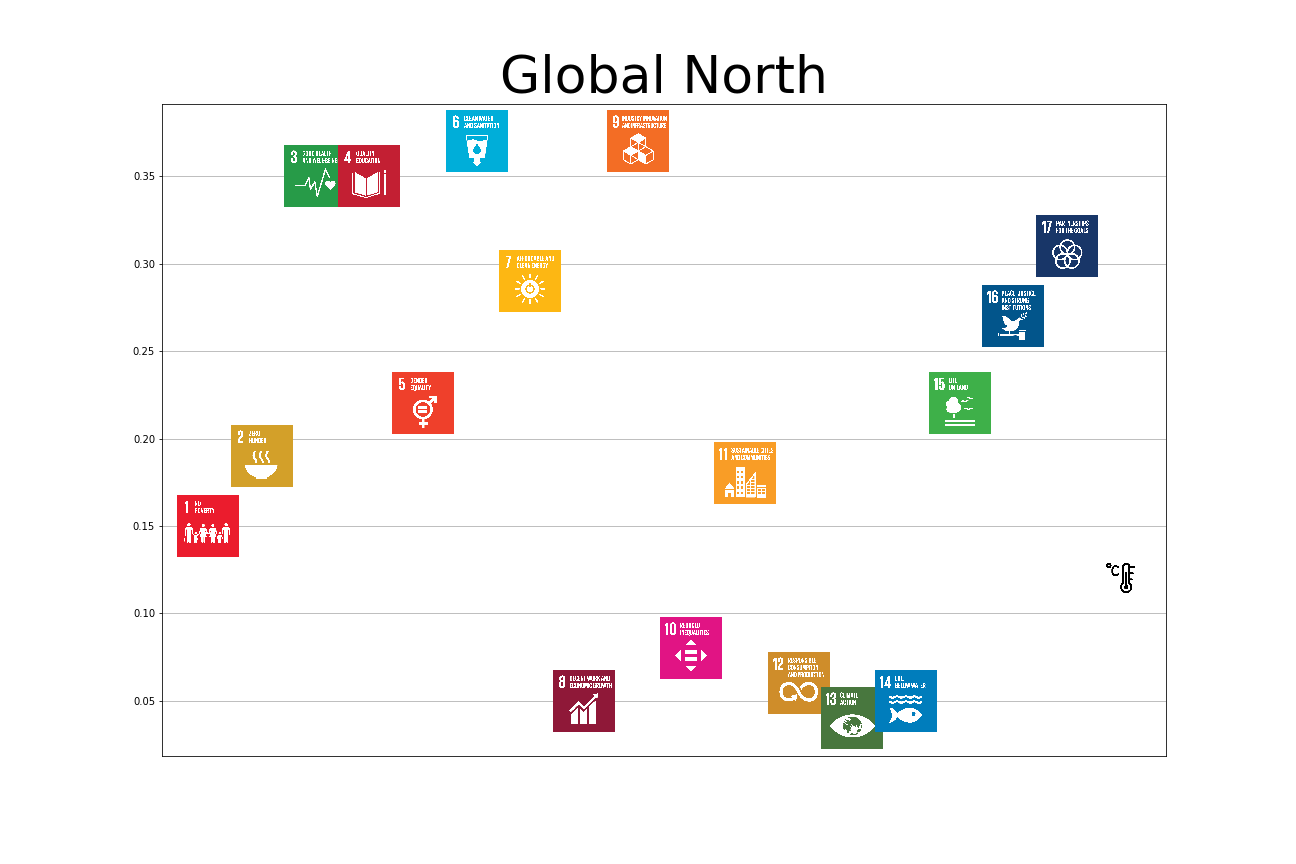}
\end{minipage}
\begin{minipage}{.47\textwidth}
  \centering
  \includegraphics[width=\linewidth]{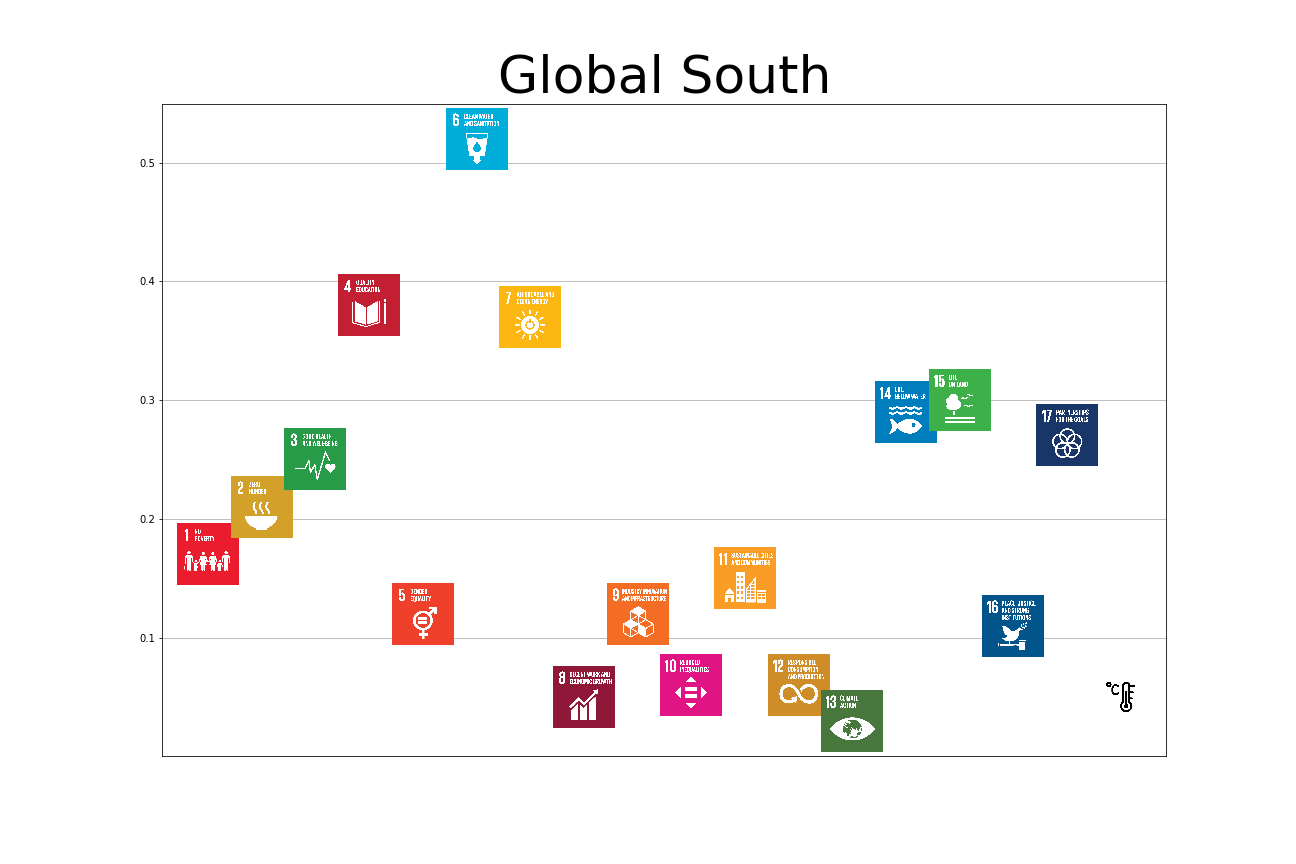}
\end{minipage}
%\end{figure}

%\begin{figure}[!h]
%\centering
\begin{minipage}{.47\textwidth}
  \centering
  \includegraphics[width=\linewidth]{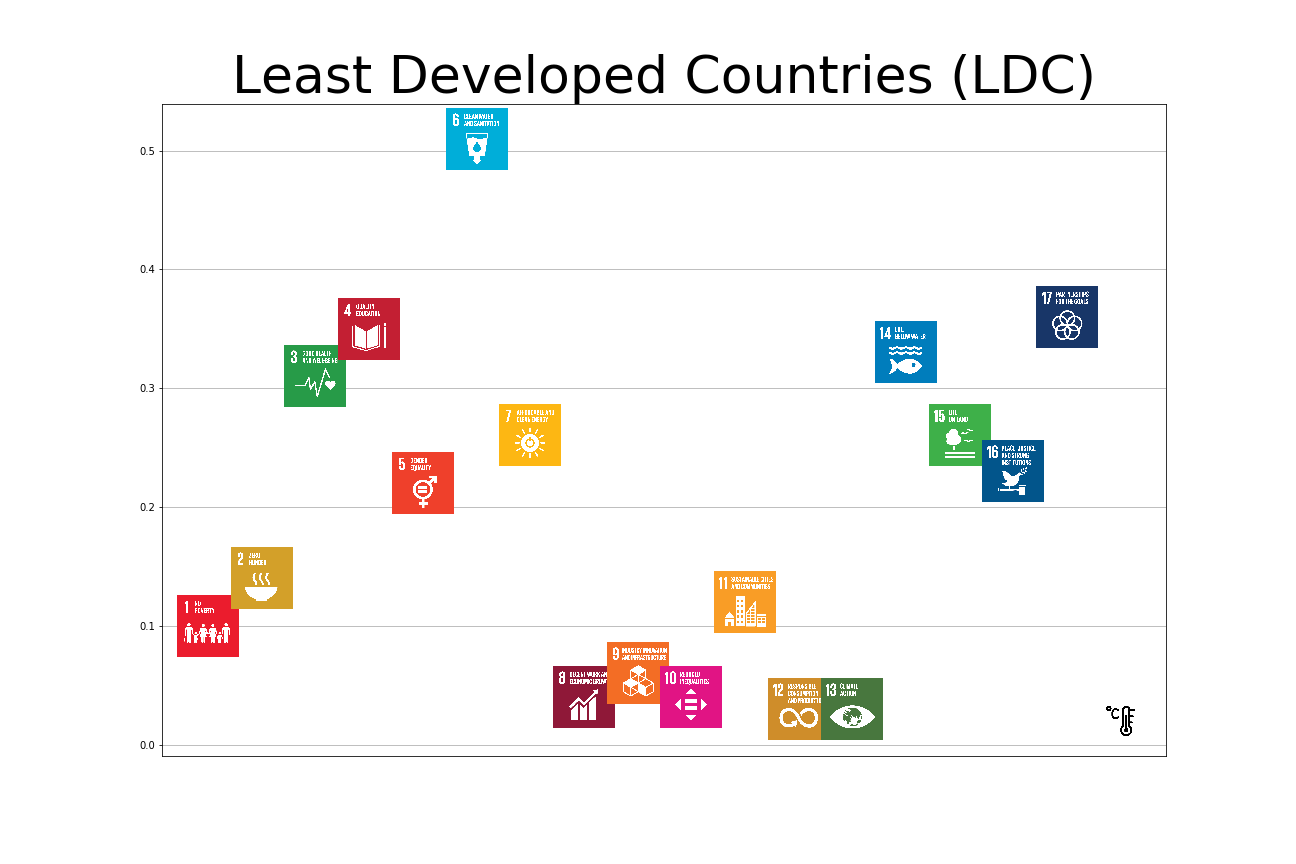}
\end{minipage}
\begin{minipage}{.47\textwidth}
  \centering
  \includegraphics[width=\linewidth]{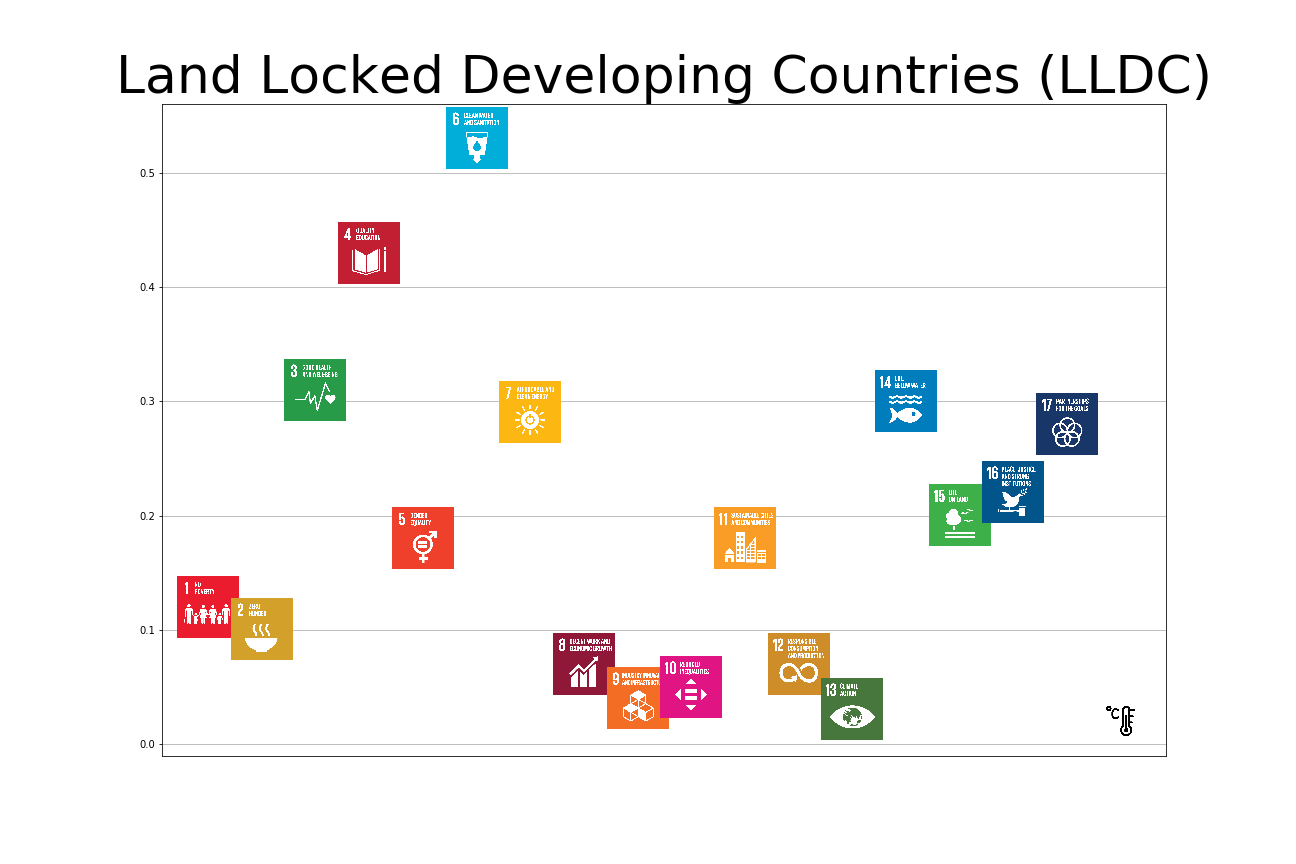}
\end{minipage}
%\end{figure}

%\begin{figure}[!h]
%\centering
\begin{minipage}{.47\textwidth}
  \centering
  \includegraphics[width=\linewidth]{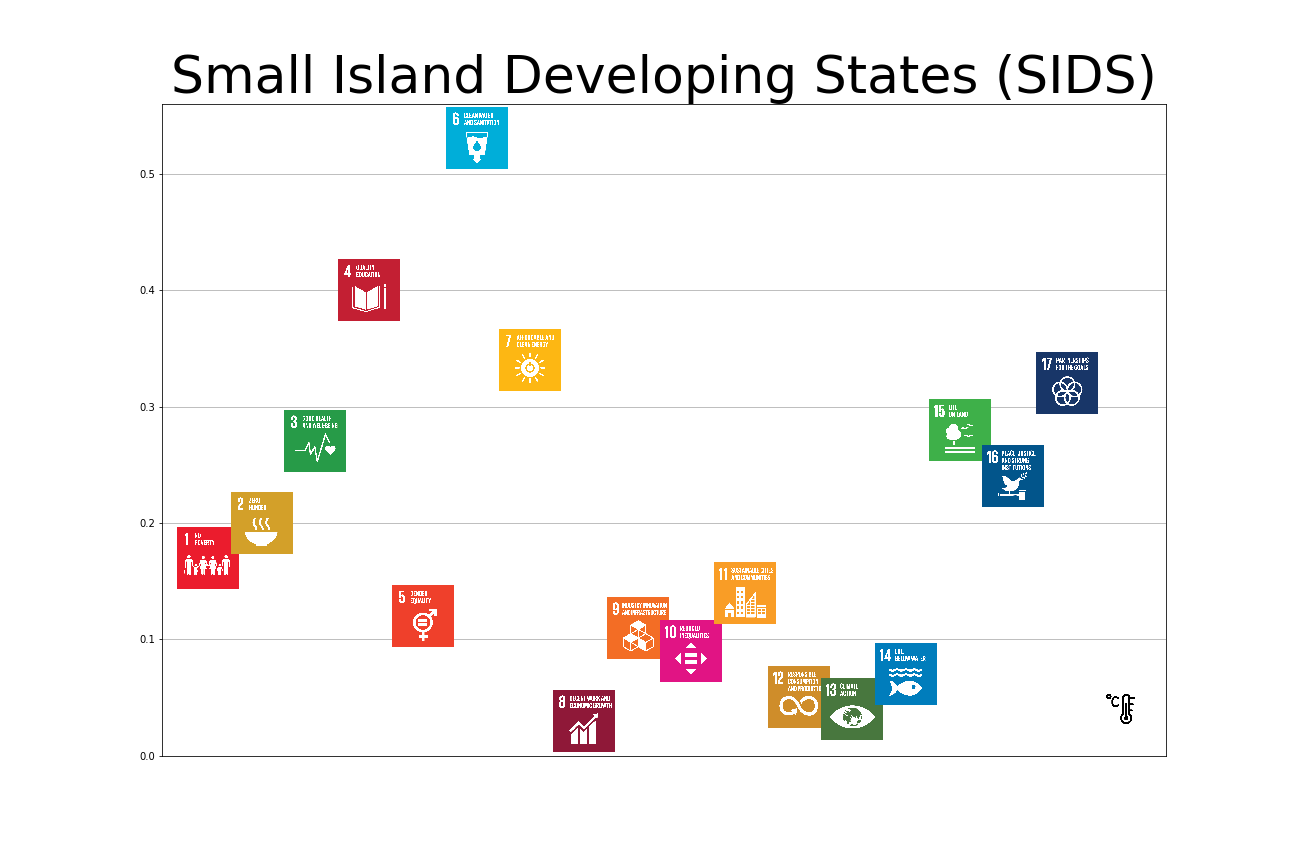}
\end{minipage}
\begin{minipage}{.47\textwidth}
  \centering
  \includegraphics[width=\linewidth]{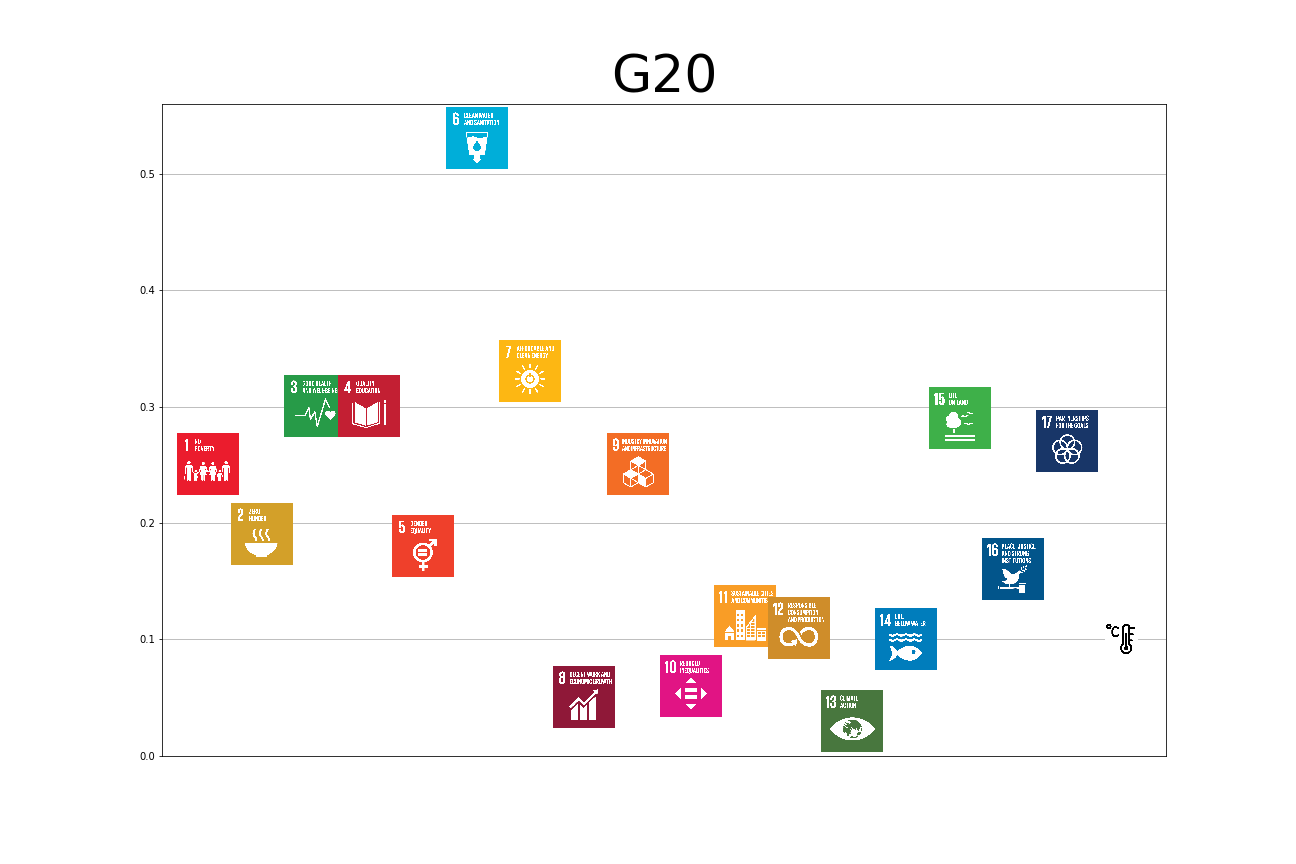}
\end{minipage}
%\end{figure}

%\begin{figure}[!h]
%\centering
\begin{minipage}{.47\textwidth}
  \centering
  \includegraphics[width=\linewidth]{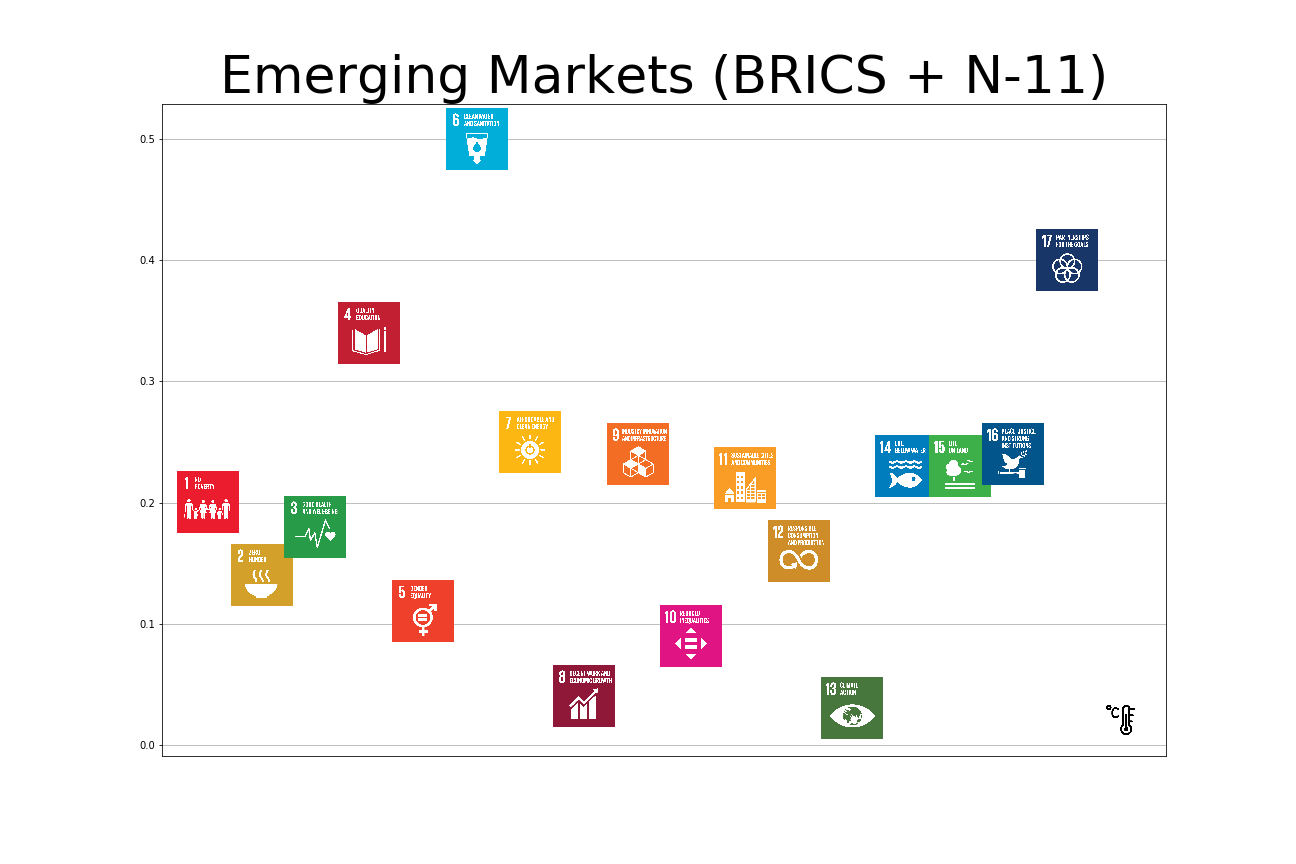}
\end{minipage}
\begin{minipage}{.47\textwidth}
  \centering
  \includegraphics[width=\linewidth]{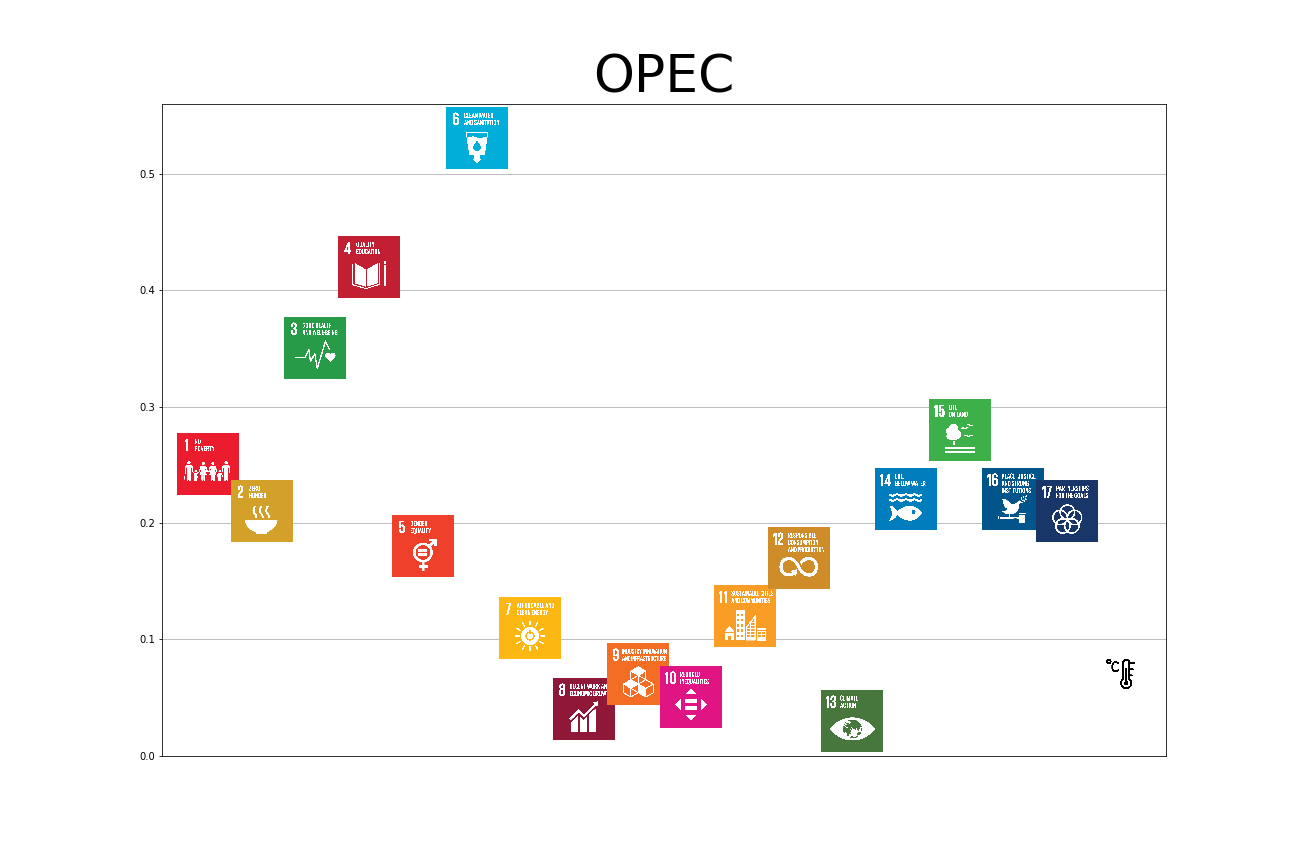}
\end{minipage}
%\end{figure}

%\begin{figure}[!h]
%\centering
\begin{minipage}{.47\textwidth}
  \centering
  \includegraphics[width=\linewidth]{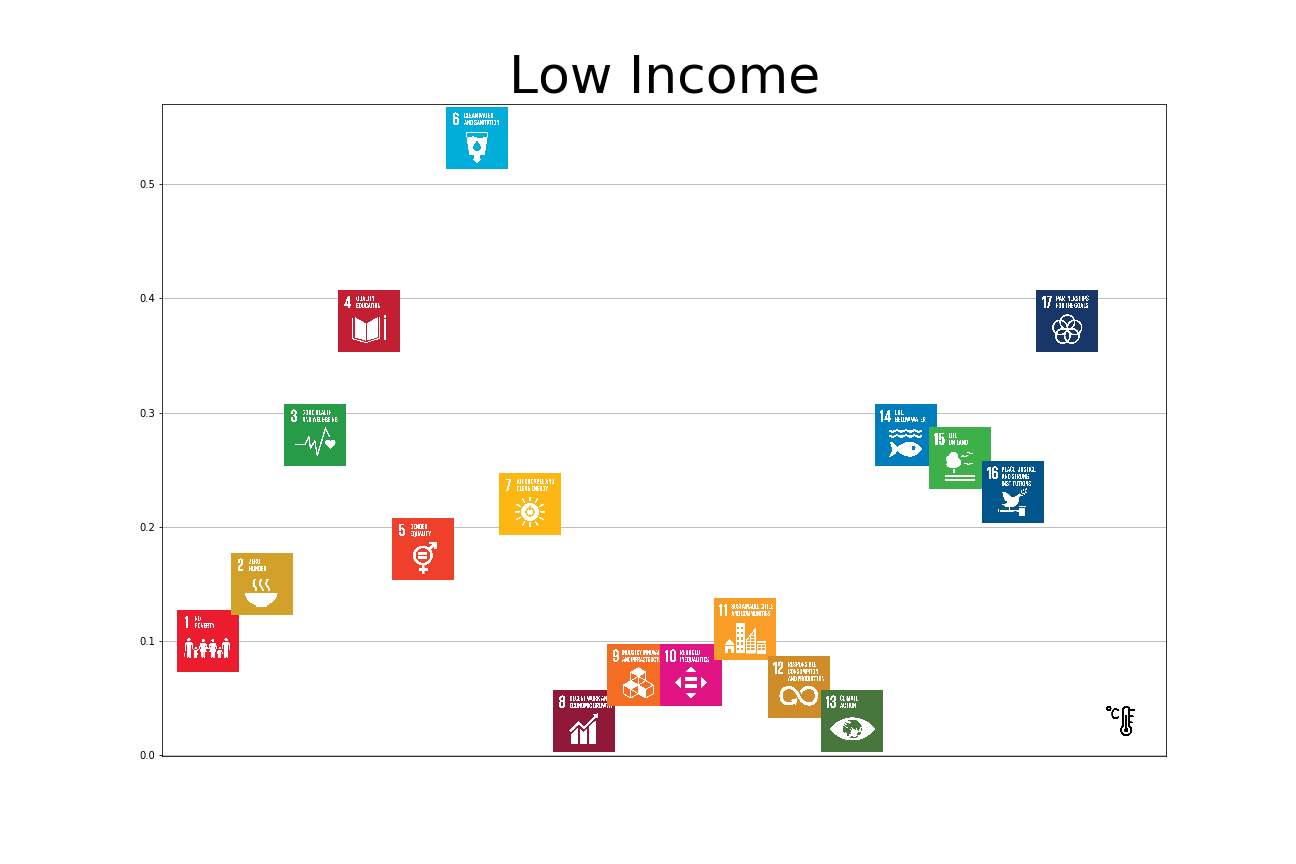}
\end{minipage}
\begin{minipage}{.47\textwidth}
  \centering
  \includegraphics[width=\linewidth]{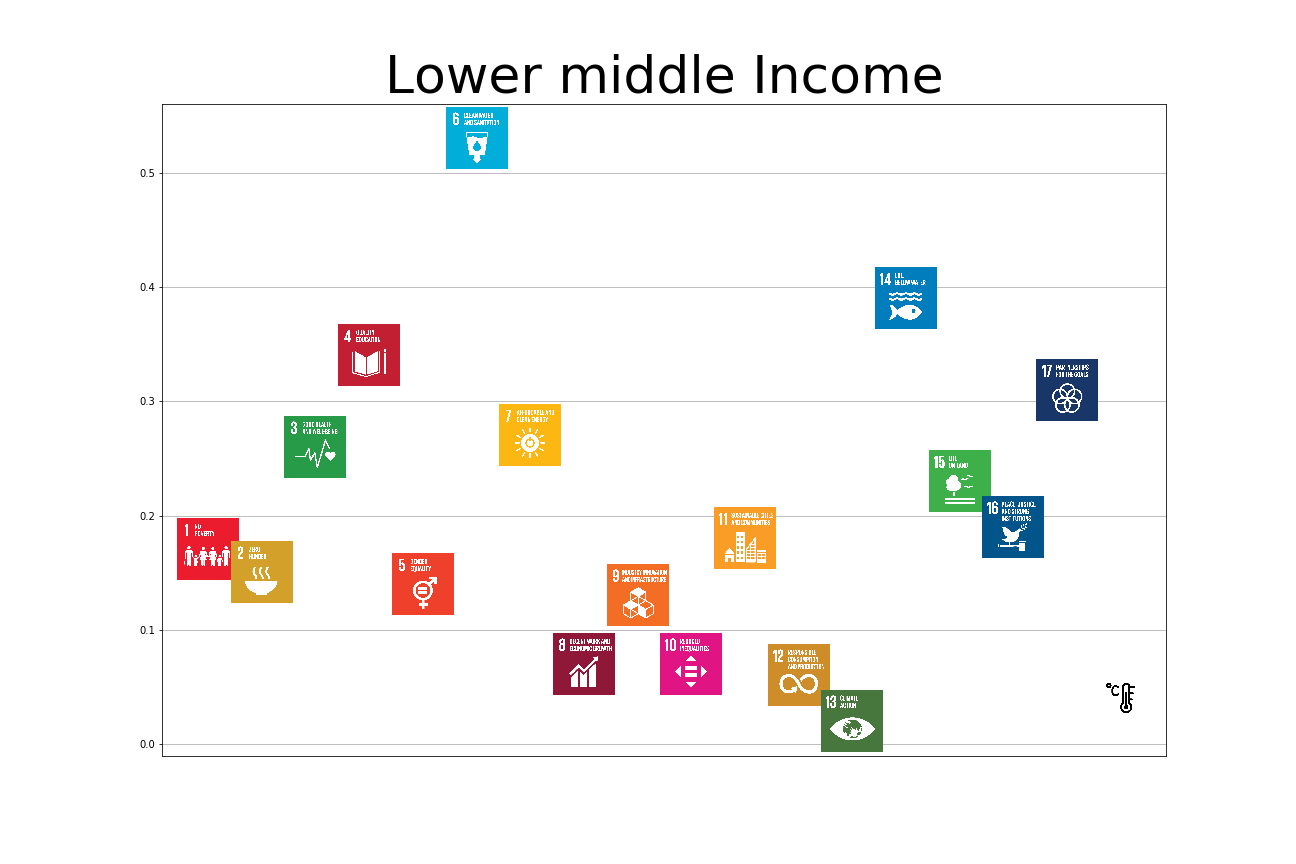}
\end{minipage}
%\end{figure}

%\begin{figure}[!h]
%\centering
\begin{minipage}{.47\textwidth}
  \centering
  \includegraphics[width=\linewidth]{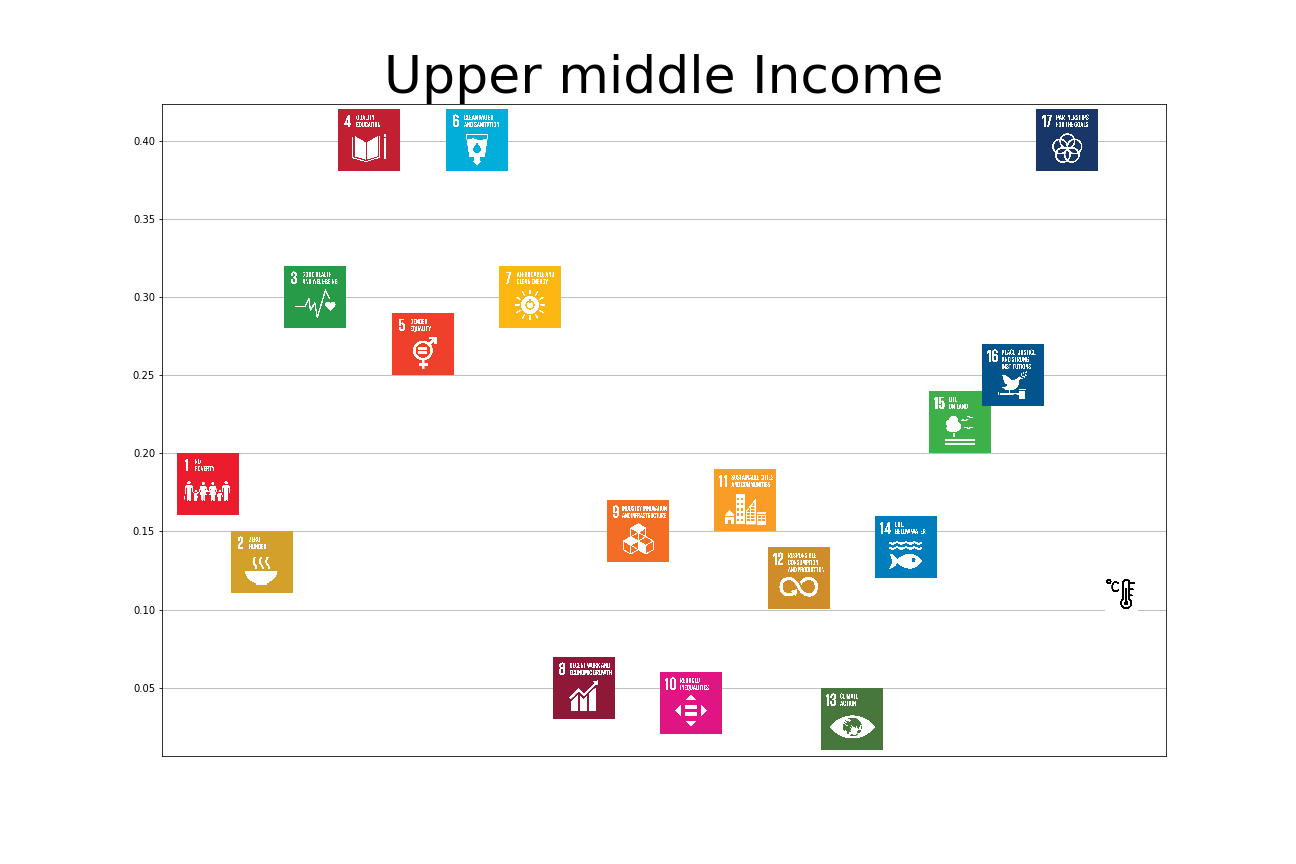}
\end{minipage}
\begin{minipage}{.47\textwidth}
  \centering
  \includegraphics[width=\linewidth]{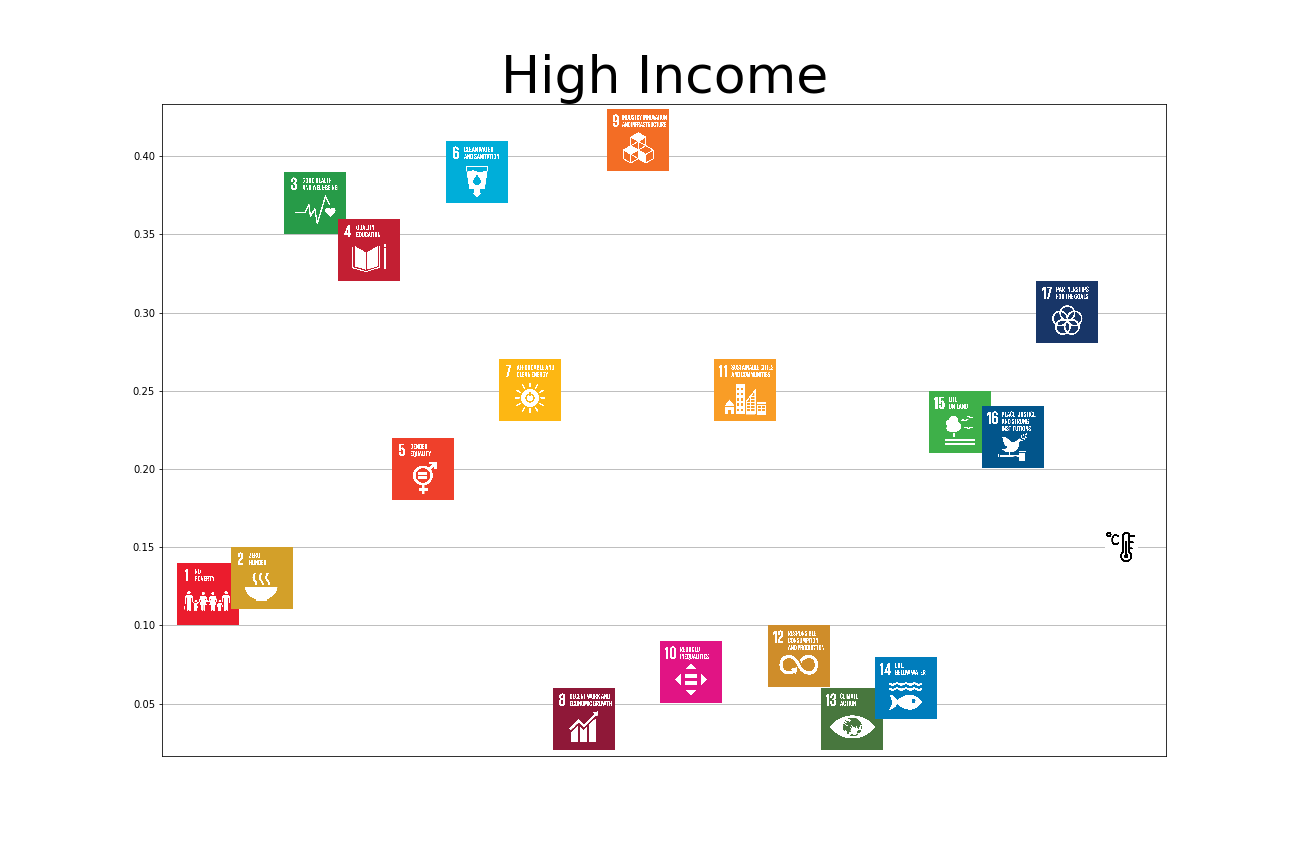}
\end{minipage}

\begin{landscape}

\subsection{Groupings of countries}
\begin{table}[h]
\centering
\scalebox{0.255}{
\renewcommand{\arraystretch}{1.2}
\begin{tabular}{l|l|l|l|l|l|l|l|l|l|l|l|l|l|l|l|l|l|l|l|l|l|l|l|l|l|l}
Northern Africa & Eastern Africa & Middle Africa & Southern Africa & Western Africa & Sub-Saharan Africa & Africa & Caribbean & Central America & South America & Latin America and the Caribbean & North America & Americas & Central Asia & Eastern Asia & South-eastern Asia & Southern Asia & Western Asia & Asia & Eastern Europe & Northern Europe & Southern Europe & Western Europe & Europe & Australia and New Zealand & Oceania (excl. AUS + NZ) & Oceania (incl. AUS + NZ) \\
\hline 
Algeria & Burundi & Angola & Botswana & Benin & Burundi & Algeria & Antigua and Barbuda & Belize & Argentina & Antigua and Barbuda & Canada & Antigua and Barbuda & Kazakhstan & China & Brunei Darussalam & Afghanistan & Armenia & Kazakhstan & Belarus & Denmark & Albania & Austria & Belarus & Australia & Fiji & Australia \\
Egypt, Arab Rep. & Comoros & Cameroon & Lesotho & Burkina Faso & Comoros & Egypt, Arab Rep. & Bahamas, The & Costa Rica & Bolivia & Bahamas, The & Greenland & Bahamas, The & Kyrgyz Republic & Korea, Dem. People's Rep. & Cambodia & Bangladesh & Azerbaijan & Kyrgyz Republic & Bulgaria & Estonia & Bosnia and Herzegovina & Belgium & Bulgaria & New Zealand & Papua New Guinea & New Zealand \\
Morocco & Djibouti & Central African Republic & Namibia & Cote d'Ivoire & Djibouti & Morocco & Barbados & El Salvador & Brazil & Barbados & United States & Barbados & Tajikistan & Japan & Indonesia & Bhutan & Bahrain & Tajikistan & Czech Republic & Finland & Croatia & France & Czech Republic &  & Solomon Islands & Fiji \\
Tunisia & Eritrea & Chad & South Africa & Gambia, The & Eritrea & Tunisia & Cuba & Guatemala & Chile & Cuba &  & Cuba & Turkmenistan & Mongolia & Lao PDR & India & Cyprus & Turkmenistan & Hungary & Iceland & Greece & Germany & Hungary &  & Vanuatu & Papua New Guinea \\
 & Ethiopia & Congo, Rep. &  & Ghana & Ethiopia & Burundi & Dominica & Honduras & Colombia & Dominica &  & Dominica & Uzbekistan &  & Malaysia & Iran, Islamic Rep. & Georgia & Uzbekistan & Poland & Ireland & Italy & Liechtenstein & Poland &  & Micronesia, Fed. Sts. & Solomon Islands \\
 & Kenya & Congo, Dem. Rep. &  & Guinea-Bissau & Kenya & Comoros & Grenada & Mexico & Ecuador & Grenada &  & Grenada &  &  & Myanmar & Maldives & Iraq & China & Moldova & Latvia & Malta & Luxembourg & Moldova &  & Palau & Vanuatu \\
 & Madagascar & Equatorial Guinea &  & Liberia & Madagascar & Djibouti & Haiti & Nicaragua & Guyana & Haiti &  & Haiti &  &  & Philippines & Nepal & Israel & Korea, Dem. People's Rep. & Romania & Lithuania & Montenegro & Netherlands & Romania &  & Kiribati & Micronesia, Fed. Sts. \\
 & Malawi & Gabon &  & Mali & Malawi & Eritrea & Jamaica & Panama & Paraguay & Jamaica &  & Jamaica &  &  & Singapore & Pakistan & Jordan & Japan & Russian Federation & Norway & Portugal & Switzerland & Russian Federation &  & Samoa & Palau \\
 & Mauritius & Sao Tome and Principe &  & Mauritania & Mauritius & Ethiopia & Puerto Rico &  & Peru & Puerto Rico &  & Puerto Rico &  &  & Thailand & Sri Lanka & Kuwait & Mongolia & Slovak Republic & Sweden & Serbia &  & Slovak Republic &  & Tonga & Kiribati \\
 & Mozambique &  &  & Niger & Mozambique & Kenya & Trinidad and Tobago &  & Suriname & Trinidad and Tobago &  & Trinidad and Tobago &  &  & Timor-Leste &  & Lebanon & Brunei Darussalam & Ukraine & United Kingdom & Slovenia &  & Ukraine &  & Tuvalu & Samoa \\
 & Rwanda &  &  & Nigeria & Rwanda & Madagascar &  &  & Uruguay & Belize &  & Belize &  &  & Vietnam &  & Oman & Cambodia &  &  & Spain &  & Denmark &  &  & Tonga \\
 & Seychelles &  &  & Senegal & Seychelles & Malawi &  &  & Venezuela, RB & Costa Rica &  & Costa Rica &  &  &  &  & Qatar & Indonesia &  &  &  &  & Estonia &  &  & Tuvalu \\
 & Somalia &  &  & Sierra Leone & Somalia & Mauritius &  &  &  & El Salvador &  & El Salvador &  &  &  &  & Saudi Arabia & Lao PDR &  &  &  &  & Finland &  &  &  \\
 & South Sudan &  &  & Togo & South Sudan & Mozambique &  &  &  & Guatemala &  & Guatemala &  &  &  &  & Syrian Arab Republic & Malaysia &  &  &  &  & Iceland &  &  &  \\
 & Uganda &  &  &  & Uganda & Rwanda &  &  &  & Honduras &  & Honduras &  &  &  &  & Turkey & Myanmar &  &  &  &  & Ireland &  &  &  \\
 & Tanzania &  &  &  & Tanzania & Seychelles &  &  &  & Mexico &  & Mexico &  &  &  &  & United Arab Emirates & Philippines &  &  &  &  & Latvia &  &  &  \\
 & Zambia &  &  &  & Zambia & Somalia &  &  &  & Nicaragua &  & Nicaragua &  &  &  &  & Yemen, Rep. & Singapore &  &  &  &  & Lithuania &  &  &  \\
 & Zimbabwe &  &  &  & Zimbabwe & South Sudan &  &  &  & Panama &  & Panama &  &  &  &  &  & Thailand &  &  &  &  & Norway &  &  &  \\
 &  &  &  &  & Angola & Uganda &  &  &  & Argentina &  & Argentina &  &  &  &  &  & Timor-Leste &  &  &  &  & Sweden &  &  &  \\
 &  &  &  &  & Cameroon & Tanzania &  &  &  & Bolivia &  & Bolivia &  &  &  &  &  & Vietnam &  &  &  &  & United Kingdom &  &  &  \\
 &  &  &  &  & Central African Republic & Zambia &  &  &  & Brazil &  & Brazil &  &  &  &  &  & Afghanistan &  &  &  &  & Albania &  &  &  \\
 &  &  &  &  & Chad & Zimbabwe &  &  &  & Chile &  & Chile &  &  &  &  &  & Bangladesh &  &  &  &  & Bosnia and Herzegovina &  &  &  \\
 &  &  &  &  & Congo, Rep. & Angola &  &  &  & Colombia &  & Colombia &  &  &  &  &  & Bhutan &  &  &  &  & Croatia &  &  &  \\
 &  &  &  &  & Congo, Dem. Rep. & Cameroon &  &  &  & Ecuador &  & Ecuador &  &  &  &  &  & India &  &  &  &  & Greece &  &  &  \\
 &  &  &  &  & Equatorial Guinea & Central African Republic &  &  &  & Guyana &  & Guyana &  &  &  &  &  & Iran, Islamic Rep. &  &  &  &  & Italy &  &  &  \\
 &  &  &  &  & Gabon & Chad &  &  &  & Paraguay &  & Paraguay &  &  &  &  &  & Maldives &  &  &  &  & Malta &  &  &  \\
 &  &  &  &  & Sao Tome and Principe & Congo, Rep. &  &  &  & Peru &  & Peru &  &  &  &  &  & Nepal &  &  &  &  & Montenegro &  &  &  \\
 &  &  &  &  & Botswana & Congo, Dem. Rep. &  &  &  & Suriname &  & Suriname &  &  &  &  &  & Pakistan &  &  &  &  & Portugal &  &  &  \\
 &  &  &  &  & Lesotho & Equatorial Guinea &  &  &  & Uruguay &  & Uruguay &  &  &  &  &  & Sri Lanka &  &  &  &  & Serbia &  &  &  \\
 &  &  &  &  & Namibia & Gabon &  &  &  & Venezuela, RB &  & Venezuela, RB &  &  &  &  &  & Armenia &  &  &  &  & Slovenia &  &  &  \\
 &  &  &  &  & South Africa & Sao Tome and Principe &  &  &  &  &  & Canada &  &  &  &  &  & Azerbaijan &  &  &  &  & Spain &  &  &  \\
 &  &  &  &  & Benin & Botswana &  &  &  &  &  & Greenland &  &  &  &  &  & Bahrain &  &  &  &  & Austria &  &  &  \\
 &  &  &  &  & Burkina Faso & Lesotho &  &  &  &  &  & United States &  &  &  &  &  & Cyprus &  &  &  &  & Belgium &  &  &  \\
 &  &  &  &  & Cote d'Ivoire & Namibia &  &  &  &  &  &  &  &  &  &  &  & Georgia &  &  &  &  & France &  &  &  \\
 &  &  &  &  & Gambia, The & South Africa &  &  &  &  &  &  &  &  &  &  &  & Iraq &  &  &  &  & Germany &  &  &  \\
 &  &  &  &  & Ghana & Benin &  &  &  &  &  &  &  &  &  &  &  & Israel &  &  &  &  & Liechtenstein &  &  &  \\
 &  &  &  &  & Guinea-Bissau & Burkina Faso &  &  &  &  &  &  &  &  &  &  &  & Jordan &  &  &  &  & Luxembourg &  &  &  \\
 &  &  &  &  & Liberia & Cote d'Ivoire &  &  &  &  &  &  &  &  &  &  &  & Kuwait &  &  &  &  & Netherlands &  &  &  \\
 &  &  &  &  & Mali & Gambia, The &  &  &  &  &  &  &  &  &  &  &  & Lebanon &  &  &  &  & Switzerland &  &  &  \\
 &  &  &  &  & Mauritania & Ghana &  &  &  &  &  &  &  &  &  &  &  & Oman &  &  &  &  &  &  &  &  \\
 &  &  &  &  & Niger & Guinea-Bissau &  &  &  &  &  &  &  &  &  &  &  & Qatar &  &  &  &  &  &  &  &  \\
 &  &  &  &  & Nigeria & Liberia &  &  &  &  &  &  &  &  &  &  &  & Saudi Arabia &  &  &  &  &  &  &  &  \\
 &  &  &  &  & Senegal & Mali &  &  &  &  &  &  &  &  &  &  &  & Syrian Arab Republic &  &  &  &  &  &  &  &  \\
 &  &  &  &  & Sierra Leone & Mauritania &  &  &  &  &  &  &  &  &  &  &  & Turkey &  &  &  &  &  &  &  &  \\
 &  &  &  &  & Togo & Niger &  &  &  &  &  &  &  &  &  &  &  & United Arab Emirates &  &  &  &  &  &  &  &  \\
 &  &  &  &  &  & Nigeria &  &  &  &  &  &  &  &  &  &  &  & Yemen, Rep. &  &  &  &  &  &  &  &  \\
 &  &  &  &  &  & Senegal &  &  &  &  &  &  &  &  &  &  &  &  &  &  &  &  &  &  &  &  \\
 &  &  &  &  &  & Sierra Leone &  &  &  &  &  &  &  &  &  &  &  &  &  &  &  &  &  &  &  &  \\
 &  &  &  &  &  & Togo &  &  &  &  &  &  &  &  &  &  &  &  &  &  &  &  &  &  &  & 
\end{tabular}}
\tiny
\begin{tablenotes}
\centering
\item{World contains all listed countries.}
\end{tablenotes}
\end{table}

\end{landscape}

\begin{table}[h]
\centering
\scalebox{0.265}{
\renewcommand{\arraystretch}{1.2}
\begin{tabular}{l|l|l|l|l|l|l|l|l|l|l|l}
Global North & Global South & LDC & LLDC & SIDS & G20 & Emerging Markets & OPEC & Low Income & Lower middle Income & Upper middle Income & High Income \\
\hline
Albania & Fiji & Yemen, Rep. & Afghanistan & Antigua and Barbuda & Australia & Bangladesh & Algeria & Afghanistan & Angola & Albania & Antigua and Barbuda \\
Austria & Micronesia, Fed. Sts. & Afghanistan & Armenia & Bahamas, The & Canada & Egypt, Arab Rep. & Angola & Benin & Bangladesh & Algeria & Australia \\
Belarus & Tonga & Burundi & Azerbaijan & Barbados & Saudi Arabia & Indonesia & Equatorial Guinea & Burkina Faso & Bhutan & Argentina & Austria \\
Belgium & Vanuatu & Angola & Bhutan & Belize & United States & Iran, Islamic Rep. & Gabon & Burundi & Bolivia & Armenia & Bahamas, The \\
Bosnia and Herzegovina & Tuvalu & Benin & Bolivia & Comoros & India & Mexico & Iran, Islamic Rep. & Central African Republic & Cambodia & Azerbaijan & Bahrain \\
Bulgaria & Solomon Islands & Mozambique & Botswana & Cuba & Russian Federation & Nigeria & Iraq & Chad & Cameroon & Belarus & Barbados \\
Croatia & Samoa & Burkina Faso & Burkina Faso & Dominica & South Africa & Pakistan & Kuwait & Congo, Dem. Rep. & Comoros & Belize & Belgium \\
Cyprus & Papua New Guinea & Niger & Burundi & Dominican Republic & Turkey & Philippines & Libya & Eritrea & Congo, Rep. & Bosnia and Herzegovina & Canada \\
Czech Republic & Palau & Central African Republic & Central African Republic & Fiji & Argentina & Turkey & Nigeria & Ethiopia & Cote d'Ivoire & Botswana & Chile \\
Denmark & Kiribati & Chad & Chad & Grenada & Brazil & Korea, Dem. People's Rep. & Saudi Arabia & Gambia & Djibouti & Brazil & Croatia \\
Estonia & Bangladesh & Lesotho & Ethiopia & Guinea-Bissau & Mexico & Vietnam & United Arab Emirates & Guinea & Egypt, Arab Rep. & Bulgaria & Malta \\
Finland & Bhutan & Liberia & Kazakhstan & Guyana & France & Brazil & Congo, Dem. Rep. & Guinea-Bissau & El Salvador & China & Cyprus \\
France & Cambodia & Congo, Dem. Rep. & Kyrgyz Republic & Haiti & Germany & Russian Federation & Venezuela, RB & Haiti & Ghana & Colombia & Czech Republic \\
Greece & China & Djibouti & Lao PDR & Jamaica & Italy & India &  & Liberia & Honduras & Costa Rica & Denmark \\
Germany & India & Togo & Lesotho & Kiribati & United Kingdom & China &  & Madagascar & India & Cuba & Estonia \\
Greenland & Indonesia & Equatorial Guinea & Malawi & Maldives & China & South Africa &  & Malawi & Indonesia & Dominica & Finland \\
Hungary & Lao PDR & Eritrea & Mali & Mauritius & Indonesia &  &  & Mali & Kenya & Dominican Republic & France \\
Iceland & Malaysia & Ethiopia & Moldova & Palau & Japan &  &  & Mozambique & Kiribati & Ecuador & Germany \\
Ireland & Myanmar & Gambia & Mongolia & Papua New Guinea & Korea, Dem. People's Rep. &  &  & Nepal & Kyrgyz Republic & Equatorial Guinea & Greece \\
Italy & Mongolia & Madagascar & Nepal & Puerto Rico &  &  &  & Niger & Lao PDR & Fiji & Greenland \\
Latvia & Nepal & Malawi & Niger & Samoa &  &  &  & Rwanda & Lesotho & Gabon & Hungary \\
Liechtenstein & Pakistan & Mali & Paraguay & Sao Tome and Principe &  &  &  & Sierra Leone & Mauritania & Georgia & Iceland \\
Lithuania & Philippines & Rwanda & Rwanda & Seychelles &  &  &  & Somalia & Moldova & Grenada & Ireland \\
Luxembourg & Sri Lanka & Senegal & South Sudan & Singapore &  &  &  & South Sudan & Mongolia & Guatemala & Israel \\
Malta & Thailand & Sierra Leone & Tajikistan & Solomon Islands &  &  &  & Syrian Arab Republic & Morocco & Guyana & Italy \\
Montenegro & Timor-Leste & Mauritania & Turkmenistan & Suriname &  &  &  & Tajikistan & Myanmar & Iran, Islamic Rep. & Japan \\
Netherlands & Vietnam & Guinea-Bissau & Uganda & Timor-Leste &  &  &  & Tanzania & Nicaragua & Iraq & Korea, Dem. People's Rep. \\
Norway & Maldives & Guinea & Uzbekistan & Tuvalu &  &  &  & Togo & Nigeria & Jamaica & Kuwait \\
Poland & Grenada & Comoros & Zambia & Vanuatu &  &  &  & Uganda & Pakistan & Jordan & Latvia \\
Portugal & Dominica & Sao Tome and Principe & Zimbabwe &  &  &  &  & Yemen, Rep. & Papua New Guinea & Kazakhstan & Liechtenstein \\
Romania & Barbados & Zambia &  &  &  &  &  &  & Philippines & Lebanon & Lithuania \\
Serbia & Antigua and Barbuda & Uganda &  &  &  &  &  &  & Sao Tome and Principe & Libya & Luxembourg \\
Slovakia & Cuba & Tanzania &  &  &  &  &  &  & Senegal & Malaysia & Netherlands \\
Slovenia & Bahamas, The & South Sudan &  &  &  &  &  &  & Solomon Islands & Maldives & New Zealand \\
Spain & Puerto Rico & Sudan &  &  &  &  &  &  & Sudan & Mauritius & Norway \\
Sweden & Jamaica & Bhutan &  &  &  &  &  &  & Timor-Leste & Mexico & Oman \\
Switzerland & Algeria & Cambodia &  &  &  &  &  &  & Tunisia & Montenegro & Palau \\
Ukraine & Angola & Bangladesh &  &  &  &  &  &  & Ukraine & Namibia & Panama \\
United Kingdom & Benin & Haiti &  &  &  &  &  &  & Uzbekistan & Paraguay & Poland \\
Canada & Botswana & Kiribati &  &  &  &  &  &  & Vanuatu & Peru & Portugal \\
United States & Burkina Faso & Lao PDR &  &  &  &  &  &  & Vietnam & Romania & Puerto Rico \\
Azerbaijan & Cameroon & Myanmar &  &  &  &  &  &  & Zambia & Russian Federation & Qatar \\
Georgia & Central African Republic & Nepal &  &  &  &  &  &  & Zimbabwe & Samoa & Saudi Arabia \\
Israel & Chad & Vanuatu &  &  &  &  &  &  &  & Serbia & Seychelles \\
Russian Federation & Congo & Tuvalu &  &  &  &  &  &  &  & South Africa & Singapore \\
Turkey & Cote d'Ivoire & Solomon Islands &  &  &  &  &  &  &  & Sri Lanka & Slovak Republic \\
Australia & Congo, Dem. Rep. & Timor-Leste &  &  &  &  &  &  &  & Suriname & Slovenia \\
New Zealand & Djibouti &  &  &  &  &  &  &  &  & Thailand & Spain \\
Korea, Dem. People's Rep. & Egypt, Arab Rep. &  &  &  &  &  &  &  &  & Tonga & Sweden \\
Japan & Equatorial Guinea &  &  &  &  &  &  &  &  & Turkey & Switzerland \\
Singapore & Eritrea &  &  &  &  &  &  &  &  & Turkmenistan & Trinidad and Tobago \\
 & Ethiopia &  &  &  &  &  &  &  &  & Tuvalu & United Arab Emirates \\
 & Gabon &  &  &  &  &  &  &  &  & Venezuela, RB & United Kingdom \\
 & Gambia, The &  &  &  &  &  &  &  &  &  & United States \\
 & Ghana &  &  &  &  &  &  &  &  &  & Uruguay \\
 & Kenya &  &  &  &  &  &  &  &  &  &  \\
 & Lesotho &  &  &  &  &  &  &  &  &  &  \\
 & Liberia &  &  &  &  &  &  &  &  &  &  \\
 & Libya &  &  &  &  &  &  &  &  &  &  \\
 & Madagascar &  &  &  &  &  &  &  &  &  &  \\
 & Malawi &  &  &  &  &  &  &  &  &  &  \\
 & Mali &  &  &  &  &  &  &  &  &  &  \\
 & Morocco &  &  &  &  &  &  &  &  &  &  \\
 & Mozambique &  &  &  &  &  &  &  &  &  &  \\
 & Namibia &  &  &  &  &  &  &  &  &  &  \\
 & Niger &  &  &  &  &  &  &  &  &  &  \\
 & Nigeria &  &  &  &  &  &  &  &  &  &  \\
 & Rwanda &  &  &  &  &  &  &  &  &  &  \\
 & Senegal &  &  &  &  &  &  &  &  &  &  \\
 & Sierra Leone &  &  &  &  &  &  &  &  &  &  \\
 & Somalia &  &  &  &  &  &  &  &  &  &  \\
 & South Africa &  &  &  &  &  &  &  &  &  &  \\
 & South Sudan &  &  &  &  &  &  &  &  &  &  \\
 & Sudan &  &  &  &  &  &  &  &  &  &  \\
 & Syrian Arab Republic &  &  &  &  &  &  &  &  &  &  \\
 & Togo &  &  &  &  &  &  &  &  &  &  \\
 & Tunisia &  &  &  &  &  &  &  &  &  &  \\
 & Uganda &  &  &  &  &  &  &  &  &  &  \\
 & Tanzania &  &  &  &  &  &  &  &  &  &  \\
 & Zambia &  &  &  &  &  &  &  &  &  &  \\
 & Zimbabwe &  &  &  &  &  &  &  &  &  &  \\
 & Seychelles &  &  &  &  &  &  &  &  &  &  \\
 & Sao Tome and Principe &  &  &  &  &  &  &  &  &  &  \\
 & Mauritius &  &  &  &  &  &  &  &  &  &  \\
 & Mauritania &  &  &  &  &  &  &  &  &  &  \\
 & Guinea-Bissau &  &  &  &  &  &  &  &  &  &  \\
 & Guinea &  &  &  &  &  &  &  &  &  &  \\
 & Comoros &  &  &  &  &  &  &  &  &  &  \\
 & Burundi &  &  &  &  &  &  &  &  &  &  \\
 & Belize &  &  &  &  &  &  &  &  &  &  \\
 & Bahamas, The &  &  &  &  &  &  &  &  &  &  \\
 & Argentina &  &  &  &  &  &  &  &  &  &  \\
 & Bolivia &  &  &  &  &  &  &  &  &  &  \\
 & Brazil &  &  &  &  &  &  &  &  &  &  \\
 & Chile &  &  &  &  &  &  &  &  &  &  \\
 & Colombia &  &  &  &  &  &  &  &  &  &  \\
 & Costa Rica &  &  &  &  &  &  &  &  &  &  \\
 & Cuba &  &  &  &  &  &  &  &  &  &  \\
 & Dominican Republic &  &  &  &  &  &  &  &  &  &  \\
 & Ecuador &  &  &  &  &  &  &  &  &  &  \\
 & El Salvador &  &  &  &  &  &  &  &  &  &  \\
 & Guatemala &  &  &  &  &  &  &  &  &  &  \\
 & Haiti &  &  &  &  &  &  &  &  &  &  \\
 & Honduras &  &  &  &  &  &  &  &  &  &  \\
 & Jamaica &  &  &  &  &  &  &  &  &  &  \\
 & Mexico &  &  &  &  &  &  &  &  &  &  \\
 & Panama &  &  &  &  &  &  &  &  &  &  \\
 & Paraguay &  &  &  &  &  &  &  &  &  &  \\
 & Peru &  &  &  &  &  &  &  &  &  &  \\
 & Puerto Rico &  &  &  &  &  &  &  &  &  &  \\
 & Suriname &  &  &  &  &  &  &  &  &  &  \\
 & Trinidad and Tobago &  &  &  &  &  &  &  &  &  &  \\
 & Uruguay &  &  &  &  &  &  &  &  &  &  \\
 & Venezuela, RB &  &  &  &  &  &  &  &  &  &  \\
 & Nicaragua &  &  &  &  &  &  &  &  &  &  \\
 & Guyana &  &  &  &  &  &  &  &  &  &  \\
 & Grenada &  &  &  &  &  &  &  &  &  &  \\
 & Dominica &  &  &  &  &  &  &  &  &  &  \\
 & Barbados &  &  &  &  &  &  &  &  &  &  \\
 & Antigua and Barbuda &  &  &  &  &  &  &  &  &  &  \\
 & Iraq &  &  &  &  &  &  &  &  &  &  \\
 & Afghanistan &  &  &  &  &  &  &  &  &  &  \\
 & Armenia &  &  &  &  &  &  &  &  &  &  \\
 & Bahrain &  &  &  &  &  &  &  &  &  &  \\
 & Iran, Islamic Rep. &  &  &  &  &  &  &  &  &  &  \\
 & Jordan &  &  &  &  &  &  &  &  &  &  \\
 & Kazakhstan &  &  &  &  &  &  &  &  &  &  \\
 & Kuwait &  &  &  &  &  &  &  &  &  &  \\
 & Kyrgyz Republic &  &  &  &  &  &  &  &  &  &  \\
 & Lebanon &  &  &  &  &  &  &  &  &  &  \\
 & Oman &  &  &  &  &  &  &  &  &  &  \\
 & Qatar &  &  &  &  &  &  &  &  &  &  \\
 & Saudi Arabia &  &  &  &  &  &  &  &  &  &  \\
 & Tajikistan &  &  &  &  &  &  &  &  &  &  \\
 & Turkmenistan &  &  &  &  &  &  &  &  &  &  \\
 & United Arab Emirates &  &  &  &  &  &  &  &  &  &  \\
 & Uzbekistan &  &  &  &  &  &  &  &  &  &  \\
 & Yemen, Rep. &  &  &  &  &  &  &  &  &  & 
\end{tabular}}
\tiny
\begin{tablenotes}
\centering
\item{LDC: Least Developed Countries}
\item{LLDC: Land Locked Developing Countries}
\item{SIDS: Small Island Developing States}
\item{Emerging Markets: BRICS + N-11}
\end{tablenotes}

\end{table}

\end{document}